\documentclass[pdflatex,sn-nature]{sn-jnl}

\usepackage{graphicx}%
\usepackage{graphicx}%
\usepackage{multirow}%
\usepackage{amsmath,amssymb,amsfonts}%
\usepackage{amsthm}%
\usepackage{mathrsfs}%
\usepackage[title]{appendix}%
\usepackage{xcolor}%
\usepackage{textcomp}%
\usepackage{manyfoot}%
\usepackage{booktabs}%
\usepackage{algorithm}%
\usepackage{algorithmicx}%
\usepackage{algpseudocode}%
\usepackage{listings}%
\usepackage{lineno}
\usepackage{booktabs}
\usepackage{subcaption}
\usepackage{microtype}
\usepackage{array}
\usepackage{float}
\usepackage{url}
\usepackage{hyperref}
\usepackage{tabularx}
\usepackage{setspace}
\usepackage{caption}
\usepackage{svg}
\usepackage{fancyvrb}
\usepackage{fvextra}

\usepackage{pifont}
\newcommand{\cmark}{\textcolor{pos}{\ding{51}}}
\newcommand{\xmark}{\textcolor{neg}{\ding{55}}}

\hypersetup{
    colorlinks=true,
    linkcolor=blue!60!black,
    citecolor=green!50!black,
    urlcolor=blue!60!black,
}

\modulolinenumbers[5]

\definecolor{pos}{RGB}{0,120,60}
\definecolor{neg}{RGB}{180,0,0}
\newcommand{\gain}[1]{{+#1}}

\begin{document}


\title[WILSON]{WILSON -- Multi-magnification composites as a foundation for \\patient-level pathology analysis and diagnostic text generation}

\author[1]{\fnm{Saghir} \sur{Alfasly}}\equalcont{The first and second authors contributed equally to this work}
\author[1]{\fnm{Wataru} \sur{Uegami}}\equalcont{The first and second authors contributed equally to this work}
\author[1]{\fnm{Sobhan} \sur{Hemati}}
\author[2]{\fnm{Wenchao} \sur{Han}}
\author[3]{\fnm{Xiaojia} \sur{Tang}}
\author[3]{\fnm{Kevin} \sur{Thompson}}
\author[1]{\fnm{Daniel} \sur{Stone}}
\author[1]{\fnm{Ghazal} \sur{Alabtah}}
\author[2]{\fnm{Chris} \sur{Hartley}}
\author[2]{\fnm{Chady} \sur{Meroueh}}
\author[2]{\fnm{John C.} \sur{Cheville}}
\author[2]{\fnm{Thomas J.} \sur{Flotte}}
\author[2]{\fnm{Saba} \sur{Yasir}}
\author[4]{\fnm{Michael R.} \sur{Lucas}}
\author[4]{\fnm{Eric W.} \sur{Klee}}
\author[4]{\fnm{Cheryl L.} \sur{Willman}}
 \author[5]{\fnm{Judy C.} \sur{Boughey}}
 \author[6]{\fnm{Matthew P.} \sur{Goetz}}
 \author[3]{\fnm{Krishna R.} \sur{Kalari}}
 \author[1,2]{\fnm{H.R.} \sur{Tizhoosh}}\email{tizhoosh.hamid@mayo.edu}

\affil[1]{\orgdiv{KIMIA Lab, Department of AI \& Informatics}, \orgname{Mayo Clinic}, \orgaddress{\city{Rochester}, \state{MN}, \country{USA}}}
\affil[2]{\orgdiv{Department of Laboratory Medicine and Pathology}, \orgname{Mayo Clinic}, \orgaddress{\city{Rochester}, \state{MN}, \country{USA}}}
\affil[3]{\orgdiv{Department of Quantitative Health Sciences}, \orgname{Mayo Clinic}, \orgaddress{\city{Rochester}, \state{MN}, \country{USA}}}
\affil[4]{Mayo Clinic Comprehensive Cancer Center, Mayo Clinic, Rochester, MN, USA}
\affil[5]{\orgdiv{Division of Breast and Melanoma Surgical Oncology, Department of Surgery}, \orgname{Mayo Clinic}, \orgaddress{\city{Rochester}, \state{MN}, \country{USA}}}
\affil[6]{\orgdiv{Department of Oncology}, \orgname{Mayo Clinic}, \orgaddress{\city{Rochester}, \state{MN}, \country{USA}}}


\abstract{Pathologists examine tissue sections across multiple magnifications and slides, integrating morphological information to establish a diagnosis. In contrast, most pathology foundation models encode thousands of image patches from a single representative (``best'') slide and aggregate their features, limiting multiscale and patient-level contextual integration. Here we introduce WILSON, a vision–language foundation model that represents whole-slide images and multi-slide cases through multi-magnification composites. Trained on approximately 189,000 slides spanning 42 organs and 838 diagnostic entities, WILSON enables patient-level analysis, multimodal retrieval and diagnostic text generation. Without task-specific training, WILSON outperformed a dedicated case-level model (macro-F1, 0.52 versus 0.38) and matched substantially larger slide-level models while requiring 400–1,350 times fewer computational operations. End-to-end fine-tuning improved breast cancer subtyping and stromal lymphocyte grading. WILSON also surpassed PRISM in image–text retrieval (recall@1, 75.6\% versus 58.1\%) and generated diagnostic descriptions that received higher ratings from five board-certified pathologists than competing models. These findings demonstrate that compact, multiscale representations can support computationally efficient, clinically grounded pathology foundation models.}


\maketitle

\begin{figure}[H]
\centering
\includegraphics[width=\textwidth]{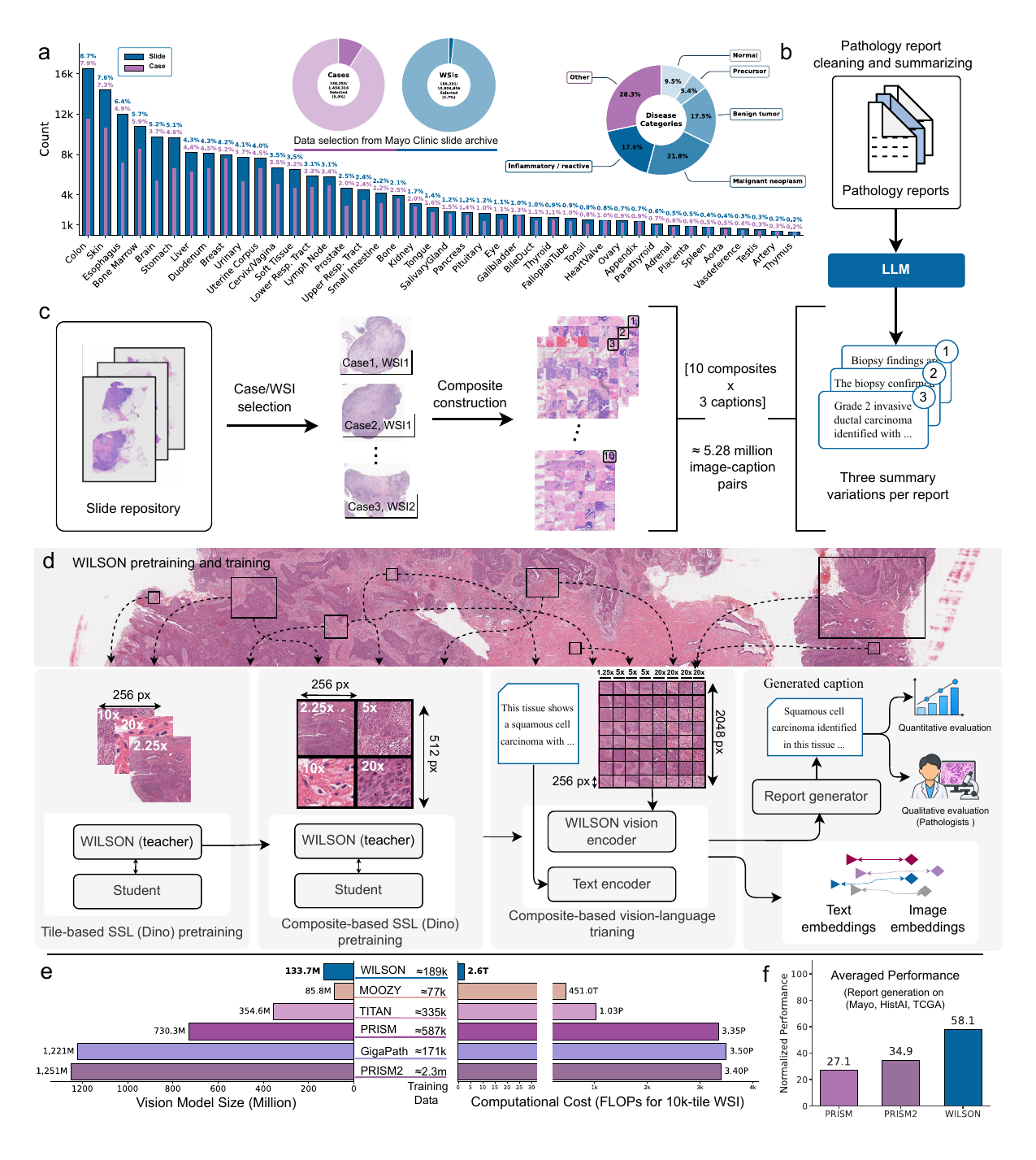}
\caption{\textbf{Overview of WILSON.} \textbf{a}, Slides and cases per organ (bars), with the cohort further grouped by organ and by diagnostic category (donuts). \textbf{b}, An LLM cleans each pathology report and generates three paraphrased caption variants per report. \textbf{c}, Composites are tiled from selected WSIs; 178,020 WSIs have both a rendered composite and a caption record, giving 1,759,787 composites ($\approx$9.9/WSI) and $1{,}759{,}787\times3=5{,}279{,}361$ ($\approx$5.28M) image--caption pairs. \textbf{d}, WILSON pretraining and training: tile-based SSL on $256\times256$ patches at $2.25\times/10\times/20\times$ $\rightarrow$ composite-based SSL on $512\times512$ four-tile composites $\rightarrow$ vision--language alignment of $2048\times2048$ WSI composites against report text (SigLIP distillation from Gemini embeddings, CLIP, and keyword regression) $\rightarrow$ CoCa-style report generator, a multimodal decoder that cross-attends over pooled WILSON visual tokens to generate a caption autoregressively. Generated captions are evaluated quantitatively (\textbf{f}) and qualitatively by pathologists. \textbf{e}, Vision-model size (tile plus slide encoder), pretraining data (WSIs) and computational cost per WSI (forward-pass FLOPs for a $22{,}400\times22{,}400$~px tissue region at $20\times$, tiled with each model's own tile size). WILSON needs 2.6~TFLOPs per WSI, $\approx$170$\times$ fewer than the next most efficient model (MOOZY) and $>$1,000$\times$ fewer than PRISM, PRISM2 and Prov-GigaPath. \textbf{f}, Report-generation performance averaged over three cohorts (Mayo, HistAI, TCGA), as a normalized score: WILSON (58.1) exceeds PRISM2 (34.9) by 23.2 points while using $\approx$9$\times$ fewer parameters and $\approx$1,300$\times$ fewer FLOPs per WSI.}
\label{fig:wilson}
\end{figure}

Histopathological diagnosis requires integration of morphological information across spatial scales and, frequently, across multiple whole-slide images (WSIs) from the same patient case. However, most computational pathology foundation models represent WSIs by independently encoding large numbers of small image patches with a tile-level encoder~\cite{chen2024uni, xu2024gigapath} and subsequently aggregating their features into slide-level representations, typically through multiple-instance learning~\cite{ilse2018abmil, lu2021clam, shao2021transmil, ding2025titan, shaikovski2024prism, kotp2026moozy}. Although this strategy has enabled substantial advances in WSI analysis, it introduces considerable computational overhead, separates local morphology from the broader tissue context during image encoding~\cite{chen2022hipt, wang2022ctranspath}, and typically treats the individual WSI rather than the multi-slide case as the primary unit of representation. These characteristics differ from clinical pathology practice, in which diagnostic interpretation integrates multiscale morphological findings across complementary tissue sections and slides, and have led to a re-evaluation of whether patch-and-aggregate pipelines are the most appropriate foundation for pathology~\cite{tizhoosh2026rethinking, mulliqi2025foundation}.\\
\indent Vision--language foundation models offer an additional opportunity to ground histopathological representations in the diagnostic semantics contained in pathology reports~\cite{radford2021learning, huang2023visual, lu2024conch, lu2024pathchat, sun2024pathgen}. Yet establishing direct correspondence between diagnostic language and gigapixel WSIs remains challenging because conventional pipelines operate through patch-level representations and subsequent aggregation, which separates the visual encoder from the report text by an intermediate aggregation stage~\cite{ding2025titan, shaikovski2024prism}. A representation that captures diagnostically informative morphology across magnifications and multiple WSIs within a unified visual input could therefore enable image--language learning at the case level while reducing dependence on computationally intensive patch-wise processing.\\
\indent Here we introduce WILSON, named in honor of Louis B. Wilson, a pioneering Mayo Clinic pathologist whose work on rapid frozen-section diagnosis helped shape modern surgical pathology \cite{Mayo1929,Dahlin1980,Gal2005}. WILSON (Fig.~\ref{fig:wilson}) is a pathology foundation model that represents single WSI and multi-WSI cases as multi-magnification composite images and processes each composite as a unified visual input. The composites integrate tissue regions sampled across spatial scales and, for case-level analysis, across multiple WSIs, enabling direct learning of WSI- and case-level representations without a separate patch-feature aggregation stage. WILSON is trained using diagnostic information derived from pathology reports, directly aligning visual representations with clinically meaningful text. This formulation also enables end-to-end adaptation of the visual encoder for downstream WSI tasks rather than restricting adaptation to an aggregator operating on frozen tile features.\\
\indent We trained WILSON using approximately 189k WSIs spanning 42 organs and 838 diagnostic entities and evaluated it across internal and external cohorts encompassing case- and WSI-level disease classification, molecular and treatment-related prediction, image--text retrieval and pathology-text generation. We show that composite images retain diagnostically relevant information compared with conventional patch-based representations, that WILSON is competitive with or outperforms specialized case- and WSI-level foundation models across multiple tasks, and that its unified representation substantially reduces computational requirements while permitting end-to-end fine-tuning. In addition, we demonstrate bidirectional retrieval between histopathology and diagnostic text and generation of pathology descriptions from composite images. Together, these findings establish multi-magnification composite representation as an alternative approach to foundation modeling that brings the computational unit of pathology AI closer to the multiscale, multi-slide structure of clinical diagnostic reasoning.

\section{Results}

\subsection{Multi-magnification composites preserve diagnostically relevant information from gigapixel whole-slide images}
\begin{figure}[H]
    \centering
    \includegraphics[width=0.94\linewidth]{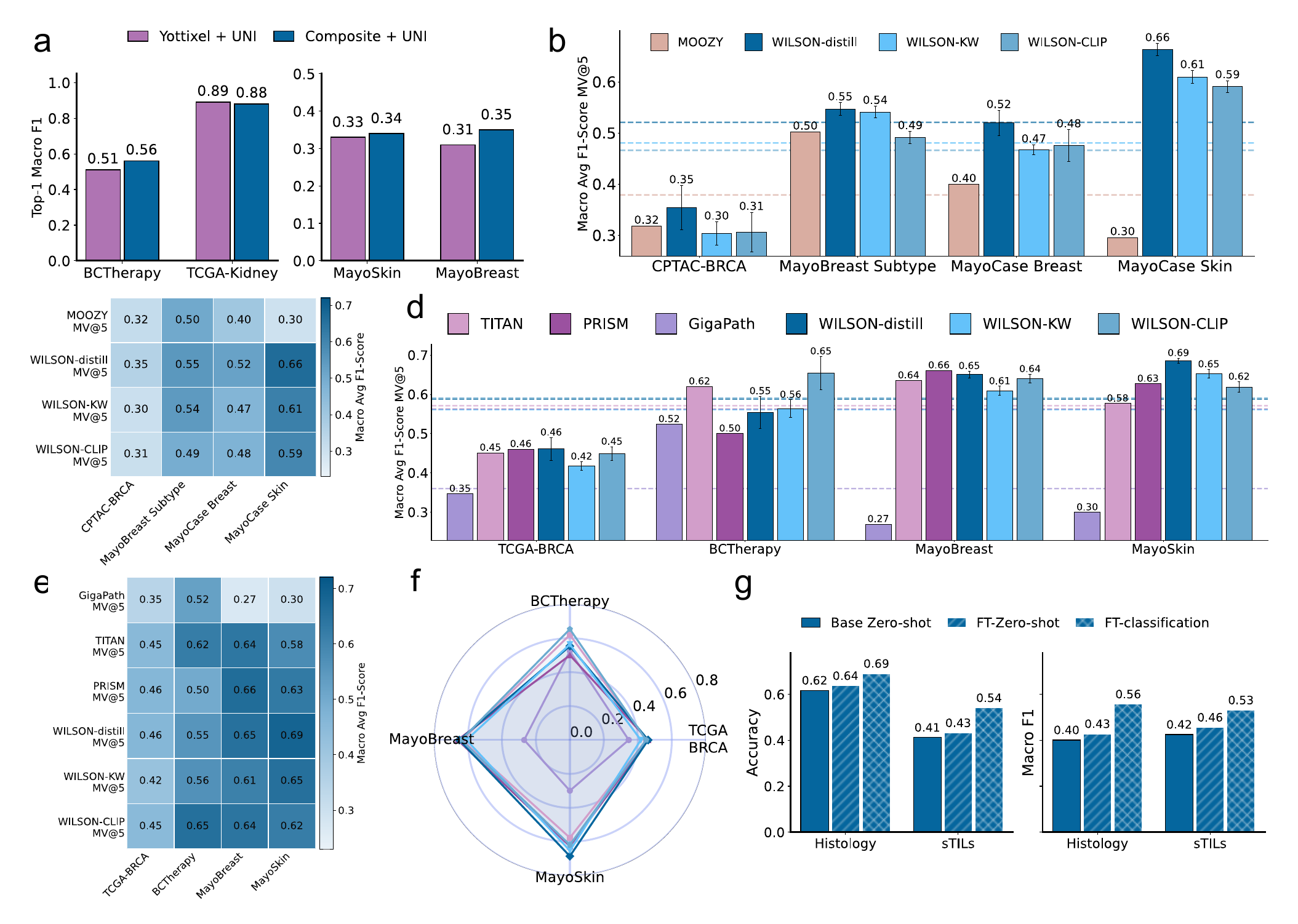}
    \caption{\textbf{WILSON matches or exceeds specialized case- and WSI-level
baselines and benefits from efficient end-to-end fine-tuning.} \textbf{a}, Multi-magnification composites versus Yottixel tile selection as alternative WSI representations, both encoded with the pretrained UNI foundation model. Top-1 macro-averaged F1 retrieval is shown for four WSI-level benchmarks, split across two sub-axes with different $y$-ranges: BCTherapy and TCGA-Kidney (left, $0$--$1$) and MayoSkin and MayoBreast (right, $0$--$0.5$). Bars are single point estimates. \textbf{b}, Case-level zero-shot retrieval performance (macro-averaged F1 under top-5 majority voting, MV@5) of three WILSON variants (WILSON-distill, WILSON-KW, WILSON-CLIP) against MOOZY, a case-level pathology foundation model, on four case-level benchmarks: CPTAC-BRCA, MayoBreastSubtype, MayoCaseBreast and MayoCaseSkin. Error bars indicate the standard deviation across the 10 composite images generated per case; MOOZY operates directly on the WSI and therefore has no associated variance. Color-matched horizontal dashed lines mark each model's mean across the four benchmarks. \textbf{c}, The same values as in \textbf{b} shown as a model~$\times$~benchmark heatmap (rows, models; columns, benchmarks; shading, macro-averaged F1). \textbf{d}, WSI-level zero-shot MV@5 retrieval performance of the same three WILSON variants against three WSI-level foundation models (GigaPath, TITAN, PRISM) on four WSI-level benchmarks: TCGA-BRCA, BCTherapy, MayoBreast and MayoSkin. Error bars indicate the standard deviation across the 10 composite images generated per WSI; the three baselines operate on frozen tile-level features and therefore have no associated variance. Color-matched dashed lines mark each model's mean across the four benchmarks. \textbf{e}, The same values as in \textbf{d} shown as a model~$\times$~benchmark heatmap. \textbf{f}, The same six models and four WSI-level benchmarks shown as a radar plot (BCTherapy, top; TCGA-BRCA, right; MayoSkin, bottom; MayoBreast, left; radial axis, macro-averaged F1 from $0.0$ to $0.8$), with colors as in \textbf{d}. \textbf{g}, Effect of end-to-end fine-tuning of WILSON on MayoTNBC, a cohort of 508 triple-negative breast cancer cases \cite{LeonFerre2018}, evaluated on histological subtyping (Histology) and stromal tumor-infiltrating lymphocyte (sTILs) grading. Accuracy (left) and macro-averaged F1 (right) are shown for three settings: the pretrained backbone evaluated zero-shot (Base Zero-shot), the same backbone after end-to-end fine-tuning and evaluated zero-shot with no task-specific head (FT-Zero-shot), and the fine-tuned backbone trained jointly with a task-specific classification head (FT-classification).}
    \label{fig:resultsCombined}
\end{figure}

Whole-slide images (WSIs) frequently contain billions of pixels, making direct end-to-end processing with contemporary vision encoders computationally challenging. We therefore asked whether diagnostically informative regions sampled across multiple magnifications could be assembled into a single, computationally manageable composite image while preserving the information required for downstream WSI analysis. Unlike conventional tile-based approaches, which independently encode hundreds to thousands of image tiles followed by feature aggregation on top of frozen embeddings, the composite representation presents tissue regions sampled across multiple spatial scales to an image encoder as a single visual input.\\
\indent To evaluate the composite representation independently of WILSON training, we used the pretrained UNI foundation model~\cite{chen2024uni}, which had not been trained on composite images, to extract embeddings from two alternative WSI representations: multi-magnification composites and tiles selected using the Yottixel framework~\cite{kalra2020yottixel}, a tile-based WSI representation method. We compared the resulting representations using top-1 retrieval across four benchmarks encompassing diagnostic classification (MayoSkin, MayoBreast, TCGA-Kidney) and treatment-response prediction (BCTherapy~\cite{sammut2022multiomic}), including two public datasets and two internal Mayo Clinic cohorts (Fig.~\ref{fig:resultsCombined}a, Table~\ref{tab:compositeVsYottixel}). Note: WILSON-distill (contrastive alignment of composite images with  embeddings of pathology report summaries), WILSON-KW (prediction of diagnostic keyword vectors from composite images), and WILSON-CLIP (joint contrastive learning of composite images and report summaries).

\indent Despite representing each WSI through a compact composite image, performance remained closely matched to patch-based WSI indexing across all four benchmarks, with absolute differences of $\leq 0.05$ in macro-F1. Composite representations performed better on three benchmarks (BCTherapy: 0.56 versus\ 0.51; MayoSkin: 0.34 versus\ 0.33; MayoBreast: 0.35 versus\ 0.31) and slightly lower on one cohort (TCGA-Kidney: 0.88 versus\ 0.89). Thus, across diagnostically and clinically distinct tasks, multi-magnification composites preserved task-relevant information at a level comparable to patch-based WSI representations, while being processed as a single visual input without requiring patch-level feature aggregation.

\subsection{WILSON enables multi-slide case-level diagnostic retrieval}
A central motivation for the composite-image framework is its
ability to integrate tissue regions sampled from multiple WSIs belonging to the
same patient into a unified case-level representation 
This extends the computational unit from an individual slide to the patient case,
reflecting the integration of information across multiple tissue sections and
slides during pathological assessment. To evaluate whether WILSON can learn
diagnostically informative representations at this level, we benchmarked the
three WILSON variants (WILSON-distill, WILSON-KW and WILSON-CLIP) against
MOOZY~\cite{kotp2026moozy}, a case-level pathology foundation model, on four
case-classification tasks spanning an external cohort (CPTAC-BRCA,
PAM50~\cite{parker2009pam50} molecular subtyping) and three internal Mayo Clinic
cohorts (MayoCaseBreast, histological subtype; MayoBreastSubtype,
breast tumor molecular subtype classification; and MayoCaseSkin, skin histological diagnosis). Performance was
assessed using zero-shot top-5 majority-voting retrieval, with macro-averaged F1
computed across the 10 composite images generated per case to quantify
variability (Fig.~\ref{fig:resultsCombined}b,c, \ref{fig:retrieval_radar}).\\
\indent WILSON exceeded MOOZY on all three internal cohorts. The largest margin
was observed on MayoCaseSkin, where WILSON-distill reached a macro-F1 of 0.66
compared with 0.30 for MOOZY (\gain{0.36}) and every WILSON variant scored at
least 0.59, followed by MayoCaseBreast (0.52 versus\ 0.40, \gain{0.12}) and
MayoBreastSubtype (0.55 versus\ 0.50, \gain{0.05}). Averaged across the four
benchmarks, WILSON-distill attained 0.52 against 0.38 for MOOZY (\gain{0.14}),
with WILSON-KW at 0.48 and WILSON-CLIP at 0.47. WILSON-distill, supervised using
dense Gemini report embeddings, was the strongest variant on three of the four
tasks, indicating that a richer text-side supervision signal transfers to
case-level retrieval more effectively than keyword regression or contrastive
alignment alone.\\
\indent Performance on the external CPTAC-BRCA cohort was both lower in absolute
terms and more variable across composites for all models. WILSON-distill scored
slightly above MOOZY (0.35 versus\ 0.32), whereas WILSON-KW (0.30) and WILSON-CLIP
(0.31) fell marginally below it; in each case the difference was within the
composite-to-composite standard deviation and should not be read as a meaningful
separation. This is the only task on the panel that requires predicting a
molecular class from morphology alone in an external cohort, and the results
indicate that such transfer is less consistent than transfer across internal 
diagnostic cohorts. Collectively, these results show that histopathological
information distributed between multiple WSIs can be integrated within a unified
composite representation that supports case-level diagnostic retrieval across
histological and dermatopathological classification tasks without task-specific
fine-tuning.

\subsection{WILSON matches dedicated WSI-level foundation models with substantially lower computational requirements}

Case-level retrieval evaluates WILSON at the representational level for which it was primarily designed. Next, we ask whether the same composite representation remains competitive at the conventional single-WSI level. We benchmarked the three WILSON variants (WILSON-distill, WILSON-KW and WILSON-CLIP) against three published WSI-level foundation models---Prov-GigaPath~\cite{xu2024gigapath}, TITAN~\cite{ding2025titan} and PRISM~\cite{shaikovski2024prism}---on four benchmarks spanning two public cohorts (TCGA-BRCA, breast histological subtype; BCTherapy, breast treatment response prediction) and two internal Mayo Clinic cohorts (MayoBreast and MayoSkin, histological subtype). All models were evaluated using the same zero-shot top-5 majority-voting retrieval protocol. For WILSON, macro-averaged F1 was computed across the 10 composites generated per WSI to quantify variability; the baseline models operate on frozen tile-level representations and therefore have no corresponding composite-to-composite variance (Fig.~\ref{fig:resultsCombined}d--f).\\
\indent Across the three diagnostic benchmarks, WILSON-distill achieved the highest mean macro-F1 (0.60), compared with 0.58 for PRISM, 0.57 for WILSON-CLIP and 0.56 for both TITAN and WILSON-KW. On the external TCGA-BRCA cohort, WILSON-distill matched PRISM (0.46) and performed similarly to TITAN (0.45). On MayoBreast, WILSON-distill reached 0.65, compared with 0.66 for PRISM and 0.64 for TITAN and WILSON-CLIP. The largest separation was observed on MayoSkin, a 20-class dermatopathology task, where WILSON-distill achieved 0.69 compared with 0.63 for PRISM and 0.58 for TITAN, and all three WILSON variants outperformed the three WSI-level baselines. On BCTherapy, WILSON-CLIP achieved the highest macro-F1 (0.65), compared with 0.62 for TITAN and 0.50 for PRISM, although variability across composites was greater on this relatively small cohort (160 WSIs).\\
\indent Performance across datasets also differed among the published WSI-level models. Prov-GigaPath achieved lower performance on three of the four benchmarks, including macro-F1 values of 0.27 and 0.30 on MayoBreast and MayoSkin, respectively. TITAN and PRISM transferred more consistently to the internal cohorts, although neither uniformly outperformed WILSON. Collectively, these results indicate that WILSON's composite representation remains competitive with dedicated WSI-level foundation models across both public and institutional cohorts, despite replacing thousands of independently encoded tiles and subsequent feature aggregation with a single composite input.\\
\indent This comparable performance is achieved with substantially lower computational requirements. WILSON represents each WSI as a single $2048\times2048$ composite processed in one forward pass and uses 2.7--9.1$\times$ fewer parameters than the three WSI-level baselines, together with two to three orders of magnitude fewer FLOPs per WSI (Section~\ref{sec:efficiency}). Thus, in all benchmarks evaluated here, explicit tile-wise encoding and feature aggregation were not necessary to achieve competitive slide-level retrieval performance. The compact composite representation therefore provides an efficient alternative for WSI-level representation while retaining the ability to adapt the complete visual encoder end-to-end, which we evaluate next.

\subsection{WILSON enables end-to-end adaptation for downstream whole-slide image tasks}
\label{sec:wsi_finetune}

Most WSI-level foundation models use a two-stage architecture in which a tile encoder generates representations for thousands of image patches and a separate slide-level aggregator operates on the resulting features. In practice, downstream adaptation commonly relies on frozen tile representations because propagating gradients through thousands of tiles per WSI makes end-to-end fine-tuning computationally challenging at WSI scale~\cite{mulliqi2025foundation,tizhoosh2026rethinking}. WILSON instead represents each WSI as a single composite image processed by a unified visual encoder, making it feasible to adapt the complete visual backbone end-to-end.\\
\indent We evaluated this capability on MayoTNBC, a cohort of 508 triple-negative breast cancer cases \cite{LeonFerre2018}, using two downstream tasks: histologic subtyping and stromal tumor-infiltrating lymphocyte (sTIL) grading. We compared three settings: (i) \emph{pretrained retrieval}, using the pretrained WILSON embedding without downstream adaptation; (ii) \emph{fine-tuned retrieval}, using embeddings from the backbone after end-to-end adaptation while retaining the retrieval-based evaluation without a task-specific classification head; and (iii) \emph{fine-tuned classification}, in which the complete backbone and a task-specific classification head were jointly optimized (Fig.~\ref{fig:resultsCombined}g, Table~\ref{tab:finetune}).\\
\indent End-to-end adaptation improved performance on both tasks. For histologic subtyping, fine-tuning the backbone increased accuracy from 0.617 to 0.637 and macro-F1 from 0.401 to 0.428. Joint optimization with a classification head further increased accuracy to 0.690 and macro-F1 to 0.557, a gain of \gain{0.156} in macro-F1 over the pretrained model. For sTIL grading, backbone adaptation increased accuracy from 0.412 to 0.432 and macro-F1 from 0.425 to 0.456; joint optimization increased these values to 0.541 and 0.531, respectively, corresponding to a \gain{0.106} gain in macro-F1 over the pretrained model.\\
\indent The ability to adapt the complete visual encoder is particularly relevant to WILSON because it is substantially smaller than current WSI-level foundation models and was pretrained on a comparatively modest cohort of approximately 189k WSIs. Rather than relying exclusively on increasing model capacity and pretraining-data scale to capture the diversity of downstream pathology tasks, WILSON provides a complementary strategy in which a compact pretrained representation can be specialized when task-specific data become available. The gains observed after end-to-end adaptation on MayoTNBC support this strategy: competitive zero-shot representations can serve as a starting point, while the computational tractability of the composite formulation allows the complete model to be further optimized for specific downstream applications.\\
\indent Together with the zero-shot comparisons above, these findings indicate that the composite representation provides two complementary properties: competitive WSI-level representations without explicit tile-feature aggregation and a practical route to end-to-end downstream adaptation. This combination of compact model size, comparatively modest pretraining scale, and full-model adaptability may be particularly useful for specialized pathology applications in which sufficiently representative data for large-scale foundation-model pretraining are unavailable, but smaller task-specific cohorts can support targeted model adaptation.

\begin{figure}[H]
\centering
\includegraphics[width=1.00 \linewidth]{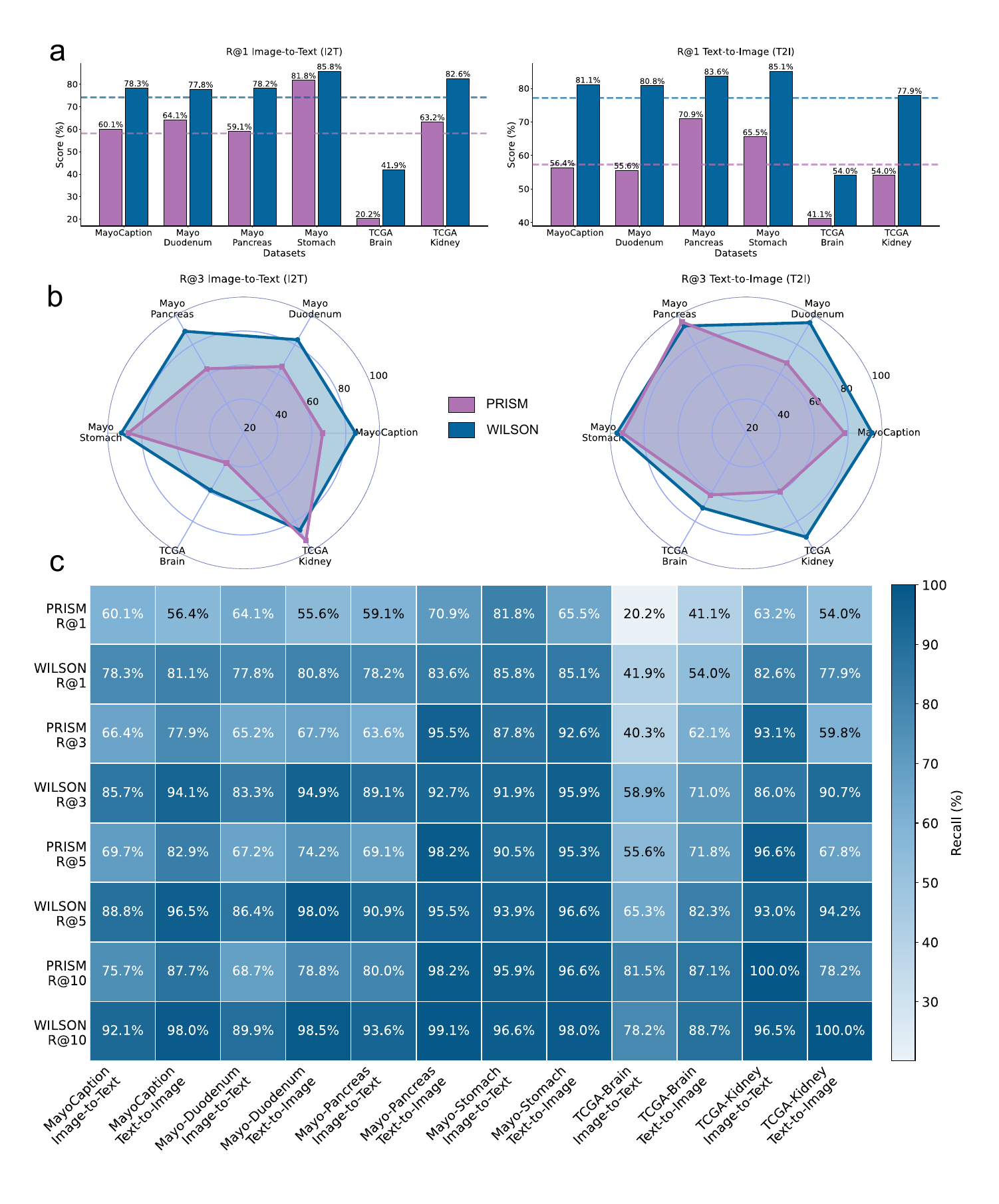}
\caption{\textbf{WILSON versus\ PRISM  across all six cohorts: MayoCaption, MayoDuodenum, MayoPancreas, MayoStomach, TCGA-Brain, TCGA-Kidney at $K\in\{1,3,5,10\}$.} \textbf{(a)} I2T recall as radar plots (one axis per cohort, R@1--R@10 left to right); \textbf{(b)} the same I2T values as grouped bars, with dashed lines marking each model's across-cohort mean (WILSON $\approx$74.1\%, PRISM $\approx$58.1\% at R@1). \textbf{(c,d)} mirror (a,b) for T2I. The WILSON envelope contains the PRISM envelope almost everywhere and both narrow toward TCGA-Brain; the orange curve extends past blue only on the TCGA-Kidney axis (I2T, R@3--R@10) and the MayoPancreas axis (T2I, R@3/R@5), the two crossover cases detailed in Table~\ref{tab:wilson_prism}.}
\label{fig:crossmodal}
\end{figure}

\subsection{WILSON enables bidirectional retrieval between whole-slide images and diagnostic text}
\label{sec:crossmodal}  The preceding experiments evaluated the visual representation independently of WILSON's text encoder. Because WILSON is trained by aligning composite images with pathology-report text, we next asked whether the resulting joint embedding space supports retrieval across modalities. We evaluated image-to-text (I2T) and text-to-image (T2I) retrieval using recall@K ($K\in\{1,3,5,10\}$), with a match defined by disease-group membership. WILSON-CoCa was compared with PRISM, a published slide-level histopathology vision--language foundation model, using the same evaluation protocol. Evaluation included an internal Mayo Clinic cohort ($n=456$ WSIs), analyzed both collectively and by organ (duodenum, pancreas and stomach), and two external cohorts: TCGA-Brain (six glioma subtypes) and TCGA-Kidney (three renal-carcinoma subtypes) (Fig.~\ref{fig:crossmodal}, Fig.~\ref{fig:SuppRetrievalCrossModal}, Tables~\ref{tab:mayo_horizontal}--\ref{tab:tcga_kidney}).\\ \indent Across the twelve cohort--direction combinations, WILSON showed its largest advantage over PRISM at the highest retrieval rank. Mean R@1 was 75.6\% for WILSON compared with 58.1\% for PRISM, an absolute difference of 17.9 percentage points. The advantage narrowed progressively with increasing $k$, to 13.5, 11.9 and 8.4 points at R@3, R@5 and R@10, respectively (Table~\ref{tab:wilson_prism}). WILSON led PRISM at R@1 in every cohort and retrieval direction evaluated. At higher values of $k$, PRISM exceeded WILSON in a small number of comparisons, including TCGA-Kidney I2T and TCGA-Brain I2T, indicating that the principal advantage of WILSON was in ranking a relevant disease-group match first rather than simply making that group retrievable within a larger candidate set.\\ \indent On the internal Mayo Clinic cohort, WILSON achieved R@1 values of 78.29\% for I2T and 81.14\% for T2I retrieval, with a median rank of 1 in both directions. Performance remained high when evaluated separately by organ: I2T R@1 was 85.81\% for stomach, 78.18\% for pancreas and 77.78\% for duodenum, while T2I R@1 ranged from 80.81\% to 85.14\%. These results indicate that the shared embedding space supports retrieval in both directions across anatomically distinct gastrointestinal tissues rather than being driven by a single organ or disease group.\\ \indent Performance on the external cohorts was more heterogeneous. On TCGA-Kidney, WILSON retained strong cross-modal retrieval, reaching 82.56\% I2T and 77.91\% T2I R@1, with a median rank of 1 in both directions. Performance was lower on TCGA-Brain, where I2T and T2I R@1 were 41.94\% and 54.03\%, respectively, although recall increased to 78.23\% and 88.71\% at R@10. PRISM showed a similar relative reduction on TCGA-Brain, suggesting that cross-modal discrimination among glioma subtypes is more challenging for both models than among the disease groups represented in the other cohorts.\\ \indent Together, these results demonstrate that aligning a single multi-magnification composite with diagnostic text produces a shared representation that supports bidirectional retrieval across internal and external cohorts. Despite its compact composite-based architecture, WILSON consistently ranked relevant disease-group matches above those retrieved by a substantially larger slide-level vision--language model at R@1, while differences between the models diminished as the retrieval set expanded.

\subsection{WILSON generates high-quality pathology captions}
\label{sec:wilsonCocaResults}
Cross-modal retrieval demonstrates that WILSON's joint embedding space captures associations between histopathology and diagnostic language, but does not establish whether visual information can be translated into diagnostically meaningful text. We therefore evaluated WILSON-CoCa, a generative extension of WILSON that decodes free-text pathology captions directly from composite images. WILSON-CoCa was compared with two published histopathology vision--language foundation models capable of text generation: PRISM~\cite{shaikovski2024prism}, which combines a Virchow~\cite{vorontsov2024virchow} tile encoder with a BioGPT-style decoder, and PRISM2~\cite{shaikovski2025prism2}, which uses Virchow2~\cite{zimmermann2024virchow2} with a Phi decoder~\cite{abdin2024phi3}. Models were evaluated on the same held-out WSIs using a fixed random seed and the same caption-quality metrics against pathologist-authored reference text. WILSON-CoCa generated captions autoregressively from the attentional-pooler representation of a single composite image per WSI.

\begin{figure}[H]
  \centering
\includegraphics[width=1.00 \linewidth]{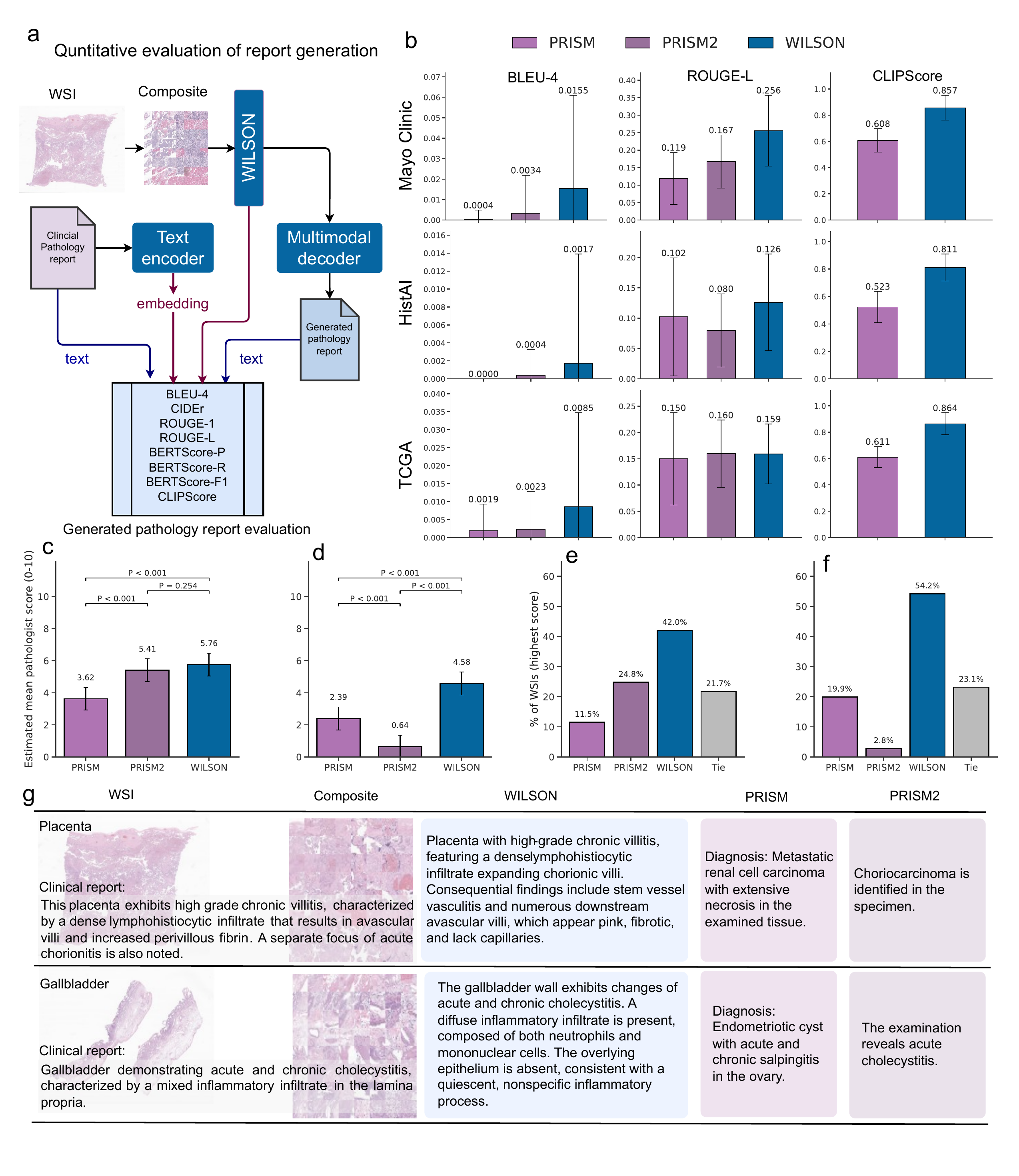}
\caption{\textbf{WILSON-CoCa generates pathology captions closer to
reference reports than PRISM and PRISM2, and preferred by pathologists,
internally and externally.} WILSON-CoCa is a CoCa-style~\cite{yu2022coca}
captioning extension of WILSON, compared against
PRISM~\cite{shaikovski2024prism} and PRISM2~\cite{shaikovski2025prism2}.
\textbf{a}, Pipeline: a composite is encoded and decoded into a caption,
scored against the reference by $n$-gram overlap (BLEU-4, CIDEr, ROUGE-1/L),
embedding similarity (BERTScore P/R/F1) and CLIPScore (model-specific
contrastive space; not reported for PRISM2, which has no text encoder).
\textbf{b}, BLEU-4, ROUGE-L and CLIPScore on three cohorts: 1{,}000 Mayo
Clinic pairs, HistAI (48) and TCGA (50; mean $\pm$ s.d., $y$-axes vary).
WILSON-CoCa leads both baselines on overlap metrics and BERTScore recall
(Table~\ref{tab:caption_metrics}; Supplementary Note~\ref{sec:supp_caption_metrics}).
\textbf{c,d}, Pathologist ratings (0--10) on Mayo (\textbf{c}) and HistAI
(\textbf{d}): five pathologists scored all models on the same 50/48 WSIs
($n=226$/$216$ ratings per model). Bars show estimated marginal means
$\pm$95\% CI from a mixed model with crossed random intercepts for slide
and pathologist; brackets give pairwise Wald $P$ values. \textbf{e,f}, Win
rate on the same ratings: \% of WSIs each model scored highest (Tie: no
winner). WILSON-CoCa leads on mean score and win rate on both cohorts.
\textbf{g}, Qualitative examples (placenta, gallbladder) with WSI, report,
composite and each model's caption (more in Fig.~\ref{fig:caption_comparison_supp}).}
  \label{fig:wilsonCaptioning}
\end{figure}

On 1,000 held-out image--caption pairs from the Mayo Clinic cohort, WILSON-CoCa produced descriptions with greater overlap with the reference than either baseline, reaching ROUGE-1 of 0.371 versus 0.142 for PRISM and 0.231 for PRISM2, ROUGE-L of 0.256 versus 0.119 and 0.167, and CIDEr of 0.224 versus 0.002 and 0.055 (Table~\ref{tab:caption_metrics}, Fig.~\ref{fig:wilsonCaptioning}a). WILSON-CoCa also achieved the highest BERTScore recall (0.878 versus 0.842 and 0.856), indicating fuller coverage of the reference findings, whereas its BERTScore precision was the lowest of the three (0.863 versus 0.881 and 0.895), consistent with longer descriptions that add content beyond the reference; BERTScore F1 was highest for PRISM2 (0.875 versus 0.861 for PRISM and 0.870 for WILSON-CoCa). 
On an external TCGA cohort of 50 image--caption pairs not seen during training, WILSON-CoCa led on BLEU-4 and CIDEr (0.009 and 0.090 versus at most 0.002 and 0.027), while ROUGE-1 and ROUGE-L were within rounding of PRISM2 (0.214 versus 0.215 and 0.159 versus 0.160) and above PRISM (0.183 and 0.150) (Table~\ref{tab:caption_metrics}, Fig.~\ref{fig:wilsonCaptioning}b). Averaged across metrics after min--max normalization within metric and cohort, WILSON-CoCa ranked first on all three cohorts (Table~\ref{tab:caption_metrics}; Fig.~\ref{fig:wilson}f).

Pathologist judgments agreed with the automated metrics. The mixed-model
estimated marginal mean rating for WILSON-CoCa on Mayo was 5.76 (95\% CI
5.05--6.46), significantly higher than PRISM (3.62, CI 2.92--4.33; $P<0.001$)
and nominally but not significantly higher than PRISM2 (5.41, CI
4.70--6.11; $P=0.254$) (Fig.~\ref{fig:wilsonCaptioning}c, Table~\ref{tab:pathologist_emm}, ~\ref{tab:pathologist_pairwise}). On HistAI,
WILSON-CoCa's advantage widened and became significant against both
baselines (4.58, CI 3.86--5.30, versus 2.39, CI 1.67--3.11, for PRISM and
0.64, CI $-0.08$--1.36, for PRISM2; both $P<0.001$; Fig.~\ref{fig:wilsonCaptioning}d),
reflecting a significant dataset $\times$ model interaction (joint Wald
test, $P<0.001$) driven chiefly by PRISM2's disproportionate drop outside
the training domain (difference-of-differences $P<0.001$) rather than by any
change in the WILSON-CoCa--PRISM gap ($P=0.905$). Win rate mirrored these
estimates: pathologists rated WILSON-CoCa the single best caption on 42.0\%
of Mayo WSIs and 54.2\% of HistAI WSIs, versus 24.8\%/2.8\% for PRISM2 and
11.5\%/19.9\% for PRISM, with the remainder tied (21.7\%/23.1\%)
(Fig.~\ref{fig:wilsonCaptioning}e,f). These independent human ratings
corroborate the automated overlap metrics above, indicating that
WILSON-CoCa's advantage reflects diagnostic content rather than surface
$n$-gram similarity alone.

These comparisons require one qualification. WILSON-CoCa's decoder and text components were adapted using Mayo Clinic data, whereas PRISM and PRISM2 were evaluated using their released pretrained models. The internal caption-similarity comparison therefore reflects both model architecture and adaptation to the target reporting domain and should not be interpreted as a controlled comparison of architecture alone. The external TCGA and HistAI evaluations provide complementary evidence that the observed captioning performance is not restricted to the Mayo Clinic cohort.

Together, these results show that diagnostic information captured within a single multi-magnification composite can be translated into free-text pathology descriptions, while the errors visible in Fig.~\ref{fig:wilsonCaptioning}c and Fig.~\ref{fig:caption_comparison_supp} confirm that the output is a draft description requiring pathologist review.

\subsection{WILSON substantially reduces WSI-level computational requirements}
\label{sec:efficiency}
We next quantified the computational advantage of replacing conventional tile-wise WSI processing with a single composite representation. WILSON processes each WSI as one $2048\times2048$ composite assembled from 64 selected $256\times256$ image regions. Its visual encoder comprises a ConvNeXt-Base backbone (87.6M parameters), followed by three stride-2 downsampling blocks (46.2M parameters) that progressively reduce spatial resolution while expanding the feature dimension from 1,024 to 3,072, followed by global average pooling to produce a 3,072-dimensional WSI representation. The resulting model contains 133.7M parameters and processes the complete composite in a single forward pass, without independently encoding thousands of tiles or requiring a separate slide-level feature aggregator.\\
\indent This formulation substantially reduced the computational requirements for WSI inference. Evaluated on a common tissue area of $22{,}400\times22{,}400$~px at $20\times$ ($\approx$125~mm$^2$), tiled at each model's native tile size, WILSON required 2.58T FLOPs, compared with 450.96T for MOOZY, 1.03P for TITAN, 3.35P for PRISM, 3.40P for PRISM2 and 3.50P for Prov-GigaPath (Table~\ref{tab:modelComparison}, Fig.~\ref{fig:wilson}e), corresponding to reductions of roughly 175-, 400-, 1,298-, 1,319- and 1,355-fold, respectively. WILSON was also smaller than every WSI- and case-level baseline except MOOZY, with 133.7M parameters against 354.6M for TITAN, 730.3M for PRISM, 1,221M for Prov-GigaPath and 1,251M for PRISM2, corresponding to 2.7-, 5.5-, 9.1- and 9.4-fold reductions in model size; MOOZY is the one comparison in which WILSON is the larger model (85.8M), although MOOZY still requires about two orders of magnitude more compute per WSI. These efficiency gains were achieved alongside the strongest averaged report-generation performance of the three generative models compared (Fig.~\ref{fig:wilson}f), so lower cost did not come at the expense of downstream capability.\\
\indent The difference in computational complexity arises primarily from the unit of WSI representation. Conventional tile-based pipelines independently encode thousands of image regions before aggregating their representations at the slide level, causing computational cost to increase with the number of tiles processed. In contrast, WILSON performs a single forward pass through a fixed-size composite, with 2.57T FLOPs attributable to the backbone and only 16.2G FLOPs, some 0.6\% of the total, to the three additional downsampling blocks and the projection to the final WSI representation. Because those blocks operate on feature maps no larger than $32\times32$, they hold 34.5\% of the encoder's parameters while contributing almost none of its compute. This fixed-size formulation therefore avoids repeated tile-level encoding and the subsequent materialization and aggregation of thousands of tile embeddings.\\
\indent These reductions are particularly relevant in light of the results above: the substantially lower computational complexity and smaller model size were achieved while maintaining competitive WSI-level retrieval performance and enabling end-to-end downstream adaptation. Although FLOPs and parameter counts do not directly measure wall-clock inference time, memory consumption or energy use, they demonstrate that multi-magnification composite representation can substantially reduce the computational complexity of WSI-level foundation modeling without requiring a conventional tile-encoder--aggregator pipeline.

\section{Discussion}
\label{sec:discussion}
Recent pathology foundation models have demonstrated impressive performance across a wide range of tasks using tile-based representations of whole-slide images. At the same time, diagnostic pathology presents several distinctive characteristics, including multiscale interpretation, hierarchical tissue organization, marked spatial heterogeneity and the integration of findings across multiple slides. These properties motivate continued exploration of how pathology images may be represented for foundation-model learning. WILSON was developed as an initial attempt to design a foundation model around these pathological requirements. Specifically, we explored whether tissue morphology could be represented as a compact multi-magnification composite that captures diagnostically informative regions across scales within a single image while remaining computationally tractable. The principal finding of this study is therefore not simply that WILSON achieved competitive performance, but that such a pathology-informed representation was sufficient to support diagnostic retrieval, vision-language alignment, report generation and downstream adaptation while remaining competitive with substantially larger pathology foundation models.

The composite formulation was inspired by how pathologists themselves summarize tissue morphology. In routine diagnosis, education and scientific communication, pathologists rarely examine or present tissue as thousands of independent fields. Instead, representative low- and high-magnification views are integrated into a coherent diagnostic impression. Composite images can be viewed as a structured and reproducible translation of this process into a machine-readable representation. Importantly, this representation is not tied to arbitrary slide boundaries or tissue placement on a glass slide. By operating on selected tissue regions rather than predefined slide units, the same framework can naturally represent individual regions, entire slides or multi-slide cases.

One consequence of this reformulation is that it enables true end-to-end foundation-model training. In practice, the computational burden of processing thousands of tiles has led most pathology foundation models to rely on frozen tile encoders and optimize only downstream aggregators. By reducing the computational unit to a single composite image, WILSON makes full-model optimization tractable while preserving multiscale tissue context. Although each composite contains only a fraction of the pixels present in the original whole-slide image, WILSON remained competitive with, and occasionally outperformed, state-of-the-art models that process orders of magnitude more visual information. This observation suggests that diagnostically relevant information may be concentrated within a relatively small number of informative regions and that representation design may be at least as important as exhaustive information preservation for pathology foundation models.

The ability to fine-tune the entire visual encoder represents a second practical implication of this design. Whereas foundation-model development is often evaluated through zero-shot performance, many pathology studies involve only a few hundred cases. In such settings, adaptation to institution-specific morphology, scanner characteristics or niche clinical tasks may be more important than maximizing zero-shot accuracy. The improvements observed on the MayoTNBC cohort suggest that compact, fully differentiable foundation models provide a complementary route to adaptation that is difficult to achieve with conventional tile-based architectures.

These findings should not be interpreted as evidence that composite representations supersede tile-based foundation models. Large-scale tile aggregation remains a powerful strategy, particularly when computational resources are abundant and maximizing zero-shot performance is the primary objective. Rather, WILSON should be viewed as an alternative point within the pathology foundation-model design space. More broadly, the fact that competitive performance was achieved using fewer than 200,000 training WSIs suggests that progress in pathology foundation models may depend not only on scaling model parameters and training data, but also on exploring representations that are more closely aligned with the biological and diagnostic structure of pathology.

Several limitations warrant consideration. First, the training corpus comprised fewer than 200,000 whole-slide images from a single institution, which remains modest relative to contemporary pathology foundation models. Although this scale was partly intentional, reflecting our goal of developing a compact and computationally tractable model, it remains unclear how the architecture will perform when trained on substantially larger and more diverse datasets. Second, the 8×8 multi-magnification composite used in this study should be regarded as one instantiation of a broader representational design space rather than as an optimal formulation. The most informative selection of regions, magnifications and layouts may vary across organs, specimen types and disease entities. Future work should therefore explore adaptive or learnable composite-generation strategies and determine whether complementary composites can be integrated without reintroducing the complexity of large-scale aggregation pipelines. Importantly, such optimization should occur at the representation-learning stage, thereby preserving the generality expected of a foundation model.

Taken together, our findings establish multi-magnification composite representation as a viable alternative to exhaustive tile-based pathology modeling and demonstrate that a compact end-to-end trainable foundation model can support diagnostic retrieval, vision-language alignment, report generation and downstream adaptation within a unified framework. The broader implication is not that the current composite design is definitive, but that pathology foundation models need not be constrained by representational assumptions inherited from natural-image computer vision. The central contribution of WILSON is therefore not a specific architecture, but the demonstration that pathology-specific representations themselves constitute an important and largely unexplored design space for foundation models.

\begin{figure}
  \centering
  \includegraphics[width=1.00\linewidth]{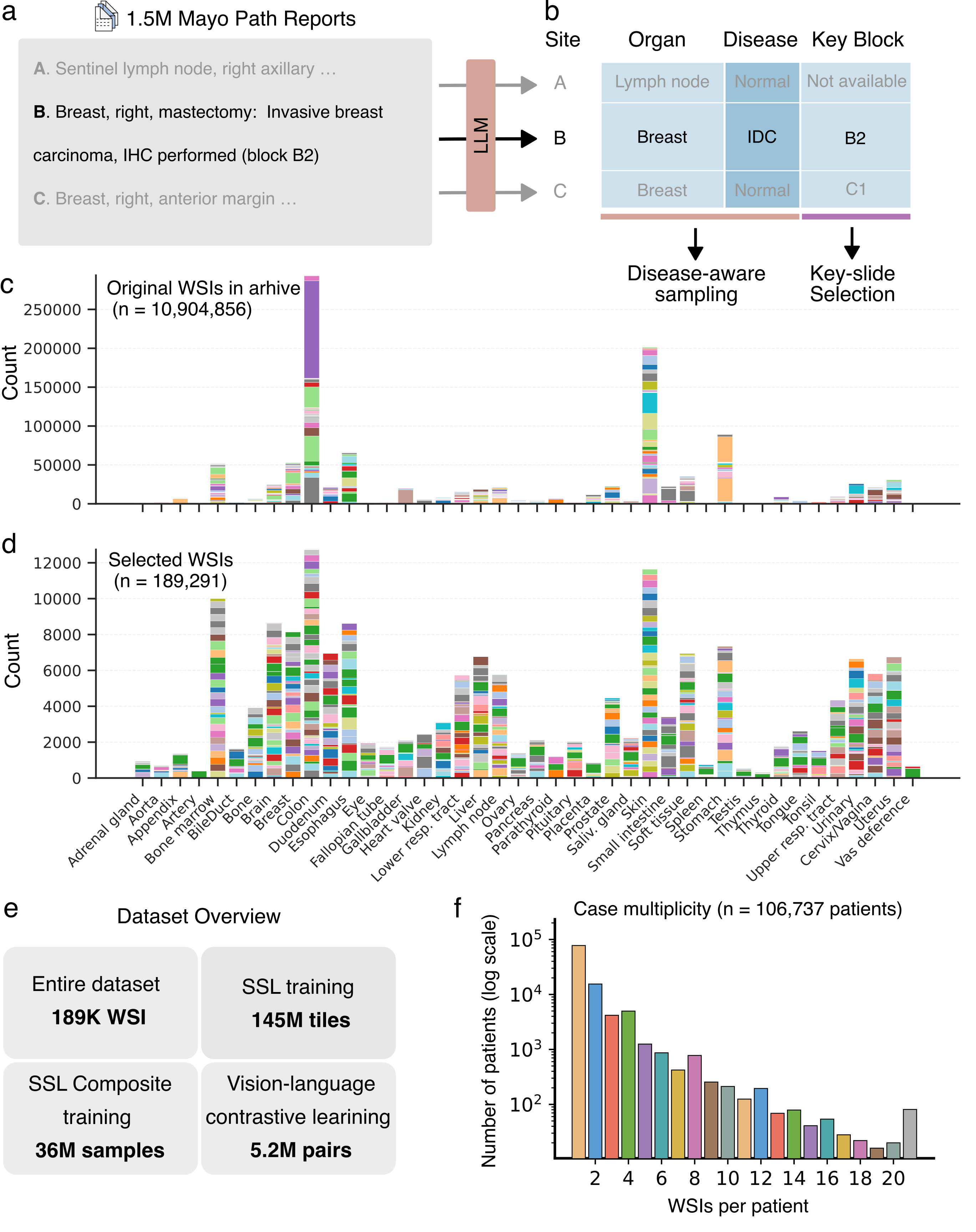}
\caption{\textbf{Data curation strategy for WILSON.}
\textbf{a, } Mayo Clinic pathology reports were segmented using an LLM. 
\textbf{b, } To maximize diagnostic diversity, the number of whole-slide images (WSIs) selected from each disease category was capped, and representative specimens were chosen from WSIs identified as key slides.
\textbf{c, } Natural institutional distribution of WSIs before sampling. Colors indicate distinct disease groups, illustrating the strong imbalance typically observed in routine clinical archives.
\textbf{d, } Distribution of the WILSON training dataset after the curation procedures described in panels \textbf{a--d}. Diversity-aware sampling substantially reduced organ- and disease-level imbalance while preserving broad diagnostic coverage.
\textbf{e,} Overview of the WILSON-DS training dataset. 
\textbf{f,} Number of patients per diagnostic category (log scale).
}
  \label{fig:WilsonData}
\end{figure}

\section{Methods}
\subsection*{Study design and ethics}
This retrospective study developed and evaluated a vision--language foundation
model for whole-slide histopathology using routinely archived clinical material
from Mayo Clinic together with public external cohorts. The study was approved
by the Mayo Clinic Institutional Review Board (protocol number 25-008701), which
waived the requirement for informed consent for the use of de-identified
archival slides and reports. All Mayo Clinic data were de-identified before
model development, and no protected health information was used for training
or evaluation. Public cohorts (TCGA, CPTAC, HistAI) were used under their
respective data-use terms. Model development used only the training partition
described below; all reported results are on held-out slides, cases or cohorts
that were never seen during any training stage.

\subsection*{Case selection from the clinical archive}

Clinical pathology archives are highly imbalanced, with a small number of
organs and diagnoses accounting for a large fraction of all cases
(Fig.~\ref{fig:WilsonData}c). To construct a training corpus that maximized
morphological, diagnostic, and linguistic diversity rather than reflecting
clinical prevalence, we performed diversity-aware sampling from a clinical
archive containing 1,454,318 pathology cases and 10,904,856 whole-slide
images (WSIs).

Pathology reports were first segmented into organ-site level diagnostic
entries. Organ sites were identified from report metadata and grouped into 42
predefined organ categories by a board-certified pathologist using
rule-based mappings. To characterize the highly heterogeneous diagnostic
landscape within each organ, diagnoses were extracted from reports using a
large language model and grouped into organ-specific disease categories using
embedding-based clustering followed by pathologist curation. These disease
categories were designed to capture diagnostically meaningful diversity and
to consolidate common terminology variants rather than to generate
case-level ground-truth labels.

Representative WSIs were selected at the organ-site level using report-derived
evidence. When multiple tissue blocks were available, a representative block
was identified from explicit report references or ancillary testing patterns;
organ sites without an identifiable representative block were excluded.
Additional details of disease-category construction, representative-block
selection, and scanner-specific rules are provided in Supplementary Note~\ref{sec:Mayo189Kselection}.

To prevent common disease entities from dominating the dataset, a maximum of
350 WSIs was sampled from each organ--disease category. This strategy reduced
more than 10 million archived WSIs to a balanced and diverse training corpus
of 189,291 WSIs spanning 838 organ--disease combinations (Mayo189K dataset).
The resulting dataset substantially increased representation of rare organs
and diagnoses relative to the source archive while preserving broad
morphological coverage (Fig.~\ref{fig:WilsonData}c,d).

\subsection*{Report-derived supervision}

For each selected organ site, organ-specific diagnostic text was extracted from
the corresponding pathology report using Gemini~2.5 Pro~\cite{geminiteam2025gemini25}. The extracted text was converted into three semantically equivalent but lexically diverse
histopathology captions, which were used as text targets for contrastive training and embedded with Gemini Embedding~2~\cite{shanbhogue2026gemini} to generate fixed text representations. (The details are described in Supplementary Note \ref{sec:report_derived_supervision})

A complementary keyword-based supervision signal was generated by assigning
curated pathology terms supported by the report text using embedding-based
retrieval and LLM verification. This produced a sparse multi-label keyword
vector for each WSI. Details of report cleaning, caption generation, keyword curation, and prompts are provided in Supplementary Note~\ref{sec:report_derived_supervision}.
 
\subsection*{Multi-magnification composite generation}
Rather than encoding a slide as thousands of independent tiles, we represent
each WSI as a fixed-size composite: an $8\times8$ grid of $256\times256$-pixel
tiles forming a single $2048\times2048$ RGB image that presents selected tissue
locations at several magnifications simultaneously. Composites are generated
by an organ-aware, abnormality-guided procedure in three phases.

\begin{figure}
  \centering
  \includegraphics[width=0.97\linewidth]{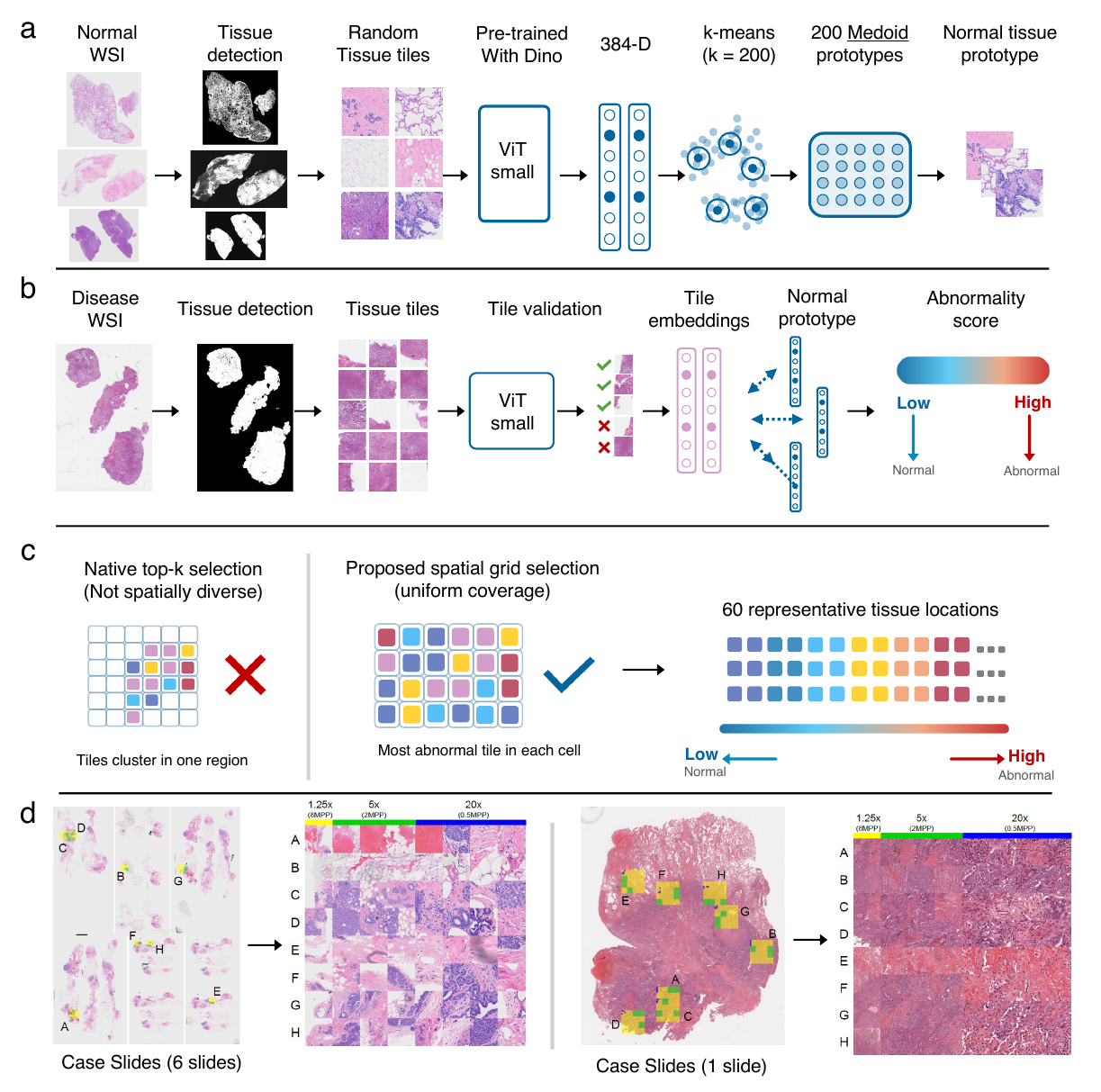}
\caption{\textbf{COMPOSITE: Clustering-Oriented Composite Assembly Pipeline.}
a, Phase 1 builds a normal-tissue reference using WSIs strategically included in the pan-organ Mayo189K cohort. Approximately 500 WSIs per organ were randomly selected from cases with diagnoses indicating no pathological findings, no malignancy, benign tissue, or similar findings, providing an organ-specific reference of normal tissue morphology. Tissue regions in healthy whole-slide
images (WSIs) are detected by Hue--Saturation--Density thresholding~\cite{bejnordi2016} and randomly
tiled; a DINO pre-trained Vision Transformer (ViT) extracts 384-dimensional
CLS-token embeddings in a single forward pass. Pooled embeddings are clustered with
MiniBatch $k$-means ($k = 200$), and each cluster mean is replaced by its medoid so
that centroids correspond to genuine tissue appearances. The 200 prototypes form a
reusable organ-specific reference. b, Phase 2 scores tiles in disease slides. The
same ViT emits both the CLS embedding and quality logits (valid/invalid) in one
pass; sub-threshold tiles (marked x) are discarded as artifacts. Remaining tiles
receive an abnormality score equal to their minimum Euclidean distance to the 200
normal centroids, higher values indicating greater deviation from normal tissue.
c, Phase 3 selects tiles with spatial diversity. Naive top-$k$ selection (red x)
concentrates tiles in a single lesion, sacrificing representativeness. Instead, the
tissue bounding box is divided into a spatial grid, the highest-scoring tile per
cell is selected, and globally best unused tiles fill empty cells---guaranteeing
coverage of all tissue zones while biasing toward diagnostically informative
locations, yielding 60 positions per slide. Ranked locations are distributed
round-robin across the 10 composites in order of abnormality score, so each
composite spans the full abnormality spectrum and all spatial zones rather than
concentrating the most atypical locations in one image. Locations per composite is
configurable: the schematic shows a six-location layout ($6 \times 6$ grid),
whereas all experiments used eight locations ($8 \times 8$ grid; panel d). Within
each composite, rows are tissue locations ordered by abnormality score and columns
encode multi-scale detail, from a 1.25$\times$ context view through 5$\times$ and
20$\times$ magnifications. d, Example $8 \times 8$ composite. Eight low-magnification
regions (A--H; $256 \times 256$ px, $8~\mu$m/pixel) were drawn from six WSIs, each contributing three mid-magnification (green boxes;
$256 \times 256$ px, $2~\mu$m/pixel) and four high-magnification regions (blue boxes;
$256 \times 256$ px, $0.5~\mu$m/pixel), assembled into a single
$2048 \times 2048$ px composite.}
  \label{fig:data}
\end{figure}


In the first phase, executed once per organ, a compact reference of normal
tissue appearance is built from histologically unremarkable slides
(Fig.~\ref{fig:data}a). Tissue tiles are embedded using a Vision
Transformer that simultaneously provides quality-control scores to remove
artifactual regions. Embeddings from normal slides are clustered to generate
an organ-specific set of representative normal tissue medoids.
 

In the second phase (Fig.~\ref{fig:data}b), tiles from each slide are
embedded with the same network and assigned an abnormality score based on
their distance from the organ-specific normal reference. To balance
diagnostic focus with spatial coverage, high-scoring tiles are selected using
a spatially aware sampling strategy. The
$N_{\mathrm{select}}$ highest-scoring candidates are retained and distributed across ten composites in round-robin order of abnormality score, with eight tissue locations per composite (the composite grid size is a configurable parameter; all experiments used an 8×8 layout) so that each composite spans the slide's full abnormality spectrum rather than concentrating the most atypical regions in a single image.

In the third phase (Fig. \ref{fig:data}d, \ref{fig:composite_example}),
the selected tissue locations are assembled into multi-magnification
composites. Each composite is an $8\times8$ grid of $256\times256$-pixel
tiles in which each row corresponds to one selected tissue location. The
first column contains a $1.25\times$ context tile covering the largest field
of view. From this context region, three representative $5\times$ tiles and
four representative $20\times$ tiles are selected, yielding seven
higher-magnification views spatially linked to the same location. The
$1.25\times$ context tile is placed in the leftmost column, while the
corresponding higher-magnification tiles occupy the remaining columns of the
same row, preserving the relationship between low- and high-magnification
observations. The resulting $8\times8$ composite therefore encodes eight
tissue locations, each represented by one contextual view and seven
higher-resolution views.

The composite serves as a compact visual summary of a whole-slide image,
jointly encoding tissue heterogeneity, multi-scale morphology, and cross-scale
spatial correspondence within a single image.
 
\subsection*{Vision encoder}
\label{sec:vision_encoder}
All WILSON variants share the same convolutional encoder
(ConvNeXtV3Extended; Supplementary Fig.~\ref{fig:wilson_arch} and Supplementary Table~\ref{tab:wilson_arch}). A ConvNeXt-Base backbone~\cite{liu2022convnext}
(stage depths $3,3,27,3$; widths $128,256,512,1024$; overall stride 32;
87.57M parameters) is followed by three additional stride-2 downsampling blocks,
each consisting of layer normalization, a $2\times2$ stride-2 convolution and
GELU, which expand the channel width from 1,024 to 3,072 (46.15M parameters)
while reducing the $64\times64$ backbone feature map of a $2048\times2048$
composite to $8\times8$. The resulting 64 spatial tokens of dimension 3,072
are used by the report-generation decoder, and their global average yields a
3,072-dimensional slide embedding matched to the dimensionality of the Gemini
text embeddings, which removes the need for a cross-modal adapter. The encoder
has 133.7M parameters and processes a composite in a single forward pass of
2.6~TFLOPs; because the extended blocks operate on feature maps no larger than
$32\times32$, they contribute 34.5\% of the parameters but 0.6\% of the
compute. When the encoder is warm-started from a backbone-only self-supervised
checkpoint, the extended-block convolutions are initialized by tiling the final
backbone downsampling filters and rescaling them to He-initialization
magnitude, preserving learned filter directions while keeping activations
within half-precision range (Extended Data Fig.~\ref{fig:wilson_arch}d).

\subsection*{Self-supervised pretraining}
The WILSON vision encoder was obtained through a two-stage self-supervised learning (SSL) curriculum built on a ConvNeXt backbone (ConvNeXt-V3 with a global-average-pooling head), initialized from DINOv3 ConvNeXt-Base weights pretrained on the LVD-1689M natural-image corpus. Both stages used a DINO-style teacher--student self-distillation objective with multi-crop augmentation and histopathology-specific augmentations (random rotation and color/stain jitter), applied to hematoxylin-and-eosin whole-slide images spanning multiple organs and diagnoses.
In the first, \emph{tile-based}, stage the encoder was trained on individual $224\times224$-pixel tissue tiles for 20 epochs on 8 GPUs (H200; total batch size 1{,}024; learning rate $2\times10^{-4}$ with a 1-epoch warm-up), using two global crops ($224$\,px) and eight local crops ($96$\,px) per tile.
In the second, \emph{composite-based}, stage four spatially co-registered $256\times256$-pixel patches captured at complementary magnifications ($2.25\times$, $5\times$, $10\times$ and $20\times$) were tiled into a single $512\times512$-pixel composite image (Fig.~\ref{fig:wilson}d), and training proceeded in three successive steps starting from the tile-based teacher checkpoint: (i) the full network was trained end-to-end at $512$-px resolution across two nodes (16 GPUs total; total batch size 512; learning rate $2\times10^{-4}$); (ii) additional SSL "extension" blocks were appended to the pretrained backbone, which was frozen, and trained alone for 3 epochs (learning rate $1\times10^{-4}$) on an expanded corpus of approximately 10 million composite tiles, allowing the new blocks to warm-start without perturbing the pretrained backbone features; and (iii) the entire extended network was then unfrozen and fine-tuned end-to-end for a further 5 epochs (learning rate $1\times10^{-4}$ with a 1-epoch warm-up) on the same corpus. The resulting composite-based teacher network constitutes the final WILSON vision encoder
used in all downstream experiments.

\subsection*{Vision--language training}
\label{sec:vlm_training}
Starting from the composite-based SSL checkpoint (\S\ref{sec:vlm_training}),
four downstream training strategies were evaluated, each targeting a
different downstream capability: (i) \emph{dense WILSON}, obtained by
SigLIP-style distillation from frozen Gemini caption embeddings, for
retrieval against free-text queries without a trainable text tower;
(ii) \emph{sparse WILSON}, obtained by multi-label regression against a
controlled pathology-keyword vocabulary, for lightweight structured tagging
and retrieval; (iii) \emph{WILSON-CLIP}, a jointly trained multi-caption
contrastive alignment followed by supervised diagnostic refinement
(\S\ref{sec:vlm_clip}), for retrieval and direct diagnosis/tissue
classification; and (iv) a \emph{generative captioning model}, obtained by
fine-tuning a CoCa decoder on top of the distilled dense WILSON encoder, for
free-text report generation. All four strategies share the same underlying
corpus of pre-rendered $2048\times2048$ multi-magnification WSI composites
and LLM-generated captions, and all initialize their vision encoder from the
same composite-based SSL checkpoint (WILSON-CLIP and CoCa are the two
exceptions in the sense that CoCa initializes one step further downstream,
from the already-distilled dense WILSON encoder --- see
\S\ref{sec:vlm_coca}).

\subsubsection*{Shared training corpus}
\label{sec:vlm_corpus}
The composite corpus contains 178,088 whole-slide images (WSIs) rendered into 1,760,461 composite images
in total ($\approx$9.9 composites per WSI on average). Of these, 178,020 WSIs
(99.96\%) have a matching LLM-generated caption record (three independent
summary variants per report, \S\ref{sec:report_derived_supervision}) together with
frozen Gemini text embeddings for the full report and for each of the three
caption variants; all 178,088 composite-bearing WSIs have a matching
multi-hot pathology-keyword annotation (a controlled vocabulary of 991
terms), regardless of whether a caption is available. Pairing every
composite with its WSI's three caption variants therefore yields
$1{,}759{,}787\times3=5{,}279{,}361$ realized image--caption training pairs
($\approx$5.28 million).

Both caption-driven stages (dense WILSON and CoCa, \S\ref{sec:vlm_gemini},
\S\ref{sec:vlm_coca}) sample this corpus with \emph{row-shuffle} sampling: at
each step, rather than iterating once over a fixed list of composites, two
composites per WSI are pooled and freshly resampled, and this resampling is
repeated 30 times per WSI over the course of an epoch (\texttt{row\_shuffle\_}
\texttt{repeats}~$=30$). This oversamples the WSI-level diversity of the
corpus --- exposing the model to many different composite/caption-variant
combinations per WSI --- rather than treating each epoch as a single static
pass over the 1.76M composites, and is the reason the realized number of
image--text pairs seen during training substantially exceeds one pass over
the static corpus size reported in Table~\ref{tab:vlm_corpus}.

\subsubsection*{WILSON Model Variants and Learning Strategies}
WILSON comprises four variants: WILSON-distill, WILSON-Keywords (WILSON-KW), WILSON-CLIP, and WILSON-CoCa, each employing a distinct learning strategy. All variants use multi-magnification composite images to represent entire whole-slide images (WSIs) or multi-slide patient cases, avoiding conventional patch-level feature aggregation. Collectively, these variants unify dense semantic embeddings, sparse diagnostic concepts, cross-modal alignment, and generative reporting within a common composite-based framework, supporting both discriminative and generative applications in computational pathology.

\textbf{WILSON-distill: SigLIP-style distillation from Gemini caption embeddings. }
\label{sec:vlm_gemini}
The WILSON-destill encoder was fine-tuned from the composite-based SSL
checkpoint using a SigLIP-style objective~\cite{Zhai_2023_ICCV} in which the
target text representations are frozen, pre-computed Gemini embeddings of
each WSI's report or one of its three caption variants, rather than
embeddings produced by a jointly trained text encoder. Unlike the symmetric
softmax (InfoNCE) contrastive loss used by WILSON-CLIP
(\S\ref{sec:vlm_clip}), a SigLIP-style loss treats every image--text pair
as an independent binary classification (match/non-match) via a sigmoid, so
it does not require the full in-batch normalization that a softmax loss
does; this makes training less sensitive to the (comparatively small) global
batch size available here. Distilling directly from Gemini's frozen
embedding space --- rather than training a text tower from scratch on
$\sim$178K captioned WSIs --- lets the vision encoder inherit Gemini's
broader clinical-language representation, and removes the need to maintain
a separate text encoder at inference time for text-conditioned retrieval.
Architecturally, this works without an adapter layer because
ConvNeXtV3Extended's native GAP output is 3072-dimensional
(\S\ref{sec:vision_encoder}), matching the Gemini embedding dimension
exactly, so image and text embeddings are compared directly in the same
3072-d space rather than through a smaller learned shared projection (as
used by WILSON-CLIP's 256-d and CoCa's 768-d joint spaces). Training used the row-shuffle sampling scheme described above
(\S\ref{sec:vlm_corpus}), drawing composite/caption pairs from the paired
corpus (178,020 WSIs) for 20 epochs on 4 nodes $\times$ 8 H200 GPUs (32
workers), with a per-GPU batch size of 4 and 16 gradient accumulation steps,
giving an effective batch of $32\times4\times16=2{,}048$ image--text pairs
per optimizer step. 2\% of WSIs were held out for validation, and
checkpoints were saved every 5,000 steps. The resulting vision encoder is the initialization used for the
CoCa captioning stage below (\S\ref{sec:vlm_coca}), making dense WILSON the
common ancestor of both the retrieval-only and the generative branches of
the pipeline.

\textbf{WILSON-KW: multi-label pathology-keyword regression. }
\label{sec:vlm_keywords}
The WILSON-KW variant fine-tunes the same composite-based SSL checkpoint
against a controlled vocabulary of pathology keywords instead of free-text
captions. Each WSI's report is reduced to a multi-hot target vector over a
991-term vocabulary, and the contrastive/captioning heads used by the
dense and CLIP variants are disabled (\texttt{keywords\_only}) so that the
vision encoder is optimized purely for multi-label keyword regression
against this sparse target (loss weight 1.0). Because keyword annotation
does not depend on a WSI having a generated caption, this strategy is
trained on the \emph{full} composite-bearing corpus (178,088 WSIs,
1,760,461 composites; \S\ref{sec:vlm_corpus}), 68 WSIs more than the
caption-paired corpus used by the dense and CoCa stages --- giving sparse
WILSON the largest usable training set of the four strategies, at the cost
of a coarser, non-generative supervisory signal. This variant provides a
computationally light structured alternative to free-text captioning: the
991-dimensional multi-hot target is cheaper to predict against than a
full caption or report embedding, and yields directly interpretable
per-keyword tags that are useful for controlled-vocabulary retrieval and as
an ablation of how much representational benefit the richer caption-level
supervision (dense WILSON, WILSON-CLIP) provides over simple tag
prediction. Training ran for 7 epochs (a shorter budget than the other three
stages, consistent with its role as a lighter-weight variant) on a single
node of 8 H200 GPUs, with a per-GPU batch size of 4 and 16 gradient
accumulation steps (effective batch $8\times4\times16=512$), holding out
10\% of WSIs for validation --- a larger validation fraction than the other
stages (2\%), chosen to obtain a more stable validation estimate given the
shorter, 7-epoch training budget.

\textbf{WILSON-CLIP: multi-caption contrastive alignment and supervised refinement. }
\label{sec:vlm_clip}
As a third strategy, a collaborator trained WILSON-CLIP, which likewise
initializes its visual encoder from the composite-based SSL checkpoint but
--- unlike dense WILSON's frozen-embedding distillation --- aligns it
against all three caption variants per composite with a \emph{jointly
trained} text encoder (PathologyBERT) under a symmetric multi-positive
InfoNCE (CLIP) objective (Phase 2). WILSON-CLIP is therefore the only one of
the four strategies in which the text encoder itself is updated during
vision--language training rather than held frozen (dense WILSON, sparse
WILSON) or trained only downstream of a frozen retrieval space (CoCa's
contrastive branch reuses dense WILSON's vision tower but trains its own
CoCa text tower). WILSON-CLIP is also the only strategy with a second,
purely supervised stage: Phase 3 fine-tunes the vision encoder alone against
tissue and fine-grained diagnosis labels while the text encoder and CLIP
projections remain frozen, giving WILSON-CLIP a direct classification
capability that the other three variants do not have. Full architectural
and optimization details, including the fixed 1,410,600/349,861 train/val
composite manifest, are given in \S\ref{sec:vlm_clip}.

\textbf{WILSON-CoCa: generative captioning. }
\label{sec:vlm_coca}
The final stage fine-tunes a generative captioning model on top of the
distilled dense WILSON encoder (\S\ref{sec:vlm_gemini}), following the CoCa
recipe~\cite{yu2022coca}: a contrastive loss is trained jointly with an
autoregressive next-token captioning loss over an open-clip CoCa
(coca-ViT-L-14) text tower and multimodal decoder. This is the
only one of the four strategies that produces free text at inference time
--- dense WILSON, sparse WILSON, and WILSON-CLIP all produce embeddings or
class logits for retrieval/classification, whereas CoCa's decoder is the
``report generator'' of Fig.~\ref{fig:wilson}d. The WILSON vision tower
(ConvNeXtV3Extended) is fine-tuned in full (backbone and extension blocks;
no frozen layers), and its $8\times8{=}64$ spatial tokens for a
$2048\times2048$ composite are summarized by a 256-query, 8-head attentional
pooler --- trained from scratch, with no pretrained analog --- before
being cross-attended by the decoder. Concretely, the decoder generates a
caption autoregressively: causal self-attention over previously generated
text tokens is combined, at every decoder layer, with cross-attention over
the 256 pooled visual tokens, and the next token is predicted at each step
until an end-of-sequence token is produced. The CoCa text tower is trained
jointly (no frozen layers); the decoder keeps its pretrained weights as
initialization but is further trained rather than kept frozen. The training objective combines a SigLIP-style sigmoid contrastive loss ---
cross-GPU gathered embeddings supplemented by an 8-step \emph{negative
bank} (a rolling queue retaining embeddings from the previous 8 optimizer
steps as additional negatives, initial logit scale 10.0 and logit bias
$-10.0$, shared embedding dimension 768) --- with the captioning
cross-entropy loss at unit weight. The negative bank compensates for the
comparatively modest number of in-batch negatives available at this scale
(32 GPUs $\times$ 4 per-GPU $=128$ composites per step, versus the
tens-of-thousands-scale batches typical of large CLIP training runs),
giving the contrastive term a larger effective negative pool without
increasing per-step memory cost. Training used the same paired corpus and
row-shuffle sampling scheme as the dense WILSON stage
(\S\ref{sec:vlm_corpus}) for 20 epochs on 4 nodes $\times$ 8 H200 GPUs (32
workers), with a per-GPU batch size of 4 and 8 gradient accumulation steps,
giving an effective batch of $32\times4\times8=1{,}024$ image--caption pairs
per optimizer step (learning rates: $10^{-5}$ for the vision tower and text
tower, $10^{-3}$ for the projection heads, 10\% warm-up over 0.1 epochs,
cosine decay to 1\% of the peak rate, weight decay $10^{-4}$, gradient
clipping at 1.0, bfloat16 mixed precision). 2\% of WSIs were held out for
validation. Initializing from the dense WILSON encoder, rather than from the
composite-based SSL checkpoint directly or from WILSON-CLIP, gives the CoCa
decoder a vision tower whose representation is already aligned to a rich
clinical-language embedding space (\S\ref{sec:vlm_gemini}), rather than one
that has been additionally specialized for closed-set diagnosis/tissue
classification (WILSON-CLIP Phase 3) or left purely self-supervised
(composite SSL) --- a better-matched starting point for free-text
generation.

\vspace{0.1in}
Collectively, these variants unify dense semantic embeddings (WILSON-distill), sparse diagnostic concepts (WILSON-KW), cross-modal alignment (WILSON-CLIP), and generative reporting (WILSON-CoCa) within a common composite-based framework, supporting both discriminative and generative applications in computational pathology.
\subsection*{Composite versus patch-based slide representation}
To test whether a composite preserves the information available in a
conventional patch-based representation independently of WILSON's training,
both representations were encoded with the same frozen tile foundation model,
UNI~\cite{chen2024uni}. For the patch-based arm, 2\% tiles were selected with
Yottixel~\cite{kalra2020yottixel} and their UNI embedding scores aggregated into a slide score; for the
composite arm, the $2048\times2048$ composite was encoded by UNI directly in a
single pass. Slide-level retrieval was evaluated
by top-1 macro-averaged F1 on BCTherapy, TCGA-Kidney, MayoSkin and
MayoBreast.

 \subsection*{Zero-shot retrieval evaluation}
 \label{subse:ZSRE}
WILSON was evaluated without any task-specific training using embedding-space
retrieval. A leave-one-patient-out protocol was employed: for each query slide,
all slides originating from the same patient were excluded from the retrieval
database to prevent patient-level information leakage, and retrieval was then
performed over slides from all remaining patients. The $k$ nearest neighbors by
cosine similarity were retrieved and the query was assigned the majority label
among them (majority voting, MV@$k$), evaluated at $k=1,3,5$ and $7$; $k=1$
reduces to direct top-1 label transfer. Performance was quantified with three
complementary metrics: accuracy, balanced accuracy (the mean per-class recall)
and macro-averaged F1-score. Because several benchmarks are strongly class
imbalanced --- most markedly TCGA-BRCA and CPTAC-BRCA, where a single subtype
dominates --- raw accuracy can be inflated by majority-class prediction, so
balanced accuracy and macro-averaged F1 are the informative measures on those
cohorts. The main text reports macro-averaged F1 at MV@5 throughout; the
complete grid of all three metrics at every $k$ is given for case-level
retrieval in Figs.~\ref{fig:retrieval_radar}--\ref{fig:retrieval_heatmap_case}
and for slide-level retrieval in
Figs.~\ref{fig:retrieval_radar_wsi}--\ref{fig:retrieval_heatmap_wsi}.

Because ten composites are generated per slide, retrieval was repeated once per
composite and the mean and standard deviation across the ten runs are reported;
the standard deviation therefore measures sensitivity to composite sampling
rather than across-patient uncertainty. Case-level composites integrate tiles
from all WSIs of a case as described above. Baselines were run under the
identical protocol using their published checkpoints:
MOOZY~\cite{kotp2026moozy} for case-level retrieval, and
TITAN~\cite{ding2025titan}, PRISM~\cite{shaikovski2024prism} and
Prov-GigaPath~\cite{xu2024gigapath} for slide-level retrieval, each operating
on its own frozen tile-level features and slide aggregator and therefore
yielding a single deterministic embedding per slide. Parameter counts (tile
encoder plus slide encoder; text towers and language decoders excluded) were
taken from the released checkpoints. Per-WSI FLOPs of the baselines were
measured by counting the operations of forward passes of the released
checkpoints on random inputs (PyTorch FlopCounterMode; 2 FLOPs per
multiply--accumulate), as the tile encoder applied to all $N$ tiles plus one
pass of the slide aggregator, where $N$ is the number of tiles covering a
$22{,}400\times22{,}400$~px tissue region at $20\times$ at each model's native
tile size ($N=10{,}000$ at 224~px for PRISM, PRISM2 and MOOZY; 1,914 at 512~px
for TITAN; 7,656 at 256~px for Prov-GigaPath). WILSON's cost is that of one
forward pass of a $2048\times2048$ composite (Table~\ref{tab:modelComparison}).

 \subsection*{Cross-modal retrieval evaluation}
Image-to-text (I2T) and text-to-image (T2I) retrieval were evaluated in the
shared image--text embedding space. For I2T, each slide embedding was used to
query the set of reference caption embeddings; for T2I, each caption embedding
was used to query the set of slide embeddings. In both directions the candidate
set was ranked by cosine similarity, and a retrieved item counted as a match if
it belonged to the same disease group as the query, so the task measures
disease-group discrimination rather than exact caption recovery. Performance
was reported as recall at $k$ ($k\in\{1,3,5,10\}$), the fraction of queries
with at least one match in the top $k$, together with the mean and median rank
of the first match; lower ranks are better, and a median rank of 1 indicates
that a same-disease-group item was retrieved first for the majority of queries.

Three cohorts were used. The internal MayoCaption cohort comprises 456 WSIs
from 397 patients across three organs and is reported both pooled
(Table~\ref{tab:mayo_horizontal}) and per organ: duodenum
($n=198$; Table~\ref{tab:mayo_duodenum}), pancreas ($n=110$;
Table~\ref{tab:mayo_pancreas}) and stomach ($n=148$;
Table~\ref{tab:mayo_stomach}). The two external cohorts are TCGA-Brain
(six glioma subtypes, $n=124$; Table~\ref{tab:tcga_brain}) and TCGA-Kidney
(three renal-carcinoma subtypes, $n=86$; Table~\ref{tab:tcga_kidney}).
PRISM~\cite{shaikovski2024prism} was evaluated under the same queries,
candidate sets and match criterion using its released image and text encoders.

\subsection*{End-to-end fine-tuning}
Because WILSON's encoder is a single differentiable network, it can be
fine-tuned end-to-end on downstream labels, which tile-encoder pipelines with
frozen feature extractors cannot do. We fine-tuned the dense (report-distilled)
WILSON encoder on MayoTNBC, an internal cohort of 508 triple-negative breast
carcinoma WSIs from 508 patients annotated with two four-class labels,
histologic subtype and sTIL grade (Table~\ref{tab:wsi_datasets}). Each task was
fine-tuned as a separate run. These slides were acquired on scanners different
from those used for the Mayo189K pretraining corpus, so the cohort also probes
robustness under scanner domain shift. Train and validation splits were fixed in
advance and are disjoint at the patient level.

A linear head was attached to the encoder's 3072-dimensional pooled output and
the network was optimized with a cross-entropy loss, with the backbone held
frozen for the first two of five epochs before full end-to-end fine-tuning.
Three settings were compared: the pretrained encoder with no adaptation
(Base Zero-shot); the fine-tuned encoder with its head discarded, assessing the
backbone representation alone (FT-Zero-shot); and the trained head evaluated
directly as a classifier (FT-classification). The first two share a single
nearest-neighbor retrieval metric and therefore isolate what fine-tuning did to
the embedding space, whereas the third measures a learned decision boundary and
is not expected to agree numerically with the second. Accuracy and
macro-averaged F1 are reported for both tasks (Table~\ref{tab:finetune}); full
optimization and evaluation details are given in the Supplementary Methods
(\S\ref{sec:supp_finetune}).
 
\subsection*{Caption evaluation}
Generated captions were compared with reference captions on three held-out
cohorts: 1{,}000 Mayo Clinic WSIs, 48 HistAI WSIs and 50 TCGA WSIs, none of
which were seen during any training stage. The three cohorts differ in how
their references were derived. Mayo Clinic references are the caption variants
generated from pathology reports with Gemini~2.5 Pro~\cite{geminiteam2025gemini25}
(\S\ref{sec:report_derived_supervision}). TCGA references were derived from the
machine-readable pathology reports of Kefeli et al.~\cite{kefeli2024tcga}
and summarized into short captions with DeepSeek V4 Pro
~\cite{deepseekai2026deepseekv4}. HistAI references are the diagnostic conclusion
field supplied with the dataset's own metadata, used verbatim without LLM
summarization; HistAI also distributes a longer full-report field, which was not
used, so that all three cohorts are scored against short-caption-style
references.

Captions were scored with eight automated metrics spanning three families.
(i) \emph{Surface $n$-gram overlap}: BLEU-4, ROUGE-1 and ROUGE-L, which credit
only literal token matches (exact 1--4-gram precision, unigram $F$-measure and
longest-common-subsequence $F$-measure, respectively). (ii) \emph{Weighted
$n$-gram overlap}: CIDEr, the cosine similarity of TF--IDF-weighted 1- to
4-gram vectors, which down-weights corpus-frequent $n$-grams so that
distinctive, content-bearing ones dominate. (iii) \emph{Embedding-based
similarity}: BERTScore precision, recall and F1, computed by greedy cosine
matching of RoBERTa-large token embeddings, and CLIPScore, the cosine
similarity between the slide's image embedding and the generated caption's text
embedding in each model's own image--text contrastive space. CLIPScore
therefore measures agreement between the caption and what the model itself
encodes from the slide rather than agreement with the reference, and is not
comparable across models because each contrastive space has its own learned
scale; it is not computable for PRISM2, which exposes no standalone text
encoder. Full metric definitions and interpretation caveats are given in the
Supplementary Methods (\S\ref{sec:supp_caption_metrics}).

PRISM~\cite{shaikovski2024prism} and PRISM2~\cite{shaikovski2025prism2} were
evaluated off the shelf on the same slides, matched by fixed random seed.

\subsection*{Pathologist evaluation of captions} 
Automated text-overlap and embedding metrics reward surface similarity to a
single reference report but cannot establish whether a generated caption
conveys diagnostically correct, clinically trustworthy information. We
therefore complemented them with a blinded human evaluation. Five board-certified senior staff pathologists, none of whom were involved in the development or training of WILSON, independently evaluated the generated captions using a dedicated web application
(Fig.~\ref{fig:pathologistInterface}) that presented each whole-slide image
alongside the captions generated by WILSON-CoCa, PRISM and PRISM2 for that
slide, labeled anonymously and in randomized order (A/B/C) so that model
identity and presentation order could not bias scoring. For each caption,
the pathologist entered an independent, absolute agreement score on an
11-point scale from 0 (completely disagree) to 10 (completely agree) with
the ground-truth, report-derived reference caption shown alongside; scoring
each caption independently rather than ranking the three allowed ties when
multiple captions were judged equally faithful. Every rating was timestamped
and written to a per-pathologist log for reproducibility and audit. The
evaluation pool combined the same 50 held-out Mayo Clinic WSIs and 48
external HistAI WSIs used for the automated metrics (98 WSIs total), giving
1{,}326 individual caption scores in total (226 Mayo and 216 HistAI
ratings per model).

Because a simple mean score conflates genuine model differences with
systematic differences in slide difficulty and in how leniently or strictly
each pathologist scores, we modeled the ratings with a linear mixed model
with crossed random intercepts for WSI and for pathologist, and fixed
effects for dataset (Mayo versus\ HistAI), model (PRISM, PRISM2, WILSON-CoCa)
and their interaction: score $\sim$ dataset $\times$ model +
(1$\mid$wsi\_id) + (1$\mid$pathologist), fit by REML. This yields, for each
model and dataset, an estimated marginal mean (EMM) rating with a 95\%
confidence interval that is adjusted for slide difficulty and rater
leniency, together with Wald $z$-tests for pairwise model comparisons within
each dataset and for the dataset $\times$ model interaction (whether the
ranking of models differs between the internal and external cohort). As a
complementary, model-free summary that does not depend on the mixed-model
assumptions, we also computed the win rate: the fraction of WSIs on which a
given model's caption received the single highest score among the three
captions for that slide, with ties (no unique highest-scoring caption)
reported separately.

\subsection*{Datasets}
All datasets used in this study are listed in Tables~\ref{tab:wsi_datasets}, \ref{tab:case_datasets}.
No evaluation slide or patient overlapped with the training corpus, and class counts are provided in Supplementary
Tables~\ref{tab:tcga_brca_distribution}--\ref{tab:mayocaseskin_distribution}.
 
At the slide level, \textbf{Mayo189K} is an internal Mayo Clinic dataset used for training the WILSON model. In addition to whole-slide images (WSIs), associated slide captions and pathology-related keywords are available for each case.
\textbf{MayoCaption} is a caption-based retrieval benchmark derived from a
validation subset of Mayo189K and was developed for cross-modal retrieval
experiments between pathology images and text descriptions. It consists of
three organ-specific datasets: \textbf{MayoDuodenum}, \textbf{MayoStomach}, and
\textbf{MayoPancreas}.

\noindent\textbf{MayoBreast} (Table~\ref{tab:dataset_detail_MayoBreast}) comprises 1,339 WSIs from 1,289
patients across twelve breast diagnoses that span invasive carcinoma (no
special type, lobular, mixed, mucinous and metaplastic), in situ disease
(ductal and lobular carcinoma in situ), benign and fibroepithelial lesions
(fibroadenoma, phyllodes tumor, benign proliferative lesion and fat necrosis),
and lymphoproliferative disease. This cohort was sampled to be approximately
balanced and has no patient-level overlap with Mayo189K.

\noindent\textbf{MayoSkin}
(Table~\ref{tab:mayoskin_distribution}) comprises 688 WSIs from 677 patients
across twenty dermatopathologic diagnoses spanning melanocytic lesions (nevus,
nevus with atypia, blue nevus, lentigo and malignant melanoma),
keratinocytic neoplasia (actinic keratosis, seborrheic keratosis, basal cell
carcinoma and squamous cell carcinoma), inflammatory dermatoses (psoriasiform
and lichenoid dermatitis), cutaneous lymphoma, and a range of benign entities
(dermatofibroma, neurofibroma, epidermal cyst, verruca, skin tag, scar,
sebaceous gland hyperplasia and chondrodermatitis nodularis helicis). This
cohort also has no patient-level overlap with Mayo189K.

\noindent\textbf{MayoTNBC} is a WSI-level cohort comprising 508 patients with
triple-negative breast cancer. WSIs are stored in Philips TIFF format and are
associated with two labels: stromal tumor-infiltrating lymphocyte (sTIL) score
and histologic subtype.

\noindent\textbf{TCGA-BRCA}, \textbf{TCGA-Brain}, and \textbf{TCGA-Kidney} are derived
from The Cancer Genome Atlas (TCGA) project~\cite{weinstein2013cancer}. For
histologic classification tasks, we adopted the AI-ready diagnostic labels
described by Uegami et al.~\cite{uegami_2026_19736866}. TCGA caption data were
generated by summarizing pathology reports extracted by Kefeli
et al.~\cite{kefeli2024tcga} using a large language model (DeepSeek V4 Pro\cite{deepseekai2026deepseekv4}).

\noindent\textbf{TCGA-BRCA}
(Table~\ref{tab:tcga_brca_distribution}) comprises 1,087 WSIs from 1,018
patients labeled with seven histological subtypes of invasive breast
carcinoma. Its class distribution mirrors clinical prevalence and is therefore strongly imbalanced:
invasive carcinoma of no special type accounts for 812 slides and invasive
lobular carcinoma for 204, whereas the remaining five subtypes (mixed ductal
and lobular, mucinous, metaplastic, medullary and invasive micropapillary
carcinoma) together contribute 71 slides, with as few as four examples in the
rarest class.

\noindent\textbf{BCTherapy}~\cite{sammut2022multiomic} (Table \ref{tab:bctherapy_distribution}) comprises 160 WSIs from 160 patients treated with neoadjuvant therapy, labeled as residual
disease ($n=119$) or pathologic complete response ($n=41$); it is the smallest
benchmark and the only binary one at the slide level. 

\noindent\textbf{HistAI}~\cite{nechaev2025histai} is a multi-organ dataset used for evaluating cross-modal retrieval and pathology caption generation. Fifty cases (five per organ across ten organs) were selected from the original HistAI-Mixed cohort to maximize histologic diversity. Two cases were excluded because of corrupted WSI files, resulting in a final cohort of 48 patients. From the publicly available metadata, the \texttt{conclusion} field was used as the caption and the \texttt{specialization} field was used as the organ label. All selected slides are listed in Table~\ref{tab:histAI_wsi_list}.

At the case level, each unit of evaluation is a patient case comprising
multiple WSIs, from which a single composite is assembled.
\textbf{CPTAC-BRCA}~\cite{CPTAC_BRCA_2020} (Table~\ref{tab:cptac_brca_distribution}) is a publicly available dataset comprising 198 cases and 653 WSIs. A case is defined as the collection of WSIs belonging to a single patient, with an average of 3.5 WSIs per case. The dataset is labeled with the five PAM50 intrinsic molecular subtypes~\cite{parker2009pam50} (luminal A, luminal B, HER2-enriched, basal-like, and normal-like) and is used for PAM50 genotype classification, making it the only benchmark that requires inferring a transcriptomic class from morphology.

\noindent\textbf{MayoCaseBreast} (Table~\ref{tab:dataset_detail_MayoCaseBreast}) comprises 494 cases and 4,361 WSIs across the same twelve breast diagnoses as MayoBreast. MayoCaseBreast is a patient-level subset of the MayoBreast dataset, with identical diagnostic categories. All available H\&E slides from each case are included. Within a given case, all WSIs originate from either a Leica or a Pramana scanner, with no mixing of scanner types within the same case.

\noindent\textbf{MayoCaseSkin} (Table~\ref{tab:mayocaseskin_distribution}) comprises 500 cases and 1,280 WSIs across the same twenty dermatopathologic diagnoses as MayoSkin. Similar to MayoCaseBreast, MayoCaseSkin is a case-level dataset derived from MayoSkin and includes all available H\&E slides from each selected patient case.

\noindent\textbf{MayoBreastSubtype} (Table~\ref{tab:mayobreastsubtype_distribution}) is an internal case-level cohort comprising 846 breast cancer cases treated with neoadjuvant chemotherapy at Mayo Clinic. A case is defined as the collection of WSIs belonging to a single patient.
Clinical subtype labels were derived from the surrogate intrinsic subtype definitions~\cite{GOLDHIRSCH20111736} and regrouped into three clinically relevant categories: hormone receptor (HR)-positive/HER2-negative (ER $\leq$10\% and HER2-negative; corresponding to luminal A and luminal B tumors),
HER2-positive (HER2-positive by IHC 3+ or FISH amplification, corresponding to both ER+/HER2+ and ER-/HER2+ tumors),
and triple-negative breast cancer (TNBC) (ER $\leq$10\%, PR $\leq$10\%, and HER2-negative). The cohort has no patient-level overlap with Mayo189K.

\subsection*{Practical Guidelines for Using WILSON}The design of WILSON was guided by the goal of developing a pathology foundation model that is not only effective for general-purpose representation learning but also sufficiently compact to enable practical task-specific fine-tuning. In contrast to many existing pathology foundation models, which contain hundreds of millions to billions of parameters and are commonly used primarily as frozen feature extractors, WILSON was intentionally designed with a substantially smaller model footprint. In fact, WILSON contains approximately 70\% fewer parameters than even some of the smaller contemporary pathology foundation models, such as UNI. This compact architecture makes full-model adaptation considerably more practical in typical computational pathology settings.\\ \indent Accordingly, although WILSON can be used directly for zero-shot inference, feature extraction, retrieval, and linear-probing experiments, we strongly recommend task-specific fine-tuning whenever appropriate training data are available. Fine-tuning allows WILSON to adapt its learned representations to the morphology, diagnostic categories, and clinical characteristics of the intended downstream application and enables more complete use of the model's representational capacity than treating it solely as a frozen feature extractor.\\ \indent Importantly, our experiments demonstrate that effective fine-tuning of WILSON does not require thousands of whole-slide images. Meaningful adaptation can be achieved using comparatively modest task-specific datasets, making fine-tuning feasible for many institutional, organ-specific, and disease-specific applications. For particularly small datasets, users may initially freeze portions of the backbone or employ parameter-efficient adaptation strategies; as more training data become available, progressively more or all of the WILSON architecture can be fine-tuned.\\ \indent Thus, WILSON should not be viewed solely as a frozen foundation model from which generic embeddings are extracted. Its relatively compact architecture was deliberately designed to make fine-tuning an integral part of its intended use. Zero-shot evaluation and linear probing remain valuable for benchmarking representation quality and for applications where labeled data are unavailable. However, for a specific diagnostic, predictive, classification, or retrieval task, we recommend fine-tuning WILSON on representative task-specific data to maximize its capabilities.

\subsection*{Statistics and reproducibility}
No statistical method was used to predetermine sample size. Error bars on
retrieval metrics denote the standard deviation across the ten composites
generated per slide or case; baselines that operate on frozen tile features
yield a single deterministic embedding per slide and are therefore shown as
single values without variance. Training used a fixed random seed, and evaluation
subsets were drawn with a single fixed seed shared across all models so that
every model was scored on identical slides. Experiments were implemented in
PyTorch with
open\_clip and run on an NVIDIA DGX cluster
of H200 GPUs, using between 8 and 32 GPUs (1--4 nodes $\times$ 8) depending on
the training stage; per-stage GPU counts are given in
Table~\ref{tab:vlm_corpus}.
 
\subsection*{Data availability}
All external data used in this study are publicly available. TCGA whole-slide
images (TCGA-BRCA, TCGA-Brain and TCGA-Kidney) were obtained from the NCI
Genomic Data Commons (\url{https://portal.gdc.cancer.gov}); the AI-ready
diagnostic labels applied to them are available from Uegami et
al.~\cite{uegami_2026_19736866}
(\url{https://doi.org/10.5281/zenodo.19736866}). The TCGA pathology-report text
from which TCGA reference captions were derived is the machine-readable corpus
of Kefeli and Tatonetti~\cite{kefeli2024tcga}, available from Mendeley
Data (\url{https://doi.org/10.17632/hyg5xkznpx.1}) with the accompanying
extraction pipeline at \url{https://github.com/tatonetti-lab/tcga-path-reports}.
CPTAC-BRCA whole-slide images are available from The Cancer Imaging
Archive~\cite{CPTAC_BRCA_2020} (\url{https://doi.org/10.7937/TCIA.CAEM-YS80}). The
HistAI dataset, including the whole-slide images and the per-case clinical
metadata from which HistAI reference captions were taken, is available from
\url{https://github.com/HistAI/HISTAI} and the associated Hugging Face
repositories
(\url{https://huggingface.co/datasets/histai/HISTAI-metadata})~\cite{nechaev2025histai}.
The BCTherapy (TransNEO) hematoxylin-and-eosin slides used for
treatment-response prediction are available from Zenodo
(\url{https://doi.org/10.5281/zenodo.6337925}), with the accompanying study
described by Sammut et al.~\cite{sammut2022multiomic}.

Mayo Clinic whole-slide images and pathology reports contain protected health
information and cannot be shared. De-identified derived data, comprising
composite tile coordinates, LLM-generated captions and keyword annotations for
the Mayo189K corpus together with the evaluation slide identifiers for the
MayoBreast, MayoSkin, MayoTNBC, MayoCaption and case-level benchmarks, are
available from the corresponding author on reasonable request under a data-use
agreement and subject to Mayo Clinic institutional approval.

\bmhead{Funding}
The authors gratefully acknowledge the F. Craig and Patricia Jilk Fund for Data Science, Predictive Modeling \& AI for Breast Cancer for supporting this study. Funding for this project was additionally provided by the National Institutes of Health grant P50CA116201 (M.P.G., KK) and the Mayo Clinic Comprehensive Cancer Center grant  P30CA015083.

\bmhead{Acknowledgments}
The authors  acknowledge the \emph{Mayo Clinic Comprehensive Cancer Center}, Rochester, MN, USA, for its ongoing support. 

\bmhead{Code Availability} Upon publication and subject to clearance by the Mayo Clinic Ethics Office, the WILSON model will be made available for download to the research community to support further development, evaluation, and validation of computational pathology methods.

\bmhead{Competing Interests} S.A., W.U., J.C.B., M.P.G., K.R.K., and H.R.T. are named inventors on a pending U.S. provisional patent application (No. 64/152,262; filed September 10, 2026), entitled ``Multi-Stage Foundation Model for Histopathology Using Composite Whole-Slide Representations and Joint Dense-Sparse Embedding Learning.'' The application concerns a foundation model architecture distinct from the model investigated in this manuscript. The patent applicant is Mayo Clinic, and the application is currently pending.\\
\indent M.P.G. reports personal fees for CME activities from Conexiant, DAVA Oncology, IDEOlogy Health, MJH Life Sciences, PeerView, Physicians’ Education Resource, Research to Practice, Total Health Conferencing; consulting fees to Mayo Clinic from 858 Therapeutics, Inc., Amneal Pharmaceuticals, AstraZeneca, AstraZeneca UK Ltd., Ataraxis AI, BeiGene USA, Biotheranostics, Biotheryx, eChinaHealth, EcoR1, Eli Lilly and Company, Genentech, Genentech-Hoffman La-Roche, Incyclix Bio, Lilly, Novartis, Pfizer, Puma, RNA Diagnostics; grant funding to Mayo Clinic from AstraZeneca, ATOSSA Therapeutics, Biotheryx, Lilly, Loxo, Pfizer, Puma, Sermonix, SimBioSys; travel support from Lilly; patent for a series of synthesized and characterized molecules (these are not licensed): Degrading PKCB1 to Treat Cancer, US Patent \#19/781,406; Intellectual Property: SimBioSys he does not own stock or have stock options; Mayo has a financial stake (shares) in the company, and he supports SymBioSys’ product development through a ‘know how’ agreement.\\  
\indent J.C.B. receives research funding paid to her institution from Eli Lilly, SimBioSys and Quantum Leap Healthcare. She is on the DSMB for Cairns Surgical. She is supported by the W.H. Odell Professorship in Individualized Medicine.\\
\indent The other authors declare no competing financial or non-financial interests.

\bibliography{ref_new}
\clearpage

\begin{appendices}

\section{Additional Results}\label{suppresults}

\begin{figure}[h]
\centering
\includegraphics[width=\textwidth]{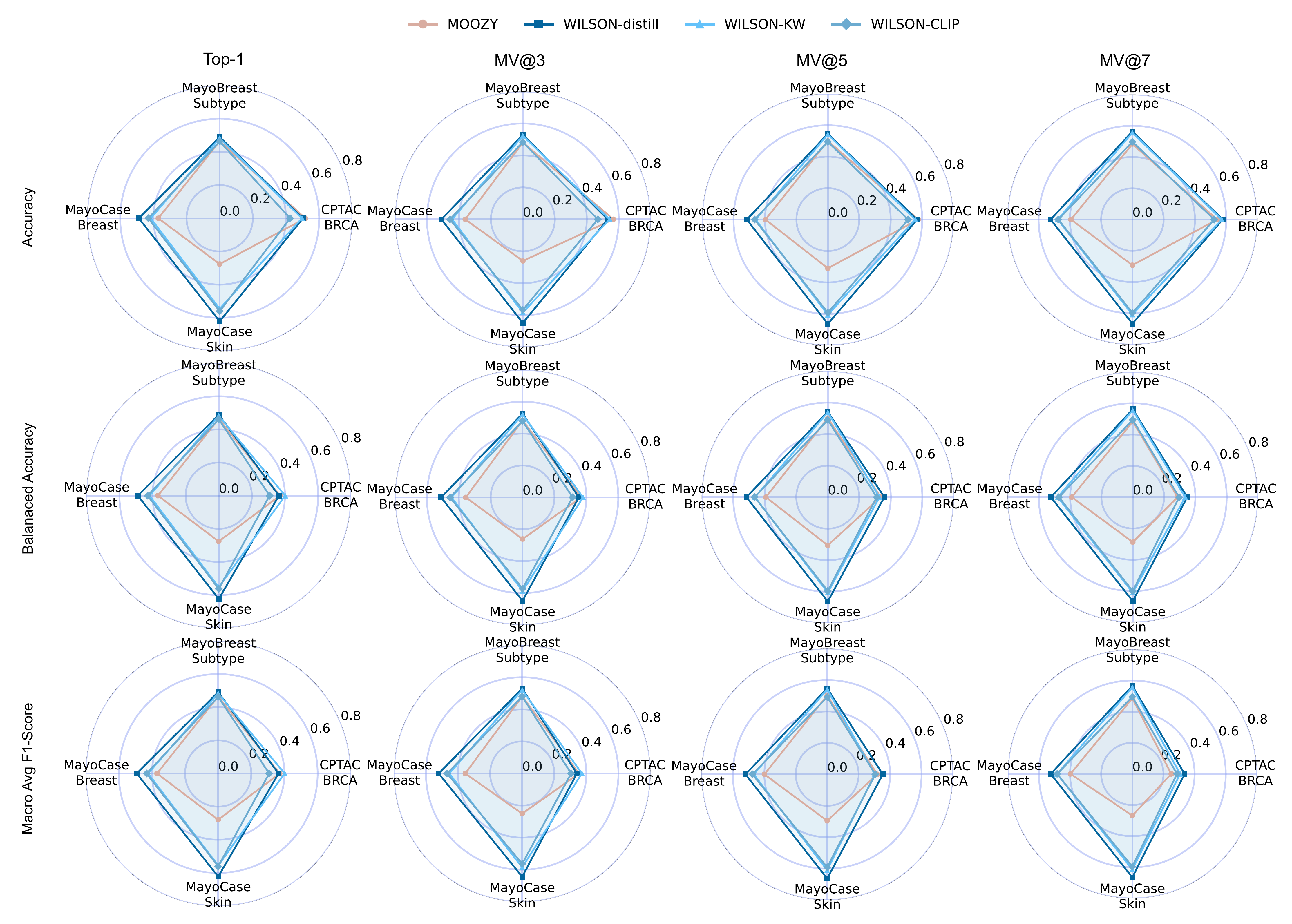}
\caption{\textbf{Case-level image retrieval performance of the vision
encoders.} Retrieval uses each model's vision encoder only (no text
encoder at inference). All whole-slide images of a case were represented in a single case-level composite before getting its corresponding embedding, and each query case was matched
against the remaining cases by cosine similarity; the label of the
retrieved neighbour(s) was transferred to the query. Columns give the
label-transfer rule (Top-1, or majority vote over the $k=3,5,7$ nearest
neighbours) and rows the resulting metric (accuracy, balanced accuracy,
macro-averaged F1-score). Axes are the four evaluation cohorts; the
radial scale spans $0$--$1$ with gridlines every $0.2$. All three WILSON
variants --- which differ only in the text supervision used during
alignment (report distillation, keywords, CLIP-style contrastive) ---
exceed MOOZY on every cohort, metric and $k$, with the largest margins on
balanced accuracy and macro-averaged F1-score. Chance level differs per
axis because class counts and prevalences differ between cohorts.}
\label{fig:retrieval_radar}
\end{figure}

\begin{figure}[h]
\centering
\includegraphics[width=\textwidth]{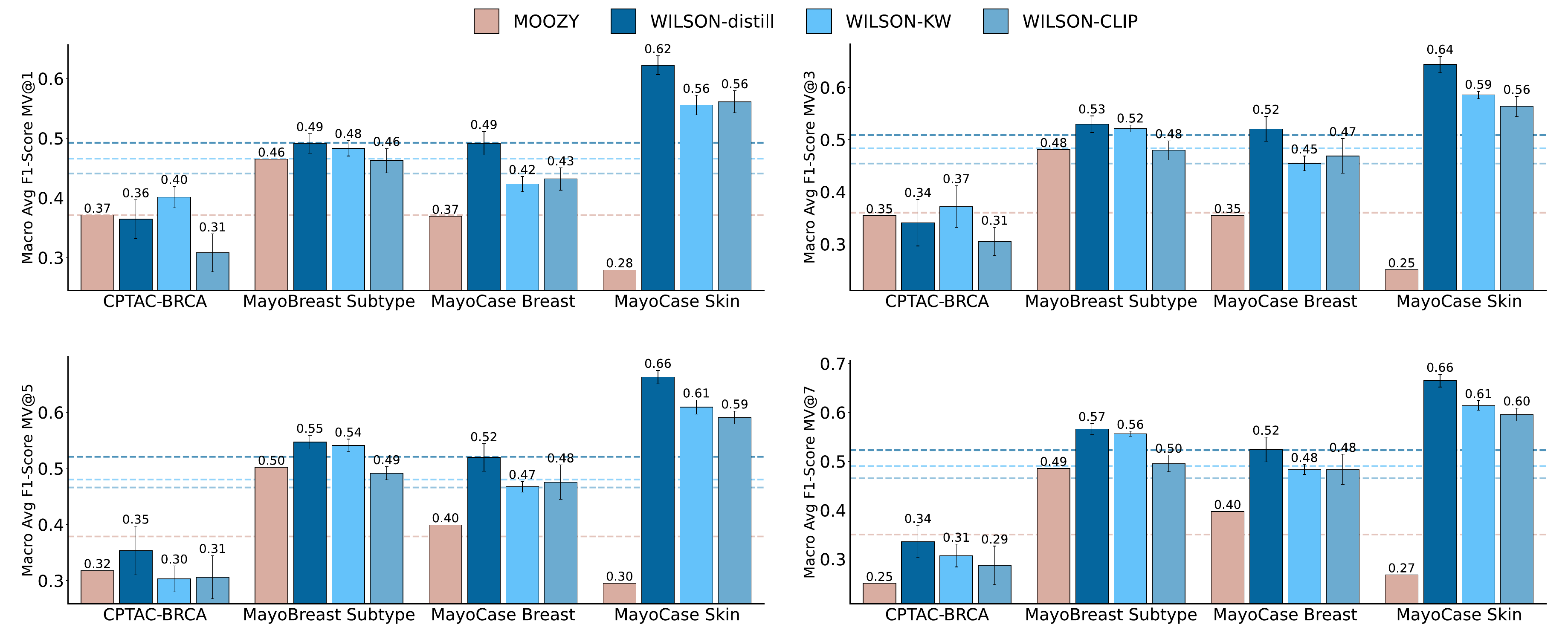}
\caption{\textbf{Case-level image retrieval accuracy by cohort,
macro-averaged F1-score.} Per-cohort breakdown of the case-level
retrieval results summarized in Fig.~\ref{fig:retrieval_radar}, using the
vision encoder of each model only. Panels correspond to the
label-transfer rule (majority vote over the $k=1,3,5,7$ nearest
neighbours); groups within each panel are the four evaluation cohorts
(CPTAC-BRCA, Mayo Breast Subtyping, Mayo Breast Diagnosis, Mayo Skin).
Bars show macro-averaged F1-score for MOOZY and three WILSON variants
differing only in the text supervision used during alignment (report
distillation, keywords, CLIP-style contrastive); error bars denote
$[\text{s.d.\ across } n \text{ folds / seeds}]$ and horizontal dashed
lines mark each model's mean across the four cohorts. Note the truncated
vertical axis. WILSON-distill achieves the highest cohort mean at every
$k$ and the largest single-cohort margin on Mayo Skin (0.62 vs.\ 0.28 for
MOOZY at $k=1$; 0.66 vs.\ 0.27 at $k=7$). Gains are smaller on the breast
cohorts, where MOOZY is comparable to WILSON-CLIP on Mayo Breast
Subtyping and exceeds it on CPTAC-BRCA at every $k$, indicating that the
advantage is cohort-dependent rather than uniform.}
\label{fig:retrieval_bar_case}
\end{figure}

\begin{figure}[h]
\centering
\includegraphics[width=\textwidth]{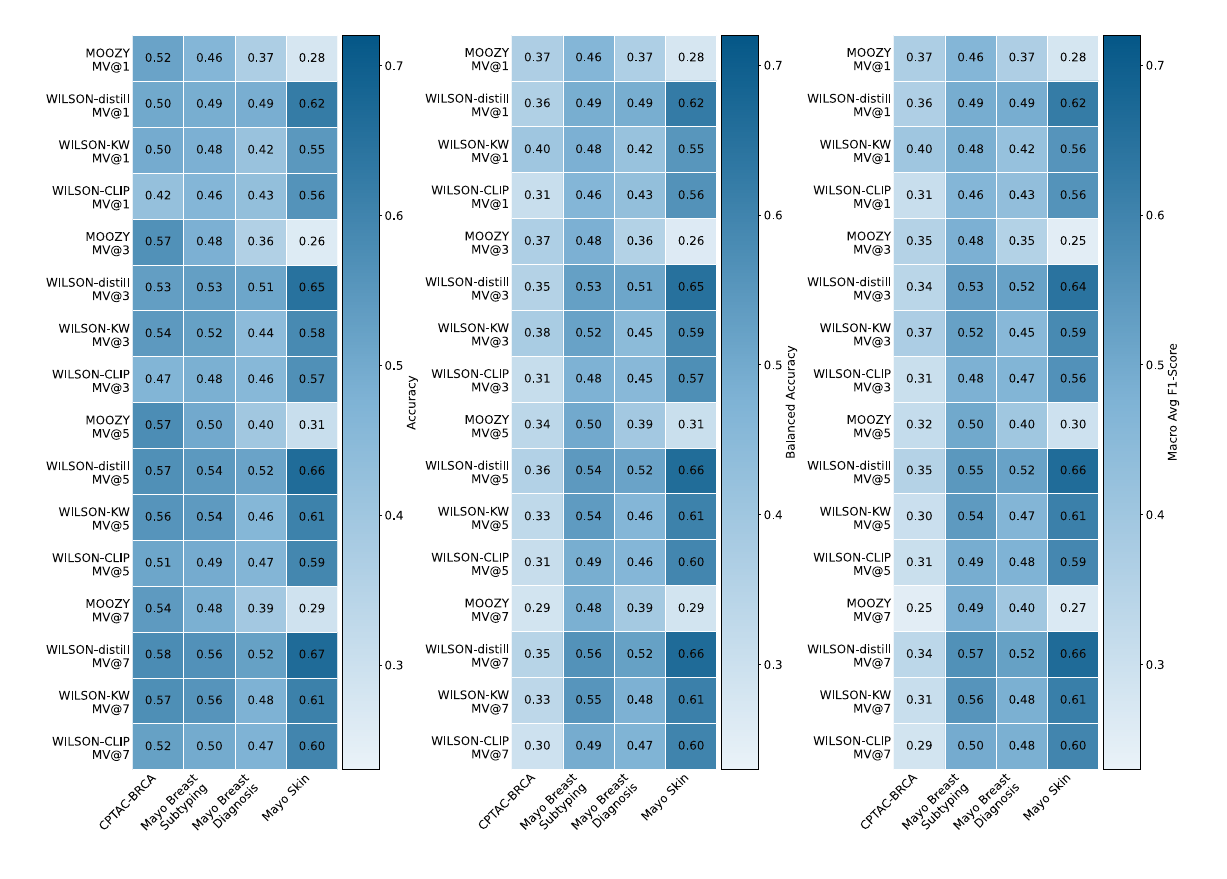}
\caption{\textbf{Complete case-level image retrieval results.} Numerical
values underlying Figs.~\ref{fig:retrieval_radar} and
\ref{fig:retrieval_bar_case}, using the vision encoder of each model only.
Panels report accuracy (left), balanced accuracy (centre) and
macro-averaged F1-score (right) on a common color scale; columns are the
four evaluation cohorts and rows every combination of model (MOOZY,
WILSON-distill, WILSON-KW, WILSON-CLIP) and label-transfer rule (majority
vote over the $k=1,3,5,7$ nearest neighbours), grouped by $k$. Three
patterns are visible. First, the WILSON encoders' advantage is largest on
Mayo Skin, where all three exceed MOOZY by roughly 0.3 on every metric and
at every $k$ (e.g.\ 0.67 vs.\ 0.29 accuracy at $k=7$). Second, on
CPTAC-BRCA the accuracy of all models (0.42--0.58) substantially exceeds
their balanced accuracy and macro-averaged F1-score (0.29--0.40),
indicating that raw accuracy there is inflated by class imbalance; MOOZY
shows the widest gap (0.57 vs.\ 0.35 at $k=3$), consistent with a bias
toward the majority class. Third, performance generally improves with
increasing $k$ for all models, with most of the gain realised by $k=5$.}
\label{fig:retrieval_heatmap_case}
\end{figure}

\begin{figure}[h]
\centering
\includegraphics[width=\textwidth]{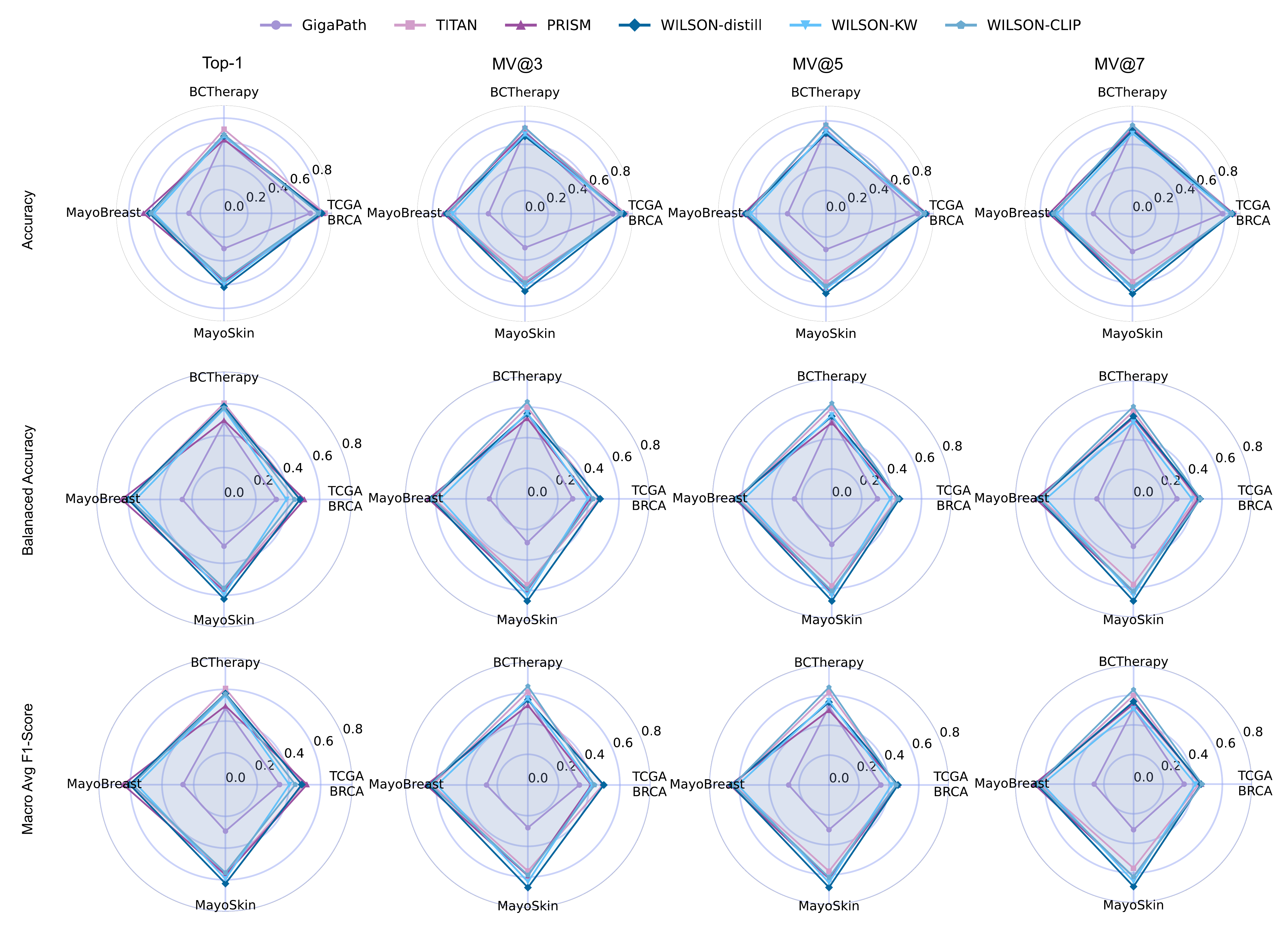}
\caption{\textbf{Slide-level image retrieval performance of the vision
encoders.} Retrieval uses each model's vision encoder only (no text
encoder at inference). Each whole-slide image was encoded into a single
slide embedding and matched against the remaining slides by cosine
similarity; the label of the retrieved neighbour(s) was transferred to the
query. Columns give the label-transfer rule (Top-1, or majority vote over
the $k=3,5,7$ nearest neighbours) and rows the resulting metric
(accuracy, balanced accuracy, macro-averaged F1-score). Axes are the four
evaluation cohorts (BCTherapy, TCGA-BRCA, Mayo Skin, Mayo Breast
Diagnosis); the radial scale spans $0$--$1$ with gridlines every $0.2$.
Six slide encoders are compared: GigaPath, TITAN, PRISM, and three WILSON
variants differing only in the text supervision used during alignment
(report distillation, keywords, CLIP-style contrastive). The WILSON
variants match or exceed the published foundation models across cohorts,
metrics and values of $k$, with the clearest separation on Mayo Skin and
Mayo Breast Diagnosis; GigaPath trails on every axis. Chance level differs
per axis because class counts and prevalences differ between cohorts.}
\label{fig:retrieval_radar_wsi}
\end{figure}

\begin{figure}[h]
\centering
\includegraphics[width=\textwidth]{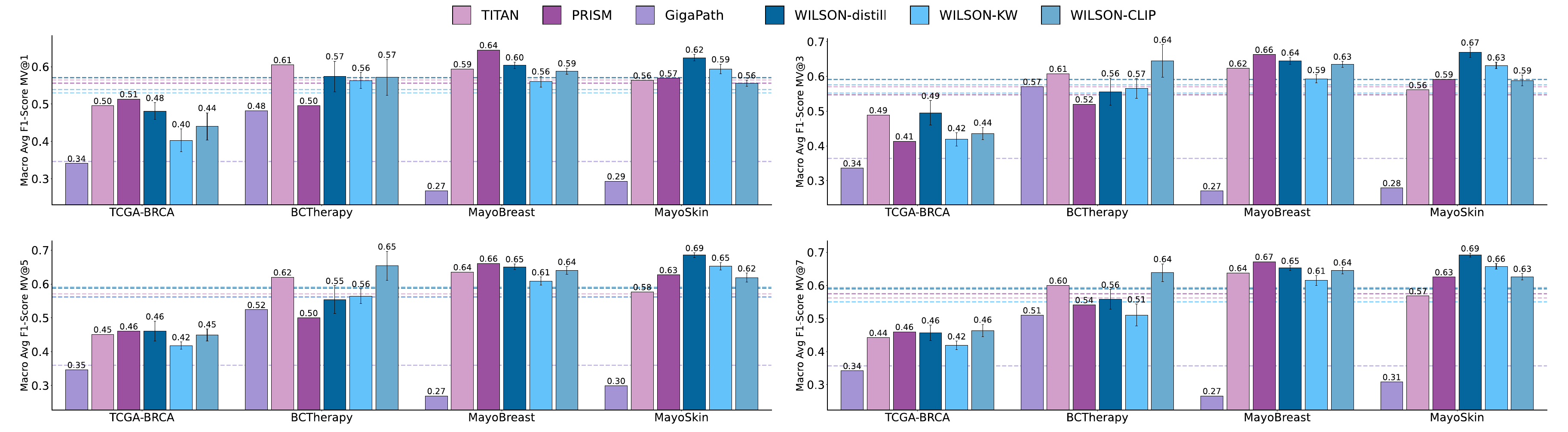}
\caption{\textbf{Slide-level image retrieval accuracy by cohort,
macro-averaged F1-score.} Per-cohort breakdown of the slide-level
retrieval results summarized in Fig.~\ref{fig:retrieval_radar_wsi}, using
the vision encoder of each model only. Panels correspond to the
label-transfer rule (majority vote over the $k=1,3,5,7$ nearest
neighbours); groups within each panel are the four evaluation cohorts
(TCGA-BRCA, BCTherapy, Mayo Breast Diagnosis, Mayo Skin). Within each
group, bars are ordered GigaPath, TITAN, PRISM, WILSON-distill,
WILSON-KW, WILSON-CLIP, the last three differing only in the text
supervision used during alignment (report distillation, keywords,
CLIP-style contrastive); error bars denote $[\text{s.d.\ across } n
\text{ folds / seeds}]$ and horizontal dashed lines mark each model's
mean across the four cohorts. Note the truncated vertical axis. The
WILSON encoders lead on Mayo Skin at every $k$ (up to 0.69 for
WILSON-distill at $k=7$ vs.\ 0.63 for PRISM and 0.57 for TITAN) and are
comparable to TITAN and PRISM on BCTherapy and Mayo Breast Diagnosis,
while TITAN and PRISM retain a small advantage on TCGA-BRCA at $k=1$.
GigaPath performs worst in every cohort and at every $k$, most markedly
on Mayo Breast Diagnosis and Mayo Skin.}
\label{fig:retrieval_bar_wsi}
\end{figure}

\begin{figure}[h]
\centering
\includegraphics[width=\textwidth]{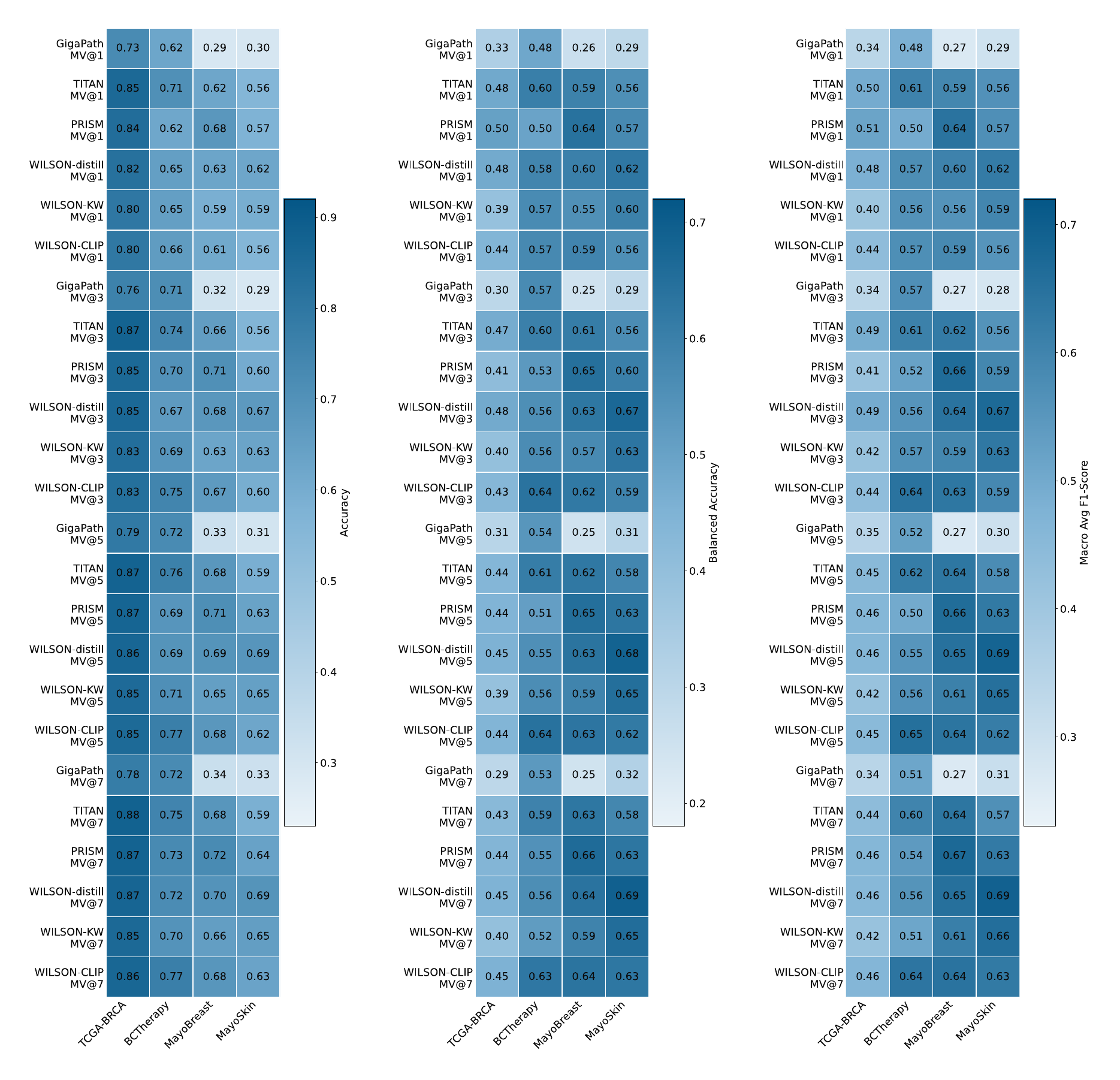}
\caption{\textbf{Complete slide-level image retrieval results.} Numerical
values underlying Figs.~\ref{fig:retrieval_radar_wsi} and
\ref{fig:retrieval_bar_wsi}, using the vision encoder of each model only.
Panels report accuracy (left), balanced accuracy (centre) and
macro-averaged F1-score (right); note that each panel has its own color
scale, so shading is comparable within but not across panels. Columns are
the four evaluation cohorts and rows every combination of model
(GigaPath, TITAN, PRISM, WILSON-distill, WILSON-KW, WILSON-CLIP) and
label-transfer rule (majority vote over the $k=1,3,5,7$ nearest
neighbours), grouped by $k$. Three patterns are visible. First, no single
encoder leads everywhere: WILSON-distill is strongest on Mayo Skin at
every $k$ (0.69 accuracy and 0.69 macro F1 at $k=7$, vs.\ 0.64 and 0.63
for PRISM), WILSON-CLIP on BCTherapy (0.77 accuracy at $k=5$), PRISM on
Mayo Breast Diagnosis (0.72 accuracy at $k=7$) and TITAN on TCGA-BRCA
(0.88 accuracy at $k=7$). Second, the gap between accuracy and balanced
accuracy is pronounced on TCGA-BRCA for all six encoders (e.g.\ 0.88
vs.\ 0.43 for TITAN at $k=7$), indicating that accuracy in that cohort is
dominated by class imbalance and that the balanced-accuracy and
macro-F1 panels are the informative ones. Third, GigaPath is the weakest
encoder throughout, and its deficit is largest on Mayo Breast Diagnosis
and Mayo Skin (balanced accuracy 0.25--0.34) while remaining competitive
on TCGA-BRCA accuracy (0.73--0.79), a pattern consistent with
majority-class prediction rather than discriminative retrieval.}
\label{fig:retrieval_heatmap_wsi}
\end{figure}

\begin{figure}[p]
  \centering
  \includegraphics[width=\linewidth]{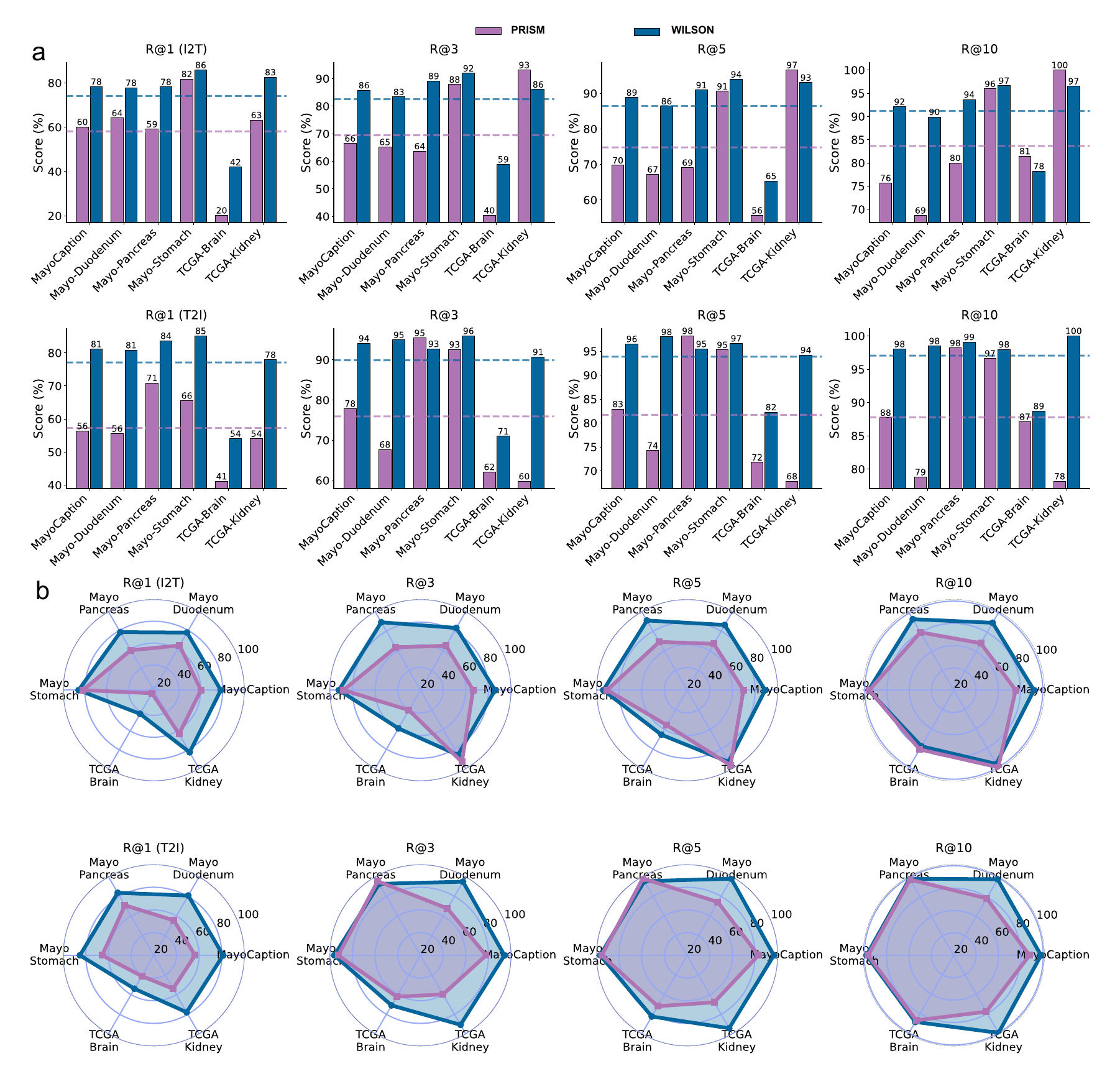}
\caption{
Comprehensive Performance Comparison of WILSON and PRISM Models on Cross-Modal Retrieval Tasks.
The figure presents recall metrics (R@1, R@3, R@5, R@10) evaluated across Mayo Clinic
(Diagnoses, Duodenum, Pancreas, Stomach) and TCGA (Brain, Kidney) datasets.
\textbf{Top section (Bar charts):} Quantitative comparison showing exact performance values
for each metric and dataset.
\textbf{Bottom section (Radar charts):} Multi-dimensional performance visualization enabling
visual assessment of relative strengths across all metrics and datasets simultaneously.
Each row represents a cross-modal retrieval direction:
\textit{Image-to-Text (I2T)} retrieves relevant diagnostic text reports given medical images;
\textit{Text-to-Image (T2I)} retrieves relevant medical images given text descriptions.
Blue elements (bars and lines) represent WILSON model performance, purple elements represent
PRISM model performance.
All metrics are reported as percentages, with bar chart values displayed above each bar for
precise quantitative reference.
}
  \label{fig:SuppRetrievalCrossModal}
\end{figure}

\begin{figure}[p]
\centering
\includegraphics[width=\textwidth]{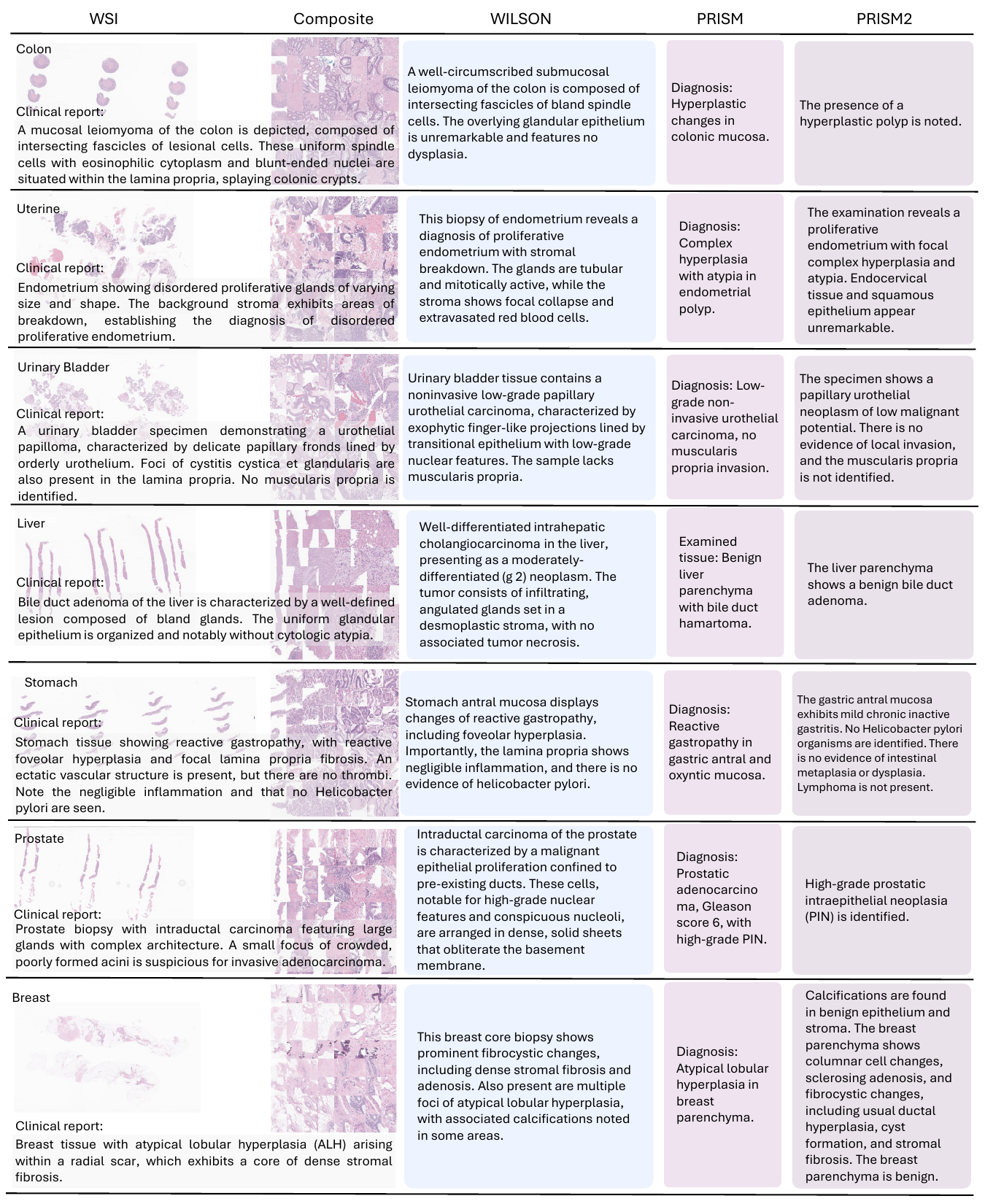}
\caption{\textbf{Additional qualitative comparison of generated slide-level
reports.} Seven further cases (colon, uterus, urinary bladder, liver, stomach,
prostate, breast), showing the whole-slide image and its clinical report, the
composite input, and the descriptions generated by WILSON, PRISM and PRISM2.
Conventions as in Fig.~\ref{fig:wilsonCaptioning}.}
\label{fig:caption_comparison_supp}
\end{figure}

\begin{figure*}[t]
\centering
\includegraphics[width=\textwidth]{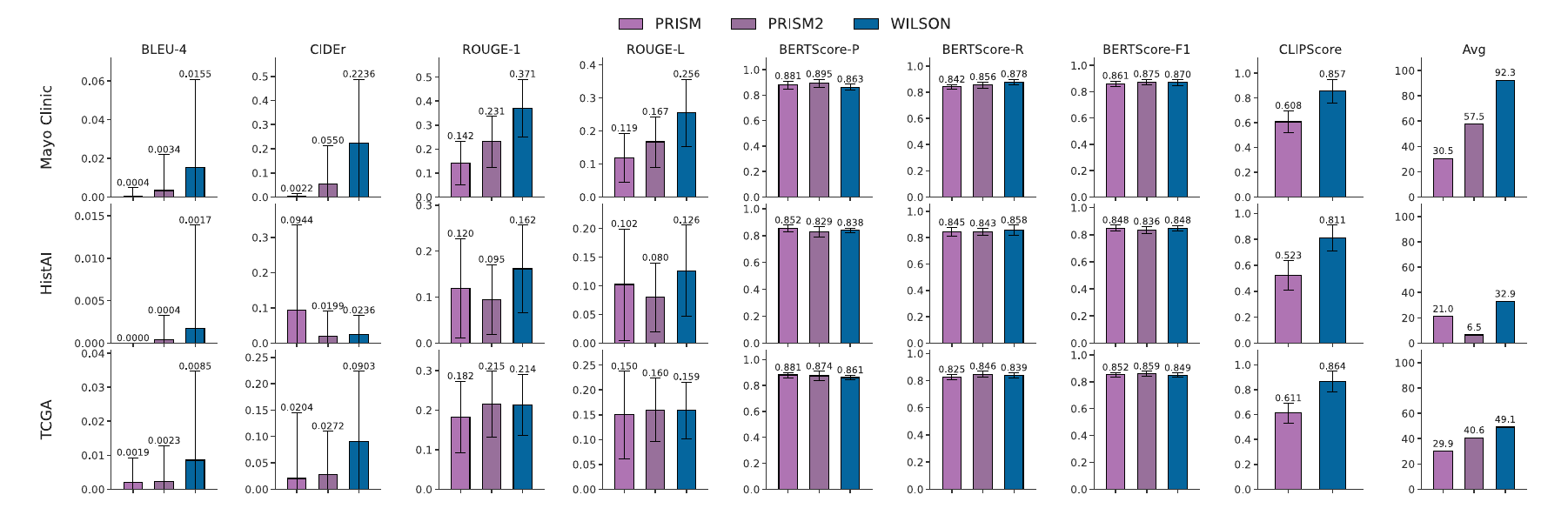}
\caption{\textbf{Caption-quality evaluation across cohorts, models, and
metrics.} Grouped bar charts of eight caption-quality metrics (columns) for
PRISM, PRISM2 and WILSON (ours) on three cohorts (rows): held-out Mayo Clinic,
HistAI and TCGA. Bars show the mean over caption pairs and error bars one
standard deviation; annotations give the mean value. Metrics are BLEU-4
($n$-gram precision), CIDEr (consensus-based TF--IDF-weighted $n$-gram
similarity), ROUGE-1 (unigram overlap), ROUGE-L (longest common subsequence),
BERTScore precision/recall/F1 (contextual similarity from BERT embeddings) and
CLIPScore (image--text similarity in a contrastive embedding space). CLIPScore
is computed within each model's own embedding space and is therefore not
comparable across models; PRISM2 exposes no general-purpose text encoder for
arbitrary strings, so no PRISM2 bar is shown. The rightmost panel (``Avg'')
summarizes each model as the mean of its metric values after min--max
normalization within metric and cohort, rescaled to 0--100
[excluding CLIPScore]. Note that $y$-axis ranges differ between panels: bar
heights are comparable within a panel but not across panels.}
\label{fig:caption_metrics_supp}
\end{figure*}

\begin{figure}[p]
  \centering
  \includegraphics[width=\linewidth]{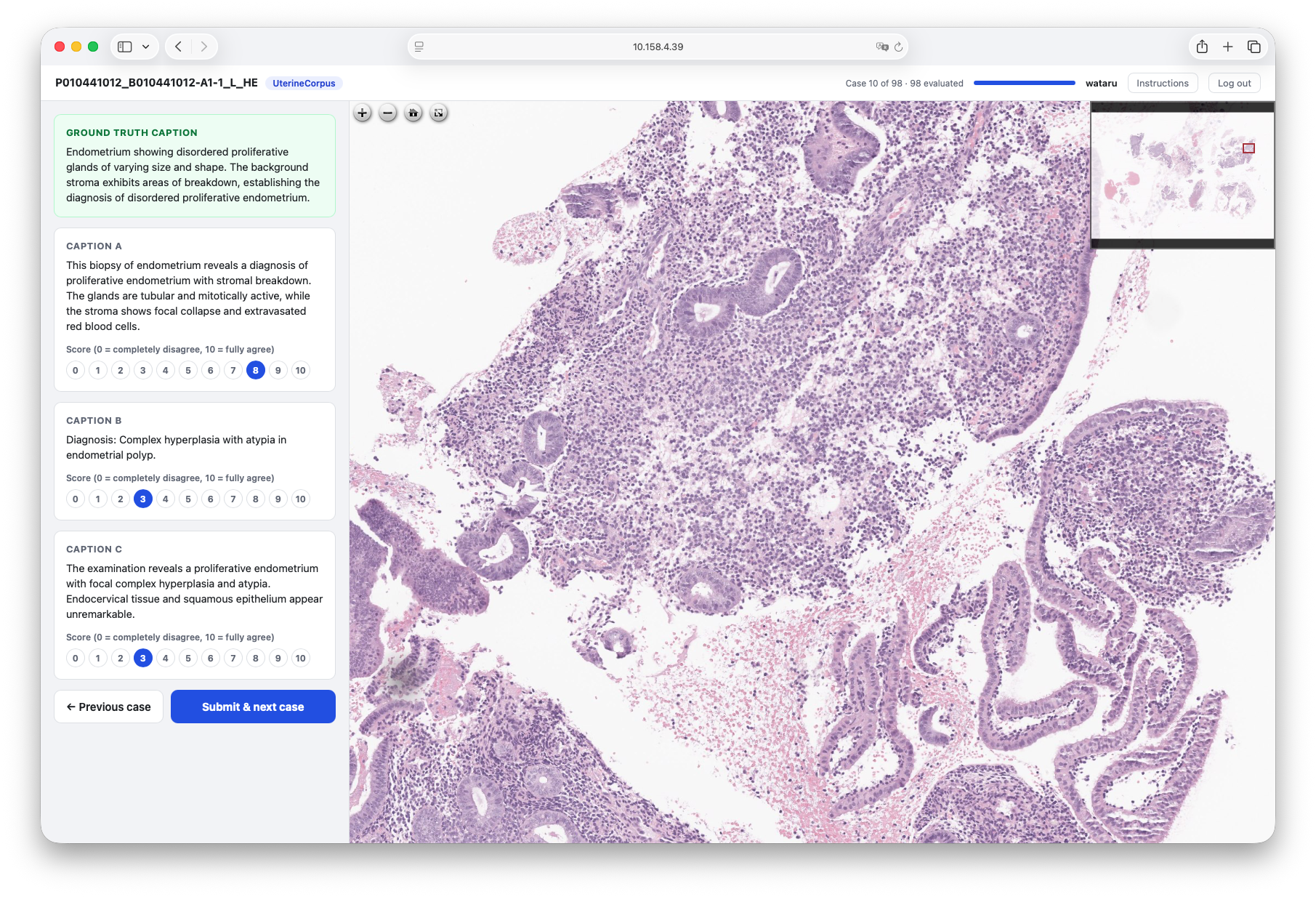}
\caption{\textbf{Blinded web interface for pathologist evaluation of
model-generated captions.} For each whole-slide image (right, with zoom,
pan and a thumbnail navigator), the pathologist rates the three candidate
captions from WILSON-CoCa, PRISM and PRISM2 (left, labeled A/B/C in an
anonymized order randomized independently for every case) against the
ground-truth, report-derived reference caption shown above them, using an
absolute 0--10 agreement scale (0 = completely disagree, 10 = fully agree)
scored independently for each caption so that ties are possible. A progress
bar tracks completion across the 98-WSI evaluation set (case 10 of 98
shown), and every submitted score is timestamped and written to a
per-pathologist log for reproducibility and audit.}
  \label{fig:pathologistInterface}
\end{figure}

\clearpage

\begin{table}[h]
\centering
\caption{Composite images vs.\ Yottixel patches as WSI representations, both
encoded with the pretrained UNI foundation model. Values are top-1 macro-F1
retrieval performance (higher is better) across two public and two internal
Mayo Clinic cohorts.}
\label{tab:compositeVsYottixel}
\begin{tabularx}{\textwidth}{lXcc}
\toprule
Dataset & Task & Yottixel + UNI & Composite + UNI \\
\midrule
BCTherapy & Treatment response & 0.51 & \textbf{0.56} \\
MayoSkin & Diagnostic classification & 0.33 & \textbf{0.34} \\
MayoBreast & Diagnostic classification & 0.31 & \textbf{0.35} \\
TCGA-Kidney & Diagnostic classification & \textbf{0.89} & 0.88 \\
\bottomrule
\end{tabularx}
\end{table}


\begin{table}[h]
\centering
\caption{End-to-end fine-tuning of WILSON (dense embedding) on MayoTNBC dataset.Accuracy (acc) and macro-averaged F1 (Macro avg) are reported for histologic subtyping and sTILs grading under three settings: the pretrained backbone evaluated zero-shot, the same backbone after end-to-end fine-tuning evaluated zero-shot, and the fine-tuned backbone with a jointly trained classification head.}
\label{tab:finetune}
\begin{tabularx}{\textwidth}{l *{6}{X}}
\toprule
& \multicolumn{2}{c}{Base zero-shot} & \multicolumn{2}{c}{Fine-tuned zero-shot} & \multicolumn{2}{c}{Fine-tuned classification} \\
\cmidrule(lr){2-3} \cmidrule(lr){4-5} \cmidrule(lr){6-7}
Task & acc & Macro avg & acc & Macro avg & acc & Macro avg \\
\midrule
Histology & 0.617 & 0.401 & 0.637 & 0.428 & 0.690 & 0.557 \\
sTILs      & 0.412 & 0.425 & 0.432 & 0.456 & 0.541 & 0.531 \\
\bottomrule
\end{tabularx}
\end{table}

\begin{table}[htbp]
\centering
\caption{Cross-modal retrieval on the MayoCaption dataset, horizontal evaluation across all diagnoses combined. $\Delta$ denotes WILSON minus PRISM.}
\label{tab:mayo_horizontal}
\begin{tabular*}{\textwidth}{@{\extracolsep{\fill}}llcccc@{}}
\toprule
Direction & Model & R@1 (\%) & R@3 (\%) & R@5 (\%) & R@10 (\%) \\
\midrule
\multirow{3}{*}{Image-to-Text} & WILSON & 78.29 & 85.75 & 88.82 & 92.11 \\
 & PRISM & 60.09 & 66.45 & 69.74 & 75.66 \\
 & $\Delta$ (W$-$P) & +18.20 & +19.30 & +19.08 & +16.45 \\
\midrule
\multirow{3}{*}{Text-to-Image} & WILSON & 81.14 & 94.08 & 96.49 & 98.03 \\
 & PRISM & 56.36 & 77.85 & 82.89 & 87.72 \\
 & $\Delta$ (W$-$P) & +24.78 & +16.23 & +13.60 & +10.31 \\
\bottomrule
\end{tabular*}
\vspace{2pt}
{\footnotesize WILSON: 456 samples; PRISM: 456 samples, 16 diagnosis labels.}
\end{table}

\begin{table}[htbp]
\centering
\caption{Cross-modal retrieval on the MayoCaption dataset, per-organ vertical evaluation for Duodenum. $\Delta$ denotes WILSON minus PRISM.}
\label{tab:mayo_duodenum}
\begin{tabular*}{\textwidth}{@{\extracolsep{\fill}}llcccc@{}}
\toprule
Direction & Model & R@1 (\%) & R@3 (\%) & R@5 (\%) & R@10 (\%) \\
\midrule
\multirow{3}{*}{Image-to-Text} & WILSON & 77.78 & 83.33 & 86.36 & 89.90 \\
 & PRISM & 64.14 & 65.15 & 67.17 & 68.69 \\
 & $\Delta$ (W$-$P) & +13.64 & +18.18 & +19.19 & +21.21 \\
\midrule
\multirow{3}{*}{Text-to-Image} & WILSON & 80.81 & 94.95 & 97.98 & 98.48 \\
 & PRISM & 55.56 & 67.68 & 74.24 & 78.79 \\
 & $\Delta$ (W$-$P) & +25.25 & +27.27 & +23.74 & +19.70 \\
\bottomrule
\end{tabular*}
\vspace{2pt}
{\footnotesize WILSON: 198 samples; PRISM: 198 samples, 5 diagnosis labels.}
\end{table}

\begin{table}[htbp]
\centering
\caption{Cross-modal retrieval on the MayoCaption dataset, per-organ vertical evaluation for Pancreas. $\Delta$ denotes WILSON minus PRISM.}
\label{tab:mayo_pancreas}
\begin{tabular*}{\textwidth}{@{\extracolsep{\fill}}llcccc@{}}
\toprule
Direction & Model & R@1 (\%) & R@3 (\%) & R@5 (\%) & R@10 (\%) \\
\midrule
\multirow{3}{*}{Image-to-Text} & WILSON & 78.18 & 89.09 & 90.91 & 93.64 \\
 & PRISM & 59.09 & 63.64 & 69.09 & 80.00 \\
 & $\Delta$ (W$-$P) & +19.09 & +25.45 & +21.82 & +13.64 \\
\midrule
\multirow{3}{*}{Text-to-Image} & WILSON & 83.64 & 92.73 & 95.45 & 99.09 \\
 & PRISM & 70.91 & 95.45 & 98.18 & 98.18 \\
 & $\Delta$ (W$-$P) & +12.73 & -2.73 & -2.73 & +0.91 \\
\bottomrule
\end{tabular*}
\vspace{2pt}
{\footnotesize WILSON: 110 samples; PRISM: 110 samples, 5 diagnosis labels.}
\end{table}

\begin{table}[htbp]
\centering
\caption{Cross-modal retrieval on the MayoCaption dataset, per-organ vertical evaluation for Stomach. $\Delta$ denotes WILSON minus PRISM.}
\label{tab:mayo_stomach}
\begin{tabular*}{\textwidth}{@{\extracolsep{\fill}}llcccc@{}}
\toprule
Direction & Model & R@1 (\%) & R@3 (\%) & R@5 (\%) & R@10 (\%) \\
\midrule
\multirow{3}{*}{Image-to-Text} & WILSON & 85.81 & 91.89 & 93.92 & 96.62 \\
 & PRISM & 81.76 & 87.84 & 90.54 & 95.95 \\
 & $\Delta$ (W$-$P) & +4.05 & +4.05 & +3.38 & +0.68 \\
\midrule
\multirow{3}{*}{Text-to-Image} & WILSON & 85.14 & 95.95 & 96.62 & 97.97 \\
 & PRISM & 65.54 & 92.57 & 95.27 & 96.62 \\
 & $\Delta$ (W$-$P) & +19.59 & +3.38 & +1.35 & +1.35 \\
\bottomrule
\end{tabular*}
\vspace{2pt}
{\footnotesize WILSON: 148 samples; PRISM: 148 samples, 6 diagnosis labels.}
\end{table}

\begin{table}[htbp]
\centering
\caption{Cross-modal retrieval on the TCGA Brain dataset. $\Delta$ denotes WILSON minus PRISM.}
\label{tab:tcga_brain}
\begin{tabular*}{\textwidth}{@{\extracolsep{\fill}}llcccc@{}}
\toprule
Direction & Model & R@1 (\%) & R@3 (\%) & R@5 (\%) & R@10 (\%) \\
\midrule
\multirow{3}{*}{Image-to-Text} & WILSON & 41.94 & 58.87 & 65.32 & 78.23 \\
 & PRISM & 20.16 & 40.32 & 55.65 & 81.45 \\
 & $\Delta$ (W$-$P) & +21.77 & +18.55 & +9.68 & -3.23 \\
\midrule
\multirow{3}{*}{Text-to-Image} & WILSON & 54.03 & 70.97 & 82.26 & 88.71 \\
 & PRISM & 41.13 & 62.10 & 71.77 & 87.10 \\
 & $\Delta$ (W$-$P) & +12.90 & +8.87 & +10.48 & +1.61 \\
\bottomrule
\end{tabular*}
\vspace{2pt}
{\footnotesize WILSON: 124 WSIs; PRISM: 124 WSIs. Disease groups: Anaplastic astrocytoma, Anaplastic oligodendroglioma, Astrocytoma, Glioblastoma, Mixed glioma, Oligodendroglioma.}
\end{table}

\begin{table}[htbp]
\centering
\caption{Cross-modal retrieval on the TCGA Kidney dataset. $\Delta$ denotes WILSON minus PRISM.}
\label{tab:tcga_kidney}
\begin{tabular*}{\textwidth}{@{\extracolsep{\fill}}llcccc@{}}
\toprule
Direction & Model & R@1 (\%) & R@3 (\%) & R@5 (\%) & R@10 (\%) \\
\midrule
\multirow{3}{*}{Image-to-Text} & WILSON & 82.56 & 86.05 & 93.02 & 96.51 \\
 & PRISM & 63.22 & 93.10 & 96.55 & 100.00 \\
 & $\Delta$ (W$-$P) & +19.34 & -7.06 & -3.53 & -3.49 \\
\midrule
\multirow{3}{*}{Text-to-Image} & WILSON & 77.91 & 90.70 & 94.19 & 100.00 \\
 & PRISM & 54.02 & 59.77 & 67.82 & 78.16 \\
 & $\Delta$ (W$-$P) & +23.88 & +30.93 & +26.37 & +21.84 \\
\bottomrule
\end{tabular*}
\vspace{2pt}
{\footnotesize WILSON: 86 WSIs; PRISM: 87 WSIs. Disease groups: Chromophobe renal cell carcinoma, Clear cell renal cell carcinoma, Papillary renal cell carcinoma.}
\end{table}

\begin{table}[htbp]
\centering
\caption{Summary of R@1 retrieval performance for WILSON and PRISM across all evaluated datasets, for both image-to-text (I2T) and text-to-image (T2I) retrieval directions.}
\label{tab:summary_r1}
\begin{tabular*}{\textwidth}{@{\extracolsep{\fill}}lcccc@{}}
\toprule
 & \multicolumn{2}{c}{Image-to-Text R@1 (\%)} & \multicolumn{2}{c}{Text-to-Image R@1 (\%)} \\
\cmidrule(lr){2-3} \cmidrule(lr){4-5}
Dataset & WILSON & PRISM & WILSON & PRISM \\
\midrule
MayoCaption (all) & 78.29 & 60.09 & 8
1.14 & 56.36 \\
Mayo -- Duodenum & 77.78 & 64.14 & 80.81 & 55.56 \\
Mayo -- Pancreas & 78.18 & 59.09 & 83.64 & 70.91 \\
Mayo -- Stomach & 85.81 & 81.76 & 85.14 & 65.54 \\
TCGA -- Brain & 41.94 & 20.16 & 54.03 & 41.13 \\
TCGA -- Kidney & 82.56 & 63.22 & 77.91 & 54.02 \\
\bottomrule
\end{tabular*}
\end{table}

\begin{table}[htbp]
\centering\footnotesize
\setlength{\tabcolsep}{4pt}
\caption{Comprehensive comparison of WILSON and PRISM cross-modal retrieval recall (R@1, R@3, R@5, R@10) across all used Mayo Clinic and TCGA dataset/organ splits, for both image-to-text (I2T) and text-to-image (T2I) retrieval.}
\label{tab:comprehensive_retrieval}
\begin{tabular*}{\textwidth}{@{\extracolsep{\fill}}llcccccccc@{}}
\toprule
 & & \multicolumn{4}{c}{WILSON (\%)} & \multicolumn{4}{c}{PRISM (\%)} \\
\cmidrule(lr){3-6} \cmidrule(lr){7-10}
Dataset & Direction & R@1 & R@3 & R@5 & R@10 & R@1 & R@3 & R@5 & R@10 \\
\midrule
\multirow{2}{*}{MayoCaption} & I2T & 78.29 & 85.75 & 88.82 & 92.11 & 60.09 & 66.45 & 69.74 & 75.66 \\
 & T2I & 81.14 & 94.08 & 96.49 & 98.03 & 56.36 & 77.85 & 82.89 & 87.72 \\
\midrule
\multirow{2}{*}{Mayo -- Duodenum} & I2T & 77.78 & 83.33 & 86.36 & 89.90 & 64.14 & 65.15 & 67.17 & 68.69 \\
 & T2I & 80.81 & 94.95 & 97.98 & 98.48 & 55.56 & 67.68 & 74.24 & 78.79 \\
\midrule
\multirow{2}{*}{Mayo -- Pancreas} & I2T & 78.18 & 89.09 & 90.91 & 93.64 & 59.09 & 63.64 & 69.09 & 80.00 \\
 & T2I & 83.64 & 92.73 & 95.45 & 99.09 & 70.91 & 95.45 & 98.18 & 98.18 \\
\midrule
\multirow{2}{*}{Mayo -- Stomach} & I2T & 85.81 & 91.89 & 93.92 & 96.62 & 81.76 & 87.84 & 90.54 & 95.95 \\
 & T2I & 85.14 & 95.95 & 96.62 & 97.97 & 65.54 & 92.57 & 95.27 & 96.62 \\
\midrule
\multirow{2}{*}{TCGA -- Brain} & I2T & 41.94 & 58.87 & 65.32 & 78.23 & 20.16 & 40.32 & 55.65 & 81.45 \\
 & T2I & 54.03 & 70.97 & 82.26 & 88.71 & 41.13 & 62.10 & 71.77 & 87.10 \\
\midrule
\multirow{2}{*}{TCGA -- Kidney} & I2T & 82.56 & 86.05 & 93.02 & 96.51 & 63.22 & 93.10 & 96.55 & 100.00 \\
 & T2I & 77.91 & 90.70 & 94.19 & 100.00 & 54.02 & 59.77 & 67.82 & 78.16 \\
\bottomrule
\end{tabular*}
\end{table}

\begin{table}[h]
\centering
\small
\caption{WILSON vs.\ PRISM at R@1 and R@10, all cohorts and directions. $\Delta$ = WILSON $-$ PRISM; bold negative values indicate PRISM leads.}
\label{tab:wilson_prism}
\begin{tabular*}{\textwidth}{@{\extracolsep{\fill}}lccc|ccc@{}}
\toprule
 & \multicolumn{3}{c}{\textbf{R@1}} & \multicolumn{3}{c}{\textbf{R@10}} \\
\textbf{Dataset / Direction} & \textbf{PRISM} & \textbf{WILSON} & \textbf{$\Delta$} & \textbf{PRISM} & \textbf{WILSON} & \textbf{$\Delta$} \\
\midrule
MayoCaption I2T & 60.1 & 78.3 & +18.2 & 75.7  & 92.1 & +16.4 \\
MayoCaption T2I & 56.4 & 81.1 & +24.7 & 87.7  & 98.0 & +10.3 \\
Mayo Duodenum I2T  & 64.1 & 77.8 & +13.7 & 68.7  & 89.9 & +21.2 \\
Mayo Duodenum T2I  & 55.6 & 80.8 & +25.2 & 78.8  & 98.5 & +19.7 \\
Mayo Pancreas I2T  & 59.1 & 78.2 & +19.1 & 80.0  & 93.6 & +13.6 \\
Mayo Pancreas T2I  & 70.9 & 83.6 & +12.7 & 98.2  & 99.1 & +0.9  \\
Mayo Stomach I2T   & 81.8 & 85.8 & +4.0  & 95.9  & 96.6 & +0.7  \\
Mayo Stomach T2I   & 65.5 & 85.1 & +19.6 & 96.6  & 98.0 & +1.4  \\
TCGA Brain I2T     & 20.2 & 41.9 & +21.7 & 81.5  & 78.2 & \textbf{-3.3} \\
TCGA Brain T2I     & 41.1 & 54.0 & +12.9 & 87.1  & 88.7 & +1.6  \\
TCGA Kidney I2T    & 63.2 & 82.6 & +19.4 & 100.0 & 96.5 & \textbf{-3.5} \\
TCGA Kidney T2I    & 54.0 & 77.9 & +23.9 & 78.2  & 100.0 & +21.8 \\
\midrule
\textbf{Mean} & 58.1 & 75.6 & +17.9 & 85.7 & 94.1 & +8.4 \\
\bottomrule
\end{tabular*}
\end{table}

\begin{table}[h]
\centering
\caption{\textbf{Parameter counts and computational cost of pathology foundation
models for a common tissue area.} Cost per WSI is measured for a
$22{,}400\times22{,}400$~px tissue region at $20\times$ ($\approx$125~mm$^2$),
tiled at each model's native tile size; for tile-encoder plus slide-aggregator
pipelines it is the tile encoder applied to all $N$ tiles plus one pass of the
slide aggregator, counted from forward passes of the released checkpoints
(2 FLOPs per multiply--accumulate). Parameters cover the vision path only (tile
encoder plus slide encoder, and MOOZY's case transformer); text towers and
language decoders (PRISM's BioGPT, PRISM2's Phi-3 Mini, TITAN's text encoder)
are excluded. Prov-GigaPath's cost includes LongNet's dilated attention as
executed. WILSON encodes one $2048\times2048$ composite of $8\times8=64$
selected tissue regions in a single forward pass, so no per-tile cost is
defined. Rows are ordered by cost per WSI.}
\label{tab:modelComparison}
\small
\newcommand{\stackL}[1]{\begin{tabular}[c]{@{}l@{}}#1\end{tabular}}
\newcommand{\stackR}[1]{\begin{tabular}[c]{@{}r@{}}#1\end{tabular}}
\begin{tabular}{@{}l l r r r r@{}}
\toprule
Model & Architecture & Params & \stackR{Tiles\\per WSI} & \stackR{FLOPs\\per tile} & \stackR{FLOPs\\per WSI} \\
\midrule
Prov-GigaPath & \stackL{ViT-g (DINOv2, 16-px patches)\\+ LongNet}          & 1,221M & \stackR{7,656\\(256 px)}  & 456.2G & 3.50P \\
\addlinespace
PRISM2        & \stackL{Virchow2 (ViT-H/14)\\+ PerceiverAR}               & 1,251M & \stackR{10,000\\(224 px)} & 340.2G & 3.40P \\
\addlinespace
PRISM         & \stackL{Virchow (ViT-H/14)\\+ Perceiver}                  & 730.3M & \stackR{10,000\\(224 px)} & 334.8G & 3.35P \\
\addlinespace
TITAN         & \stackL{CONCH v1.5 (ViT-L/16, 448 px)\\+ slide encoder}   & 354.6M & \stackR{1,914\\(512 px)}  & 538.5G & 1.03P \\
\addlinespace
MOOZY         & \stackL{ViT-S/8 (Lunit DINO)\\+ ViT 6L\\+ case transformer} & 85.8M & \stackR{10,000\\(224 px)} & 44.8G  & 450.96T \\
\addlinespace
WILSON        & \stackL{ConvNeXt-Base\\+ extended head (single pass)}     & 133.7M & \stackR{1\\composite}     & \textemdash & 2.58T \\
\bottomrule
\end{tabular}
\end{table}

\begin{table*}[t]
\centering
\small
\caption{Caption-quality evaluation across three models and three cohorts.
BLEU-4 ($n$-gram overlap), CIDEr (consensus-based), ROUGE-1 (unigram overlap),
ROUGE-L (longest common subsequence), BERTScore-P/R/F1 (contextual similarity
from BERT embeddings) and CLIPScore (CLIP-based semantic similarity) are
reported as raw scores. ``Avg'' is the mean performance across metrics after
min--max normalization within each metric, rescaled to 0--100. Best model per
cohort is bold. $^{a}$CLIPScore is computed in each model's own contrastive
embedding space and is therefore not comparable across models; PRISM2 exposes
no general-purpose text encoder for arbitrary strings. Excluded from bolding
and from Avg.}
\label{tab:caption_metrics}
\begin{tabularx}{\textwidth}{l*{9}{>{\centering\arraybackslash}X}}
\toprule
& \multicolumn{3}{c}{Mayo Clinic} & \multicolumn{3}{c}{HistAI} & \multicolumn{3}{c}{TCGA} \\
\cmidrule(lr){2-4} \cmidrule(lr){5-7} \cmidrule(lr){8-10}
Metric & PRISM & PRISM2 & WILSON & PRISM & PRISM2 & WILSON & PRISM & PRISM2 & WILSON \\
\midrule
BLEU-4        & 0.0004 & 0.0034 & \textbf{0.0155} & 0.0000 & 0.0004 & \textbf{0.0017} & 0.0019 & 0.0023 & \textbf{0.0085} \\
CIDEr         & 0.0022 & 0.0550 & \textbf{0.2236} & \textbf{0.0944} & 0.0199 & 0.0236 & 0.0204 & 0.0272 & \textbf{0.0903} \\
ROUGE-1       & 0.1424 & 0.2307 & \textbf{0.3711} & 0.1196 & 0.0948 & \textbf{0.1617} & 0.1825 & \textbf{0.2148} & 0.2136 \\
ROUGE-L       & 0.1189 & 0.1670 & \textbf{0.2556} & 0.1023 & 0.0801 & \textbf{0.1262} & 0.1497 & \textbf{0.1596} & 0.1592 \\
BERTScore-P   & 0.8812 & \textbf{0.8946} & 0.8627 & \textbf{0.8519} & 0.8290 & 0.8383 & \textbf{0.8808} & 0.8740 & 0.8609 \\
BERTScore-R   & 0.8421 & 0.8560 & \textbf{0.8781} & 0.8450 & 0.8434 & \textbf{0.8579} & 0.8252 & \textbf{0.8461} & 0.8386 \\
BERTScore-F1  & 0.8611 & \textbf{0.8745} & 0.8702 & \textbf{0.8480} & 0.8356 & 0.8476 & 0.8520 & \textbf{0.8592} & 0.8495 \\
CLIPScore$^{a}$ & 0.6077 & N/A & 0.8570 & 0.5232 & N/A & 0.8111 & 0.6112 & N/A & 0.8638 \\
\midrule
Avg           & 30.5 & 57.5 & \textbf{92.3} & 21.0 & 6.5 & \textbf{32.9} & 29.9 & 40.6 & \textbf{49.1} \\
\bottomrule
\end{tabularx}
\end{table*}

\begin{table}[htbp]
\centering
\caption{\textbf{Estimated marginal mean pathologist ratings by dataset and model.} Estimated marginal means (EMMs) and 95\% confidence intervals were derived
from a linear mixed-effects model with fixed effects for dataset, model, and
their interaction and crossed random intercepts for WSI and pathologist:
score $\sim$ dataset $\times$ model + (1$\mid$wsi\_id) +
(1$\mid$pathologist).}
\label{tab:pathologist_emm}

\begin{tabularx}{\textwidth}{lXcc}
\toprule
Dataset & Model & EMM & 95\% CI \\
\midrule
\multirow{3}{*}{Mayo}
 & WILSON-CoCa & 5.76 & 5.05--6.46 \\
 & PRISM2      & 5.41 & 4.70--6.11 \\
 & PRISM       & 3.62 & 2.92--4.33 \\
\midrule
\multirow{3}{*}{HistAI}
 & WILSON-CoCa & 4.58 & 3.86--5.30 \\
 & PRISM       & 2.39 & 1.67--3.11 \\
 & PRISM2      & 0.64 & $-0.08$--1.36 \\
\bottomrule
\end{tabularx}
\end{table}

\begin{table}[htbp]
\centering
\caption{\textbf{Pairwise model contrasts in pathologist ratings within each dataset.} Differences represent pairwise contrasts between estimated marginal means
from the linear mixed-effects model. Positive values indicate a higher rating
for the first model listed in each contrast. Wald $z$-tests were two-sided
and unadjusted for multiple comparisons.}
\label{tab:pathologist_pairwise}

\begin{tabularx}{\textwidth}{lXcccc}
\toprule
Dataset & Contrast & Difference & 95\% CI & Wald $z$ & $P$ value \\
\midrule
\multirow{3}{*}{Mayo}
 & WILSON-CoCa $-$ PRISM  & 2.13 & 1.53--2.73     & 6.96 & $<0.001$ \\
 & WILSON-CoCa $-$ PRISM2 & 0.35 & $-0.25$--0.95  & 1.14 & 0.254 \\
 & PRISM2 $-$ PRISM       & 1.78 & 1.18--2.38     & 5.82 & $<0.001$ \\
\midrule
\multirow{3}{*}{HistAI}
 & WILSON-CoCa $-$ PRISM  & 2.19 & 1.57--2.80     & 6.97  & $<0.001$ \\
 & WILSON-CoCa $-$ PRISM2 & 3.94 & 3.33--4.55     & 12.57 & $<0.001$ \\
 & PRISM $-$ PRISM2       & 1.75 & 1.14--2.37     & 5.60  & $<0.001$ \\
\bottomrule
\end{tabularx}
\end{table}


\begin{table}[t]
\centering
\small
\caption{\textbf{Design comparison of WILSON with existing slide- and case-level pathology foundation models.}
Every row is a binary feature for which \cmark\ is the favorable outcome.
Text generation released: a text generator is publicly available to the research community (TITAN describes text generation, but its generator has not been released).
Compact encoder: $<$200M parameters. Low compute: $<$10 TFLOPs per WSI ($22{,}400\times22{,}400$~px tissue region at $20\times$).
Underlying values are given in Fig.~\ref{fig:wilson}e: parameters (MOOZY 85.8M, TITAN 354.6M, PRISM 730.3M, PRISM2 1,251M, Prov-GigaPath 1,221M, WILSON 133.7M) and FLOPs per WSI (451.0T, 1.03P, 3.35P, 3.40P, 3.50P, 2.6T).
Entries are based on the published descriptions of each model.}
\label{tab:featureComparison}
\begin{tabular}{@{}lcccccc@{}}
\toprule
Feature & MOOZY & TITAN & PRISM & PRISM2 & Prov-GigaPath & \textbf{WILSON} \\
\midrule
Native multi-slide (patient-level) input & \cmark & \xmark & \xmark & \xmark & \xmark & \cmark \\
Multi-magnification input in one image & \xmark & \xmark & \xmark & \xmark & \xmark & \cmark \\
End-to-end fine-tuning of full encoder & \xmark & \xmark & \xmark & \xmark & \xmark & \cmark \\
Works without pre-extracted tile features & \xmark & \xmark & \xmark & \xmark & \xmark & \cmark \\
Text generation released & \xmark & \xmark & \cmark & \cmark & \xmark & \cmark \\
Compact encoder ($<$200M parameters) & \cmark & \xmark & \xmark & \xmark & \xmark & \cmark \\
Low compute ($<$10 TFLOPs per WSI) & \xmark & \xmark & \xmark & \xmark & \xmark & \cmark \\
\bottomrule
\end{tabular}
\end{table}

\clearpage

\section{Supplementary Data}\label{data}

\begin{table}[t]
\centering
\caption{Summary of training and validation WSI datasets. The MayoTNBC dataset contains two tasks, histological subtype classification and stromal tumor-infiltrating lymphocyte (sTIL) prediction, each comprising four classes. The MayoCaption dataset was constructed by combining the MayoDuodenum, MayoStomach, and MayoPancreas datasets for horizontal cross-modal retrieval. Histological subtype annotations for the TCGA-derived datasets were assigned according to Uegami et al.~\cite{uegami_2026_19736866}.}
\label{tab:wsi_datasets}
\small
\begin{tabular}{l p{2.8cm} p{2.8cm} c c c c}
\toprule
Dataset & Organ & Task & Classes & Patients & Slides & Format \\
\midrule
Mayo189K &
Various &
WILSON Training &
N/A &
112,296 &
189,291 &
DICOM \\

MayoBreast &
Breast (tumor, nontumor) &
Histological subtype classification &
12 &
1,289 &
1,339 &
DICOM \\

MayoSkin &
Skin (tumor, nontumor) &
Histological subtype classification &
20 &
677 &
688 &
DICOM \\

MayoTNBC &
Breast carcinoma (triple negative) &
Fine tuning &
4/4 &
508 &
508 &
Philips Tiff \\

MayoCaption &
Three organs &
Cross-modal retrieval &
18 &
397 &
456 &
DICOM \\

MayoDuodenum &
Duodenum &
Cross-modal retrieval &
7 &
172 &
198 &
DICOM \\

MayoStomach &
Stomach &
Cross-modal retrieval &
6 &
127 &
148 &
DICOM \\

MayoPancreas &
Pancreas &
Cross-modal retrieval &
5 &
106 &
110 &
DICOM \\

BCTherapy~\cite{sammut2022multiomic} &
Breast carcinoma &
Treatment response prediction &
2 &
160 &
160 &
SVS \\

TCGA-BRCA~\cite{weinstein2013cancer} &
Breast carcinoma &
Histologic subtype classification &
7 &
1,018 &
1,087 &
SVS \\

TCGA-Brain~\cite{weinstein2013cancer} &
Brain tumor &
Cross-modal retrieval &
6 &
879 &
1,703 &
SVS \\

TCGA-Kidney~\cite{weinstein2013cancer} &
Kidney &
Cross-modal retrieval &
3 &
885 &
927 &
SVS \\

HistAI~\cite{nechaev2025histai} &
10 organs &
Cross-modal retrieval &
N/A &
48 &
48 &
generic Tiff \\

\bottomrule
\end{tabular}
\end{table}

\begin{table}[t]
\centering
\caption{Summary of case-level validation WSI datasets.}
\label{tab:case_datasets}
\small
\begin{tabular}{l p{2.8cm} p{2.8cm} c c c c}
\toprule
Dataset & Organ & Task & Classes & Cases & WSIs & Format \\
\midrule
MayoCaseBreast &
Breast (tumor, nontumor) &
Histological subtype classification &
12 &
494 &
4,361 &
DICOM \\

MayoCaseSkin &
Skin (tumor, nontumor) &
Histological subtype classification &
20 &
500 &
1,280 &
DICOM \\

MayoBreastSubtype &
Breast carcinoma &
Molecular subtype classification &
3 &
846 &
2,254 &
DICOM \\

CPTAC-BRCA~\cite{CPTAC_BRCA_2020} &
Breast carcinoma &
PAM50 subtype classification &
5 &
198 &
653 &
SVS \\
\bottomrule
\end{tabular}
\end{table}

\begin{table}[t]
\centering
\caption{Class distribution of the TCGA-BRCA dataset.~\cite{weinstein2013tcga}}
\label{tab:tcga_brca_distribution}
\begin{tabularx}{\textwidth}{Xr}
\toprule
Class & Count \\
\midrule
Invasive carcinoma of no special type & 812 \\
Invasive lobular carcinoma & 204 \\
Ductal and lobular carcinoma & 27 \\
Mucinous adenocarcinoma & 20 \\
Metaplastic carcinoma & 13 \\
Medullary carcinoma & 7 \\
Invasive micropapillary carcinoma & 4 \\
\midrule
Total & 1{,}087 \\
\bottomrule
\end{tabularx}
\end{table}

\begin{table}[t]
\centering
\caption{Class distribution of the BCTherapy dataset~\cite{sammut2022multiomic}.}
\label{tab:bctherapy_distribution}
\begin{tabularx}{\textwidth}{Xr}
\toprule
Class & Count \\
\midrule
RD & 119 \\
pCR & 41 \\
\bottomrule
\end{tabularx}
\end{table}

\begin{table}[t]
\centering
\caption{Class distribution of the MayoBRCA dataset.}
\label{tab:mayobrca_distribution}
\begin{tabularx}{\textwidth}{Xr}
\toprule
Class & Count \\
\midrule
Invasive breast carcinoma of no special type (ductal) & 350 \\
Invasive lobular carcinoma & 70 \\
Mixed ductal and lobular carcinoma & 70 \\
Mucinous carcinoma & 70 \\
Metaplastic carcinoma & 70 \\
Ductal carcinoma in situ & 210 \\
Lobular carcinoma in situ & 70 \\
Fibroadenoma & 70 \\
Phyllodes tumor & 79 \\
Lymphoma or lymphoproliferative lesion & 70 \\
Benign proliferative lesion & 140 \\
Fat necrosis & 70 \\
\bottomrule
\end{tabularx}\label{tab:dataset_detail_MayoBreast}
\end{table}

\begin{table}[t]
\centering
\caption{Class distribution of the MayoSkin dataset.}
\label{tab:mayoskin_distribution}
\begin{tabularx}{\textwidth}{Xr}
\toprule
Class & Count \\
\midrule
Lentigo & 41 \\
Actinic keratosis & 40 \\
Malignant melanoma & 40 \\
Blue nevus & 39 \\
Dermatofibroma & 38 \\
Verrucae & 38 \\
Epidermal cyst & 38 \\
Basal cell carcinoma & 37 \\
Neurofibroma & 37 \\
Psoriasiform dermatitis & 36 \\
Nevus & 36 \\
Nevus with atypia & 36 \\
Chondrodermatitis nodularis helicis & 35 \\
Squamous cell carcinoma & 35 \\
Seborrheic keratosis & 33 \\
Sebaceous gland hyperplasia & 33 \\
Scar & 32 \\
Skin tag & 31 \\
Lichenoid dermatitis & 29 \\
Lymphoma & 20 \\
\bottomrule
\end{tabularx}\label{tab:dataset_detail_MayoSkin}
\end{table}

\begin{table}[t]
\centering
\caption{Class distribution of the MayoBreastSubtype dataset.}
\label{tab:mayobreastsubtype_distribution}
\begin{tabularx}{\textwidth}{Xr}
\toprule
Class & Count \\
\midrule
Hormone receptor-positive/HER2-negative & 269 \\
HER2-positive & 303 \\
Triple-negative & 274 \\
\bottomrule
\end{tabularx}
\end{table}



\begin{table}[t]
\centering
\caption{Class distribution of the CPTAC-BRCA dataset~\cite{CPTAC_BRCA_2020}.}
\label{tab:cptac_brca_distribution}
\begin{tabularx}{\textwidth}{Xr}
\toprule
Class & Count \\
\midrule
Luminal A & 57 \\
Basal-like & 27 \\
Luminal B & 17 \\
HER2-enriched & 14 \\
Normal-like & 5 \\
\bottomrule
\end{tabularx}
\end{table}

\begin{table}[t]
\centering
\caption{Class distribution of the MayoCaseBreast dataset.}
\label{tab:dataset_detail_MayoCaseBreast}
\begin{tabularx}{\textwidth}{Xr}
\toprule
Class & Count \\
\midrule
Invasive carcinoma of no special type (ductal) & 44 \\
Invasive lobular carcinoma & 42 \\
Mixed ductal and lobular carcinoma & 40 \\
Mucinous carcinoma & 41 \\
Metaplastic carcinoma & 38 \\
Ductal carcinoma in situ & 40 \\
Lobular carcinoma in situ & 45 \\
Fibroadenoma & 37 \\
Phyllodes tumor & 43 \\
Lymphoma or lymphoproliferative disease & 42 \\
Benign proliferative lesion & 40 \\
Fat necrosis & 48 \\
\bottomrule
\end{tabularx}
\end{table}



\begin{table}[t]
\centering
\caption{Class distribution of the MayoCaseSkin dataset.}
\label{tab:mayocaseskin_distribution}
\begin{tabularx}{\textwidth}{Xr}
\toprule
Class & Count \\
\midrule
Sebaceous gland hyperplasia & 31 \\
Dermatofibroma & 31 \\
Malignant melanoma & 31 \\
Skin tag & 29 \\
Nevus with atypia & 28 \\
Seborrheic keratosis & 27 \\
Blue nevus & 27 \\
Scar & 26 \\
Squamous cell carcinoma & 26 \\
Lentigo & 25 \\
Basal cell carcinoma & 25 \\
Lichenoid dermatitis & 24 \\
Actinic keratosis & 24 \\
Chondrodermatitis nodularis helicis & 23 \\
Epidermal cyst & 23 \\
Neurofibroma & 23 \\
Nevus & 22 \\
Psoriasiform dermatitis & 21 \\
Verruca & 20 \\
Cutaneous lymphoma & 14 \\
\bottomrule
\end{tabularx}
\end{table}


\begin{table}[ht] \centering \caption{HistAI~\cite{nechaev2025histai} whole-slide images (WSIs) and corresponding organ categories.} \begin{tabular}{ll} \hline \textbf{WSI ID} & \textbf{Organ} \\ \hline case\_00760/slide\_H\&E\_0.tiff & Bone \\ case\_03070/slide\_H\&E\_0.tiff & Bone \\ case\_09511/slide\_H\&E\_0.tiff & Bone \\ case\_11368/slide\_H\&E\_0.tiff & Bone \\ case\_11593/slide\_H\&E\_0.tiff & Bone \\ case\_01871/slide\_H\&E\_0.tiff & Breast \\ case\_04775/slide\_H\&E\_0.tiff & Breast \\ case\_04858/slide\_H\&E\_0.tiff & Breast \\ case\_12828/slide\_H\&E\_0.tiff & Breast \\ case\_00107/slide\_H\&E\_0.tiff & Endocrine \\ case\_03889/slide\_H\&E\_0.tiff & Endocrine \\ case\_04347/slide\_H\&E\_0.tiff & Endocrine \\ case\_13008/slide\_H\&E\_0.tiff & Endocrine \\ case\_14497/slide\_H\&E\_0.tiff & Endocrine \\ case\_03357/slide\_H\&E\_0.tiff & Gastrointestinal \\ case\_04754/slide\_H\&E\_0.tiff & Gastrointestinal \\ case\_13674/slide\_H\&E\_0.tiff & Gastrointestinal \\ case\_14249/slide\_H\&E\_0.tiff & Gastrointestinal \\ case\_16527/slide\_H\&E\_0.tiff & Gastrointestinal \\ case\_05879/slide\_H\&E\_0.tiff & Genitourinary \\ case\_07247/slide\_H\&E\_0.tiff & Genitourinary \\ case\_10189/slide\_H\&E\_0.tiff & Genitourinary \\ case\_14971/slide\_H\&E\_0.tiff & Genitourinary \\ case\_01259/slide\_H\&E\_0.tiff & Gynecologic \\ case\_03934/slide\_H\&E\_0.tiff & Gynecologic \\ case\_05088/slide\_H\&E\_0.tiff & Gynecologic \\ case\_12091/slide\_H\&E\_0.tiff & Gynecologic \\ case\_15397/slide\_H\&E\_0.tiff & Gynecologic \\ case\_01585/slide\_H\&E\_0.tiff & Head and Neck \\ case\_02038/slide\_H\&E\_0.tiff & Head and Neck \\ case\_08881/slide\_H\&E\_0.tiff & Head and Neck \\ case\_13539/slide\_H\&E\_0.tiff & Head and Neck \\ case\_16658/slide\_H\&E\_0.tiff & Head and Neck \\ case\_01705/slide\_H\&E\_0.tiff & Skin \\ case\_12553/slide\_H\&E\_0.tiff & Skin \\ case\_17219/slide\_H\&E\_0.tiff & Skin \\ case\_17492/slide\_H\&E\_0.tiff & Skin \\ case\_17619/slide\_H\&E\_0.tiff & Skin \\ case\_00315/slide\_H\&E\_0.tiff & Soft tissue \\ case\_05077/slide\_H\&E\_0.tiff & Soft tissue \\ case\_06560/slide\_H\&E\_1.tiff & Soft tissue \\ case\_08261/slide\_H\&E\_0.tiff & Soft tissue \\ case\_11651/slide\_H\&E\_0.tiff & Soft tissue \\ case\_03457/slide\_H\&E\_0.tiff & Thorax \\ case\_06572/slide\_H\&E\_0.tiff & Thorax \\ case\_12689/slide\_H\&E\_0.tiff & Thorax \\ case\_15640/slide\_H\&E\_0.tiff & Thorax \\ case\_16730/slide\_H\&E\_0.tiff & Thorax \\ \hline \end{tabular}  \label{tab:histAI_wsi_list}\end{table}


\begin{table}[ht] \centering \caption{TCGA~\cite{weinstein2013cancer} whole-slide images (WSIs) and corresponding organ categories.} \begin{tabular}{ll} \hline \textbf{WSI ID} & \textbf{Organ} \\ \hline TCGA-06-0143-01Z-00-DX3.e9011249-11f6-454b-98f1-7f2bcfea228c.svs & Brain \\ TCGA-06-0154-01Z-00-DX1.e52d5443-e663-451f-8a1f-83ec5bb9a401.svs & Brain \\ TCGA-DB-A75P-01Z-00-DX1.11655479-5A76-4689-B8F6-319F882B5EB7.svs & Brain \\ TCGA-DU-A6S3-01Z-00-DX1.08CE6930-D4E8-4084-A8C0-F2911B069E72.svs & Brain \\ TCGA-HT-7680-01Z-00-DX3.C157242E-F4D2-4246-8016-B0A44C4C4AAD.svs & Brain \\ TCGA-AC-A2QH-01Z-00-DX1.00B8BFFF-F1E2-4F99-A969-8DD7EE4F8E0B.svs & Breast \\ TCGA-AR-A24K-01Z-00-DX1.3A56BAEC-484E-4F9B-BCB4-360ABF6DDB4B.svs & Breast \\ TCGA-BH-A1F2-01Z-00-DX1.17E2FD6F-0DCF-425B-864B-21ADCDAE734B.svs & Breast \\ TCGA-BH-A1F8-01Z-00-DX1.8BB026F7-35CB-483F-B665-4C3A3EF47E1B.svs & Breast \\ TCGA-MS-A51U-01Z-00-DX1.490DE85A-ECE5-4E2A-9657-841BE6FFCCA0.svs & Breast \\ TCGA-DS-A1OC-01Z-00-DX1.D3ECE6F5-90DC-4CF4-A203-AC9F28E89959.svs & Cervix \\ TCGA-IR-A3LC-01Z-00-DX1.D5E87530-B079-4374-9DFE-4100CE6A4CDA.svs & Cervix \\ TCGA-MY-A5BF-01Z-00-DX3.CF669DE8-217E-47A6-839F-ABE9657BA566.svs & Cervix \\ TCGA-VS-A8QF-01Z-00-DX1.D8DB41E0-8C0F-4394-A2A2-8721BE35644B.svs & Cervix \\ TCGA-ZJ-AB0H-01Z-00-DX1.1F0584E3-1E50-4755-BF3D-EB62502CB975.svs & Cervix \\ TCGA-A6-5665-01Z-00-DX1.3ad2c249-d138-4037-a59b-4747ce2b789a.svs & Colorectum \\ TCGA-AA-3538-01Z-00-DX1.60d0b039-25d6-4b71-a36f-5b2764a983ef.svs & Colorectum \\ TCGA-AA-A02E-01Z-00-DX1.04D47621-9DCF-437C-A4D6-44D17579FE6D.svs & Colorectum \\ TCGA-CK-6748-01Z-00-DX1.1dd76660-7858-470c-a27b-36586b788125.svs & Colorectum \\ TCGA-G4-6628-01Z-00-DX1.d67973d1-9544-47e1-9ecb-e9d8d7f310e6.svs & Colorectum \\ TCGA-B9-4113-01Z-00-DX1.91e4a568-8f5b-4726-8352-94adf056f912.svs & Kidney \\ TCGA-B9-4617-01Z-00-DX1.d890e850-cc29-49f8-b854-cbe9a9a73c95.svs & Kidney \\ TCGA-CZ-4861-01Z-00-DX1.c36faa05-5e3b-4711-897e-c91f84405870.svs & Kidney \\ TCGA-CZ-5468-01Z-00-DX1.e2bbe417-a24c-4511-934e-674221855695.svs & Kidney \\ TCGA-EU-5905-01Z-00-DX1.f83b5cb0-7295-4f67-acac-cc1768b07c60.svs & Kidney \\ TCGA-33-4582-01Z-00-DX1.629AEDB6-E9AA-4615-92E8-5DDAAFF6103E.svs & Lung \\ TCGA-44-6144-01Z-00-DX1.604b3c7c-92e8-474a-bae8-e48415ea6196.svs & Lung \\ TCGA-49-6744-01Z-00-DX4.a3d7995d-399f-4c53-aab8-adc4ea4dbfa8.svs & Lung \\ TCGA-66-2785-01Z-00-DX1.b9439ee1-d22b-4ccd-b53b-ce7717a37a17.svs & Lung \\ TCGA-XC-AA0X-01Z-00-DX1.61A34BE0-F16B-4EC1-8E7F-7BF94F6629F4.svs & Lung \\ TCGA-DX-A3LS-01Z-00-DX1.32E89F90-9C89-4EBE-9AD3-DF1D2457F6C8.svs & Soft Tissue \\ TCGA-QQ-A8VF-01Z-00-DX1.708A1A71-F284-4FBE-A2A1-E4037788A9F8.svs & Soft Tissue \\ TCGA-SI-AA8C-01Z-00-DX2.BD0020ED-4F05-477E-AC29-6C0DB4DF62C6.svs & Soft Tissue \\ TCGA-WK-A8XO-01Z-00-DX5.45E4CD45-C8B2-4686-861C-5E7264FBA983.svs & Soft Tissue \\ TCGA-X6-A7WB-01Z-00-DX3.F2662EFE-A72B-41D7-9E0F-55C40E01BB06.svs & Soft Tissue \\ TCGA-BR-7715-01Z-00-DX1.30aff8b3-bb1b-4aff-9f6b-2dde40122359.svs & Stomach \\ TCGA-D7-A4YT-01Z-00-DX1.645AF39D-1B70-4C46-AB80-7B532994EE66.svs & Stomach \\ TCGA-FP-7735-01Z-00-DX1.97aac46a-898e-4c43-8c93-1496f40c43ed.svs & Stomach \\ TCGA-VQ-A8P5-01Z-00-DX1.F3FDCB47-3426-400C-82A9-750012F95B3B.svs & Stomach \\ TCGA-VQ-A91N-01Z-00-DX1.4F7CF3DD-8AAA-43DE-98C1-E44B2B5FB019.svs & Stomach \\ TCGA-BJ-A45I-01Z-00-DX1.29C9DC34-228A-4712-A028-07A396B4D1BD.svs & Thyroid \\ TCGA-DJ-A2PW-01Z-00-DX1.57FC77A8-DDCC-40DC-9B08-79DE5EFD02F0.svs & Thyroid \\ TCGA-EM-A2OZ-01Z-00-DX1.3146F96A-8BFE-46D2-AFFA-8593A56B58CD.svs & Thyroid \\ TCGA-ET-A3DR-01Z-00-DX1.702323F9-77A9-4D26-8236-F009C6CFADE4.svs & Thyroid \\ TCGA-FE-A3PC-01Z-00-DX1.EBF30F16-EDC4-4324-BB4E-2F5A9F9AF05E.svs & Thyroid \\ TCGA-AP-A1DR-01Z-00-DX1.59D56F74-930F-4B0D-91D4-B3C0B4A92073.svs & Uterus \\ TCGA-AX-A062-01Z-00-DX2.3B3151D7-83C9-4932-9A23-D6FE4D490F8B.svs & Uterus \\ TCGA-D1-A17S-01Z-00-DX1.8EBE62E2-B0D1-480D-8A1C-97E7359FAA79.svs & Uterus \\ TCGA-DF-A2KY-01Z-00-DX1.AE1DDE62-EA95-4DC0-9783-849591B6B656.svs & Uterus \\ TCGA-FI-A2EW-01Z-00-DX1.93D21AD2-076D-4685-A027-623C294974A5.svs & Uterus \\ \hline \end{tabular} \label{tab:tcga_wsi_list} \end{table}

\begin{table}[h]
\centering
\small
\caption{Training corpus and compute per vision--language stage.
\textsuperscript{*}WILSON-CLIP uses a fixed 1,410,600/349,861 train/val
composite manifest (\S\ref{sec:vlm_clip}); all other rows are approximate
WSI-level splits computed from each script's \texttt{val\_frac} argument
applied to the paired corpus above, since exact per-sample assignment is
determined at run time by the data loader. ``Eff.\ batch'' is
GPUs~$\times$~per-GPU batch size~$\times$~gradient-accumulation steps.}
\label{tab:vlm_corpus}
\begin{tabularx}{\textwidth}{Xrrcccc}
\toprule
Stage & Paired WSIs & Composites & Val.\ frac. & Epochs & GPUs & Eff.\ batch  \\
\midrule
WILSON-distill & 178,020 & 1,759,787 & 0.02 & 20 & 32 (4$\times$8) & 2,048 \\
WILSON-KW (keyword regr.) & 178,088 & 1,760,461 & 0.10 & 7 & 8 (1$\times$8) & 512  \\
WILSON-CLIP, Phase 2 & 1,760,461\textsuperscript{*} & 1,760,461\textsuperscript{*} & 0.199\textsuperscript{*} & 20 & 32 (4$\times$8) & 576 \\
WILSON-CLIP, Phase 3 & 1,760,461\textsuperscript{*} & 1,760,461\textsuperscript{*} & 0.199\textsuperscript{*} & 15 & 32 (4$\times$8) & 768 \\
WILSON-CoCa captioning & 178,020 & 1,759,787 & 0.02 & 20 & 32 (4$\times$8) & 1,024  \\
\bottomrule
\end{tabularx}
\end{table}
\clearpage

\section{Supplementary Method}\label{supp_method}

\subsection*{Construction of the Mayo189K training corpus}
\label{sec:Mayo189Kselection}

\paragraph{Organ-site extraction and organ classification}

Pathology reports were parsed into organ-site level diagnostic entries to enable linkage between report text and specific tissue specimens. Organ information was extracted from report metadata and mapped to predefined anatomical categories using rule-based assignment rules developed and reviewed by a board-certified pathologist. A total of 42 organ categories were defined to cover the spectrum of surgical pathology specimens encountered in the archive.

Because pathology reports frequently contain multiple specimens from different organs within a single case, organ-site level decomposition was performed before downstream processing. This allowed disease categorization and sampling to be conducted independently for each organ site rather than at the case level.

\paragraph{Construction of organ-specific disease categories}

Diagnostic terminology in routine pathology reports is highly heterogeneous, with substantial variation in nomenclature, reporting style, and diagnostic granularity. To organize this diagnostic landscape while preserving disease diversity, we constructed organ-specific disease categories using a combination of large language models, embedding-based clustering, and expert curation.

For each organ site, the primary diagnosis was extracted from pathology reports using Gemma~3~27B~\cite{gemmateam2025gemma3}. Extracted diagnoses were embedded using EmbeddingGemma~\cite{vera2025embeddinggemma} and clustered with HDBSCAN. This procedure grouped many lexical variants and closely related diagnostic expressions into common clusters while preserving distinct disease concepts.

The resulting clusters were subsequently reviewed by a board-certified pathologist independently for each organ. During this curation process, diagnostically equivalent or closely related entities were merged into unified disease categories. For example, terminology such as \textit{pituitary adenoma} and \textit{pituitary neuroendocrine tumor (PitNET)} may appear in separate clusters despite referring to the same biological entity and were therefore consolidated into a single disease category. Conversely, clusters containing diagnostically distinct entities were separated when necessary.

The resulting organ-specific disease categories were designed to facilitate diversity-aware sampling rather than to provide strict case-level diagnostic labels. Their primary purpose was to ensure broad representation of morphologic patterns and disease entities across the training corpus.

To assign all organ-site entries to curated disease categories, the category list for each organ was provided to a large language model, which classified each extracted diagnosis into the most appropriate disease category.

\paragraph{Identification of histologically unremarkable tissue}

Histologically unremarkable specimens are inconsistently described in pathology reports and are frequently not labeled explicitly as ``normal''. To ensure representation of non-neoplastic and histologically unremarkable tissue within the dataset, a large language model classifier was used to identify specimens lacking significant pathological abnormalities. These specimens were subsequently included as distinct disease categories during sampling.

\paragraph{Representative block selection}

Because lesion-level annotations were unavailable, a single representative tissue block was selected for each organ-site entry using report-derived evidence.

When only a single tissue block was available, that block was selected directly. For specimens containing multiple tissue blocks, the pathology report was searched using a large language model to identify explicit references to a key block, target block, or diagnostically representative block. When such a designation was present, the referenced block was selected.

If no explicit representative block could be identified, the block most frequently submitted for immunohistochemical (IHC) studies was selected as a surrogate representative block, based on the assumption that diagnostically important tissue regions are more likely to undergo ancillary testing.

Organ sites for which no representative block could be confidently identified were excluded from dataset construction.

\paragraph{Whole-slide image selection and scanner handling}

Only hematoxylin-and-eosin (H\&E)-stained permanent-section slides were considered for inclusion. Frozen-section slides were excluded using slide-level metadata.

Slides generated from the same representative tissue block were assumed to contain substantially equivalent tissue content. When multiple digitizations of the same specimen were available, scanner-specific selection rules were applied to ensure consistent representation while avoiding redundant inclusion of repeated scans. For specimens scanned on both Leica GT450 and Pramana SpectralHT platforms, the earliest scanned slide from the Leica GT450 series and the latest scanned slide from the Pramana SpectralHT series were retained.

\paragraph{Dataset filtering}

Starting from 1,454,318 pathology cases and 10,904,856 WSIs, pathology reports were segmented into 2,782,862 organ-site diagnostic entries. We retained only H\&E-stained permanent sections and excluded frozen sections using slide metadata. Organ categories represented by fewer than 4,000 tissue blocks were removed, leaving 36 of the 42 predefined organ categories. After filtering, the candidate pool comprised 817,205 cases, 1,543,499 organ sites, and 5,973,856 WSIs.

\paragraph{Diversity-aware sampling}

Clinical archives are strongly enriched for a small number of common organs and diagnoses. Consequently, uniform random sampling would produce a training corpus dominated by a limited set of disease entities.

To maximize morphological, diagnostic, and linguistic diversity, sampling was performed independently within each organ-specific disease category. For every organ-disease combination, a maximum of 350 WSIs was selected.

When the number of eligible specimens exceeded this limit, reports belonging to that organ-disease category were embedded using EmbeddingGemma and partitioned into $k=10$ clusters. Sampling was then performed approximately uniformly across clusters to preserve diversity in reporting language and diagnostic presentation.

This strategy intentionally reduced the influence of highly prevalent diseases while increasing representation of less common entities. Starting from an archive containing more than 10 million WSIs, the final Mayo189K training corpus comprised 189,291 WSIs spanning 838 organ-disease combinations. The resulting dataset substantially increased representation of rare organs and diagnoses relative to the source archive while preserving broad morphological coverage.

\subsection*{Report-derived supervision}
\label{sec:report_derived_supervision}

\paragraph{Overview}

Each selected organ site was paired with report-derived supervision signals
generated from the corresponding pathology report. Because routine pathology
reports frequently contain multiple specimens within a single accession,
template text, procedural comments, references, disclaimers, and physician
signatures, substantial preprocessing was required before report text could be
used for vision-language training.

Two complementary supervision signals were generated for each whole-slide
image (WSI): (1) natural-language captions describing histopathologic findings
and (2) a sparse pathology keyword vector. All report-processing and caption
generation steps were performed using Gemini~2.5 Pro~\cite{geminiteam2025gemini25}.

\paragraph{Extraction of organ-specific diagnostic text}

A single pathology report may contain diagnostic information for multiple
specimens originating from different organs. Consequently, the complete report
cannot be used directly as a textual description of an individual organ-site
WSI.

To isolate text relevant to the target organ site, Gemini~2.5~\cite{geminiteam2025gemini25} Pro was prompted
to extract only the diagnostic content corresponding to the designated specimen
part (Supplementary Prompt~\ref{prompt:report_cleaning}). The extraction procedure preserved original
diagnostic wording while removing content unrelated to tissue morphology,
including physician names, institutional information, dates, references,
technical processing details, disclaimers, and other administrative text.

When reports contained multiple labeled specimens (e.g., A, B, C), only the
text corresponding to the target specimen was retained. If no specimen labels
were present, the full report was returned. Additional organ-specific
interpretive information, including immunohistochemistry and molecular results,
was retained when explicitly linked to the target specimen.

The resulting cleaned text served as the canonical report representation for
all subsequent supervision-generation steps.

\paragraph{Generation of caption supervision}

The cleaned organ-specific report text was converted into natural-language
captions using Gemini~2.5 Pro (Supplementary Prompt~\ref{prompt:prompts_three_caption}).

The objective of caption generation was to create textual descriptions that
align closely with histomorphologic features observable on H\&E-stained whole-
slide images. Therefore, the model was instructed to describe only findings
supported by the report and visually observable on H\&E sections, while
avoiding clinical interpretation, prognostic statements, etiologic
speculation, or information not represented in tissue morphology.

For each report, three semantically equivalent but lexically diverse captions
were generated. The model was instructed to vary wording, sentence structure,
and descriptive perspective while preserving the underlying diagnostic content.
This multi-caption strategy was intended to reduce sensitivity to specific
phrasing and improve linguistic diversity during contrastive vision-language
training.

Each caption explicitly included the organ and diagnosis and preserved all
negative findings reported in the original text (e.g., ``no dysplasia'',
``absent mitotic activity'').

\paragraph{Text embedding generation}

The three generated captions were embedded independently using the Gemini
Embedding~2 model~\cite{shanbhogue2026gemini}, producing three fixed 3072-dimensional text embeddings for
each WSI. These embeddings were computed offline and used as text supervision
during model training.

\paragraph{Construction and assignment of pathology keywords}

In addition to free-text captions, we generated a structured keyword-based
supervision signal for each whole-slide image (WSI).

A board-certified pathologist manually curated a vocabulary consisting of 991 pathology keywords. The vocabulary was designed to cover the most frequently encountered diagnostic entities, morphologic findings, immunophenotypic features, and diagnostic category terms present in routine pathology reports.
The keyword list intentionally included concepts at multiple levels of
granularity, ranging from specific diagnoses (e.g., \textit{serous carcinoma})
to broader diagnostic categories (e.g., \textit{carcinoma} and
\textit{malignancy}).

Directly assigning keywords from pathology reports is challenging because
reports may express the same concept using diverse terminology, abbreviations,
or descriptive phrases. To address this issue, keyword assignment was
performed using a two-stage retrieval and verification framework.

First, both the cleaned pathology report and all 991 keywords were embedded
using the Gemini Embedding~2 model. Cosine similarity was then calculated
between the report embedding and each keyword embedding. The 50 highest-ranked
keywords were retrieved as candidate concepts for the report.

Second, the cleaned report together with the 50 candidate keywords was
provided to Gemini~2.5 Pro. The model was instructed to select only those
keywords that were explicitly supported by the report content while rejecting
semantically related but unsupported candidates. Multiple overlapping keywords
were allowed when clinically appropriate. For example, a report describing
\textit{serous carcinoma} could simultaneously activate broader concepts such
as \textit{carcinoma} and \textit{malignancy}, reflecting hierarchical
relationships that naturally exist within pathology terminology.

The final selected keywords were encoded as a 991-dimensional multi-hot vector
for each WSI. Across the Mayo189K dataset, keyword vectors contained 5--7
active terms on average.

Compared with using diagnosis labels alone, keyword-based supervision provides
additional semantic structure and allows the model to learn relationships
between morphologic findings, immunophenotypic features, diagnostic entities,
and higher-level disease concepts. Because multiple related concepts can be
assigned simultaneously, the resulting sparse representation captures
hierarchical and compositional pathology knowledge that is difficult to encode
using mutually exclusive class labels.

\subsection*{Composite Generation}
\label{sec:composite_generation}
Rather than feeding a single tile to the histopathology model, we represent each whole-slide image (WSI) as a \emph{composite}: a structured set of tiles that collectively cover the slide while prioritizing the regions most likely to carry diagnostically relevant information. The pipeline proceeds in two phases. Phase~1 is executed once per organ using healthy (normal) slides and produces a compact reference of normal tissue appearances. Phase~2 runs on every disease slide and uses that reference to select the most unusual tissue tiles for the composite.

\subsection*{Phase 1: Building the Normal Reference}
\label{sec:phase1}

\textbf{Tissue detection.}
For each normal WSI the lowest-resolution pyramid level is retrieved and converted to the Hue–Saturation–Density (HSD) color space~\cite{bejnordi2016}. HSD operates in optical-density space and is well suited to H\&E slides because it explicitly separates hematoxylin and eosin staining from slide background, correctly retaining lightly stained structures (adipose tissue, loose stroma) that simpler thresholding methods tend to discard. A pixel is labeled tissue if the smoothed cy channel of the HSD representation exceeds a fixed threshold; the resulting binary mask is used to enumerate non-overlapping $256 \times 256$\,px tile positions at the target magnification. Up to 2\,000 tile positions are randomly sampled per slide to keep memory usage bounded. Both phases share a single small Vision Transformer (ViT) pre-trained with self-supervision~\cite{caron2021emerging} and adapted for binary tile quality classification. At inference, a single forward pass yields both the 384-dimensional CLS embedding, used for abnormality scoring, and a valid-tile probability $p_{\text{valid}}$, used to discard artifacts (ink marks, air bubbles, out-of-focus regions, tissue-edge debris) at no additional computational cost.

\textbf{Embedding extraction.}
Each sampled tile is read from the WSI pyramid, resized to $64 \times 64$\,px, and passed through a Vision Transformer (ViT) that was pre-trained with DINO~\cite{caron2021emerging} self-supervised learning and subsequently fine-tuned as a two-class (valid/invalid) quality-control classifier. The 384-dimensional class token (CLS) produced by the ViT backbone is used as the tile embedding; the classification head is \emph{not} used in this phase.

\textbf{Centroid estimation.}
All embeddings collected from the normal slides of one organ are pooled and
clustered with MiniBatch $k$-Means ($k = 200$). To ensure that each centroid corresponds to a real tissue appearance rather than an uninterpretable average in embedding space, every mean centroid $\mu_c$ is replaced by its \emph{medoid}---the embedding $\mathbf{e}^*$ in cluster $c$ that minimizes the Euclidean distance to the mean:
\begin{equation}
    \mathbf{e}^*_c
    = \underset{\mathbf{e}_i \in \mathcal{C}_c}{\arg\min}
      \left\| \mathbf{e}_i - \boldsymbol{\mu}_c \right\|_2,
    \label{eq:medoid}
\end{equation}
where $\mathcal{C}_c$ denotes the set of embeddings assigned to cluster $c$.
The resulting $200 \times 384$ medoid matrix is saved as the organ's normal
reference and reused without modification in Phase~2.

\subsection*{Phase 2: Abnormality-Guided Tile Selection}
\label{sec:phase2}

\textbf{Quality filtering.}
For each disease WSI, tissue tile coordinates are obtained using the same HSD mask as in Phase~1. Every tile is processed in a \emph{single} forward pass through the quality-control ViT backbone, which yields the 384-dim CLS embedding while the ViT head simultaneously produces two-class logits from which the valid-tile probability $p_\text{valid}$ is derived via softmax. Tiles with $p_\text{valid} < 0.5$ are discarded, removing artifacts such as ink marks, air bubbles, out-of-focus regions, and tissue-edge debris before any scoring is performed.

\textbf{Abnormality scoring.}
For each remaining tile $i$ with embedding $\mathbf{e}_i$, the abnormality
score is defined as the minimum Euclidean distance to any of the $K = 200$
normal medoids $\{\mathbf{m}_k\}_{k=1}^{K}$:
\begin{equation}
    s_i = \min_{k=1,\ldots,K}
          \left\| \mathbf{e}_i - \mathbf{m}_k \right\|_2.
    \label{eq:abn_score}
\end{equation}
A high score indicates that the tile's appearance deviates substantially from all known normal tissue patterns and is therefore a candidate for pathological content. The distance matrix is computed efficiently via the identity
$\|\mathbf{a} - \mathbf{b}\|^2 = \|\mathbf{a}\|^2 + \|\mathbf{b}\|^2 - 2\,\mathbf{a}^\top\mathbf{b}$,
avoiding the construction of the full $(N \times K \times D)$ difference tensor.

\textbf{Spatial-grid selection.}
Selecting the globally top-scoring tiles would cluster the chosen patches within a single lesion region, sacrificing the spatial representativeness needed for slide-level modeling. To balance abnormality and coverage, the bounding box of all valid tissue tiles is divided into an $n_g \times n_g$ spatial grid with $n_g = \lceil\sqrt{N_\text{select}}\rceil$ (for $N_\text{select} = 60$, $n_g = 8$,
giving 64 candidate cells). Within each cell the tile with the highest abnormality score $s_i$ is selected. Cells that contain no tissue tile are filled from the globally highest-scoring unused tiles. From the resulting up to $n_g^2$ candidates, the final $N_\text{select} = 60$ tiles with the largest scores are retained. This procedure guarantees that the composite samples from all spatial zones of the tissue while still biasing strongly towards the most diagnostically informative locations.

\textbf{Output.}
The selected tiles are stored as a JSON record per slide containing:
the WSI filename; the OpenSlide pyramid level and magnification (2.5$\times$); and, for each tile, the level-0 pixel coordinate $(x, y)$, the abnormality score $s_i$, the valid-tile probability $p_{\text{valid},i}$, and the index of the nearest normal centroid $k^* = \arg\min_k \|\mathbf{e}_i - \mathbf{m}_k\|_2$.

\subsection*{Phase 3: Multi-Magnification Composite Assembly}
\label{sec:phase3}

Phase~3 reads the JSON coordinate files produced by Phase~2 and assembles the
final multi-magnification composite images. Each composite is an
$8 \times 8$ grid of $256 \times 256$\,px tiles in which each row represents
one selected tissue location across multiple magnifications.

\textbf{Composite generation.}
Phase~2 produces a pool of $N_{\mathrm{select}} = 80$ candidate tissue
locations represented by $1.25\times$ context tiles.

To generate multiple valid realizations of the composite representation from
the same WSI, a small random perturbation is added to the abnormality score of
each candidate location,
\begin{equation}
    \tilde{s}_i = s_i + \epsilon_i,
\end{equation}
where $s_i$ is the abnormality score defined in Equation~\ref{eq:abn_score} and $\epsilon_i$ is a zero-mean random noise term.

For each composite, candidate locations are ranked according to $\tilde{s}_i$, and the eight highest-ranked locations are selected. This procedure is repeated independently to generate $C=10$ composites per WSI.
The perturbation serves two purposes. First, it enables assessment of the robustness of downstream predictions to modest variations in the tile-selection process, reducing the likelihood that performance depends on a single deterministic set of selected locations. Second, during WILSON pre-training and fine-tuning, the resulting diversity of composite realizations acts as a form of data augmentation, exposing the model to multiple valid multi-magnification summaries of the same slide and thereby reducing overfitting to specific tissue locations.
Importantly, the perturbation is not intended to identify more informative regions than those selected by the original abnormality ranking. Rather, it is used to sample alternative yet diagnostically plausible composite representations from the same candidate pool.

\textbf{Representative multi-magnification region selection.}
For each selected tissue location, a $1.25\times$ context tile is extracted, covering the largest field of view and providing architectural context.
To identify representative higher-magnification regions within this context, the $1.25\times$ tile is partitioned into a coarse $4\times4$ grid and a finer $16\times16$ grid. Grid regions are characterized by their RGB color features and grouped using unsupervised clustering. Representative regions are then selected from the resulting clusters to capture the dominant visual and morphological diversity present within the context field while minimizing redundant sampling of similar tissue appearances.
Using these representative locations, three regions are extracted at $5\times$ magnification and four regions at $20\times$ magnification. Together with the original $1.25\times$ context tile, this yields eight tiles associated with the same tissue location. All tiles are resampled to $256\times256$\,px.

\textbf{Composite layout.}
Each row of the composite corresponds to one tissue location. The leftmost column contains the $1.25\times$ context tile. Columns~2--4 contain the three representative $5\times$ tiles, and columns~5--8 contain the four representative $20\times$ tiles. This arrangement preserves the relationship between broad tissue architecture and higher-magnification morphological detail while maintaining their spatial correspondence within the same context field. Each composite therefore contains eight tissue locations represented by eight tiles each, resulting in a total of 64 tiles arranged in an $8\times8$ layout.

\textbf{Output.}
Phase~3 produces $C = 10$ composite PNG images per WSI, each of size
$2048\times2048$\,px ($8\times256$ per axis). The resulting representation
encodes two complementary dimensions of information: rows correspond to
distinct tissue locations selected from the abnormality-guided candidate pool,
whereas columns correspond to contextual and representative
higher-magnification observations derived from the same location.

This structured representation provides a compact visual summary of the whole
slide, jointly encoding tissue heterogeneity, multi-scale morphology, and the
correspondence between low- and high-magnification observations within a
single image.

\begin{figure}
    \centering
    \includegraphics[width=0.75\linewidth]{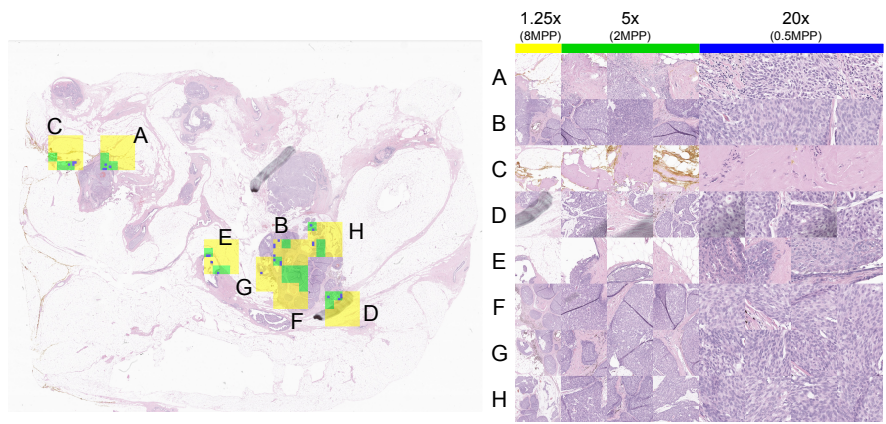}
    \caption{Example of composite image (single WSI). MayoBreast dataset.}
    \label{fig:composite_example}
\end{figure}

\subsection*{LLM Prompts 1}\label{prompt:report_cleaning}
\begin{Verbatim}[breaklines=true, breaksymbol=]
You are an expert system for extracting structured text from pathology reports.

INPUT:
- Pathology report:
% [Pathology Report]

- Target part label:
B

GOAL:
Extract ONLY the text corresponding to the specified part, preserving the original wording and formatting exactly.

INSTRUCTIONS:

1. IDENTIFY PART STRUCTURE
- Reports may contain labeled sections such as: A., B., C., etc.
- A part may appear:
  - As a single label (e.g., "B.")
  - Within a range (e.g., "A–C", "A-C", "A to C")

2. EXTRACTION RULES
- If B appears as an explicit label:
  → Extract ALL text under that label until the next label begins.

- If B falls within a label range:
  → Extract the ENTIRE text corresponding to that range.

- Include:
  - Diagnosis text
  - Description
  - Immunohistochemistry (IHC)
  - Molecular findings
  - Any interpretation clearly linked to the target part

3. INCLUDE RELATED GLOBAL SECTIONS
Include additional sections ONLY if they are clearly related to the target part, such as:
- Receptor status (ER, PR, HER2, etc.)
- IHC result blocks referencing the target part (e.g., "block B1")
- Molecular findings tied to the part

Exclude general sections if they are purely methodological or unrelated to interpretation.

4. NO LABELS CASE
If NO part labels (A, B, etc.) exist in the report:
→ Return the ENTIRE report.

5. STRICT EXCLUSIONS
Do NOT include:
- Doctor names
- Institutional or company names
- Dates
- Specimen processing details (fixation method, technical notes)
- Scanner or equipment info
- References to guidelines, journals, or protocols
- Any ".DNR" or irrelevant trailing artifacts

6. OUTPUT FORMAT
- Return ONLY the extracted text
- Do NOT add explanations, comments, or formatting
- Preserve line breaks, indentation, and wording as-is
- Merge lines only if they are clearly part of the same sentence split by line breaks

7. EDGE CASE HANDLING
- If boundaries are ambiguous, prefer including slightly MORE context rather than truncating relevant information
- If multiple mentions of the same part exist, include all of them

OUTPUT:
Return only the extracted text.
\end{Verbatim}

\subsection*{LLM Prompts 2}\label{prompt:prompts_three_caption}
\begin{Verbatim}[breaklines=true, breaksymbol=]
You are a board-certified pathologist generating training captions for H\&E-stained histopathology images. The goal is to align text with visual features for a vision-language model.
Write exactly 3 captions, each under 50 words.

Global rules:
- Each caption MUST explicitly mention the organ and the diagnosis.
- ONLY describe features that are visually observable on H\&E
- Do NOT infer clinical history, causation, prognosis, or unseen findings.
- Do NOT add any findings beyond the provided reports.
- PRESERVE NEGATIONS strictly (e.g., 'no dysplasia', 'absent mitoses'). These must remain negative and must not be rephrased as positive findings.

Diversity requirements (CRITICAL):
- Each caption must express a DISTINCT perspective.
- Use different lexical choices (including medically valid synonyms where appropriate).
- Avoid repeating the same phrase structure or keyword grouping.
- Avoid trivial rewording — ensure meaningful semantic variation.

Grounding rules:
- Integrate all assigned keywords naturally into each caption.
- Ensure the wording reflects how the feature appears microscopically.
- Prefer concrete morphologic descriptors.

Style constraints:
- Keep sentences concise, formal, and descriptive.
- Each caption must start differently.
- Avoid meta language (no 'this image shows' or 'there is').

Return ONLY valid JSON: {{"sentences": ["caption1", "caption2", "caption3"]}}

% [Pathology Report]
{report}
\end{Verbatim}

\clearpage

\begin{figure}[t]
\centering
\includegraphics[width=\textwidth]{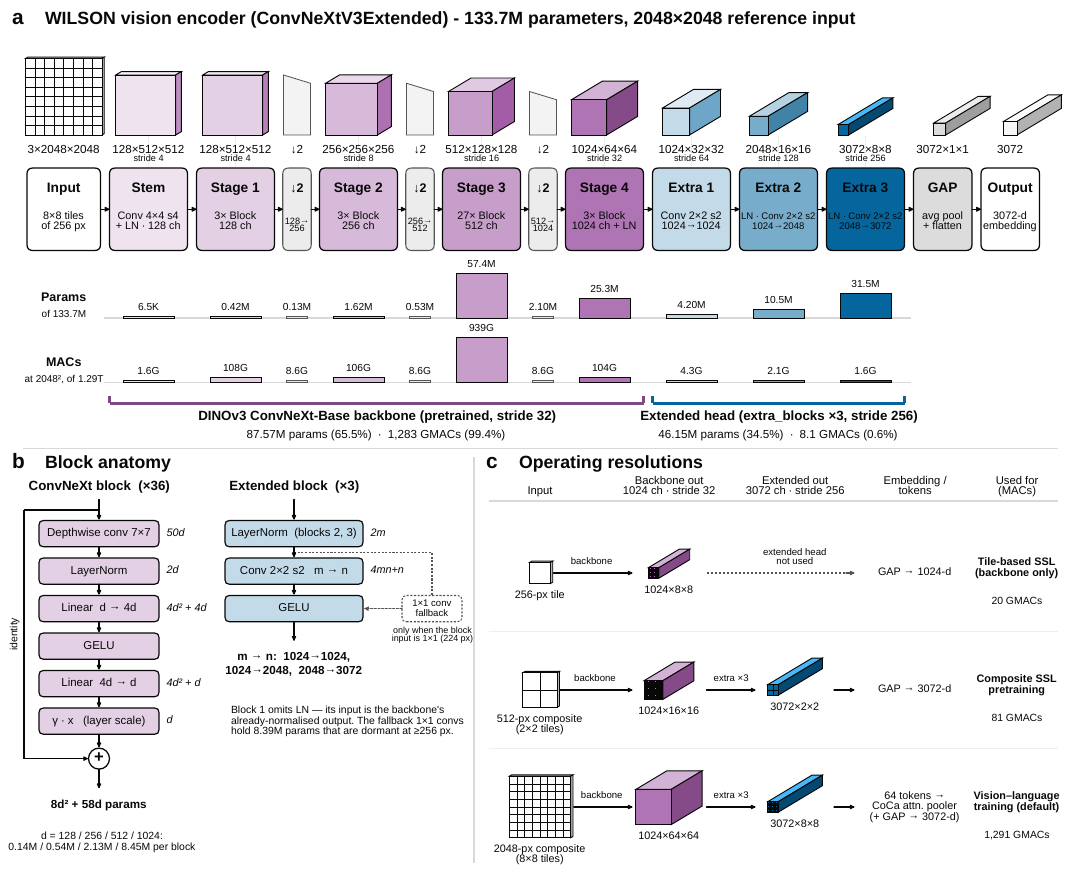}
\caption{\textbf{WILSON vision encoder (\texttt{ConvNeXtV3Extended}).}
\textbf{a}, Stage-by-stage architecture for the $2048\times2048$ reference input (an $8\times8$
grid of 256-px tiles). Tensor glyphs encode each feature map (face $\propto\log_2$ spatial size,
depth $\propto\log_2$ channels), making the progression from a broad, shallow
$128\times512\times512$ map to a narrow, deep $3072\times8\times8$ map visible at a glance. A
DINOv3-pretrained ConvNeXt-Base backbone (depths $3,3,27,3$; widths $128,256,512,1024$;
stride~32; 87.57M params) is followed by three stride-2 downsampling blocks (extra\_blocks
$\times3$; 46.15M params) that extend the width to 3072 (stride~256) and global average pooling,
for 133.7M parameters. Column-aligned bars show where parameters and compute (conv/linear MACs)
reside: Stage~3 dominates both (57.4M params, 939~GMACs), whereas the extended head holds 34.5\%
of the parameters but only 0.6\% of the compute.
\textbf{b}, Anatomy of the two block types with symbolic parameter counts; the first extended
block omits its LayerNorm (its input is already normalized by the backbone), and a $1\times1$
convolutional fallback replaces the stride-2 convolution only when a block's input is already
$1\times1$ (e.g.\ 224-px inputs).
\textbf{c}, The three operating resolutions. At 224--256~px (tile-based SSL) only the backbone is
used and its 1024-d GAP output is taken directly (20~GMACs); at $512\times512$ (composite-based
SSL pretraining, \S\ref{sec:vlm_training}) the final map is $3072\times2\times2$ (81~GMACs); at
$2048\times2048$ (composite-based vision--language training, \S\ref{sec:vlm_training}) the final
$3072\times8\times8$ map provides 64 spatial tokens to the CoCa attentional pooler
(\S\ref{sec:vlm_coca}) (1,291~GMACs).
\textbf{d}, Warm-start of the extra-block convolutions when the backbone is loaded from a
domain-specific SSL checkpoint: the $1024\times512\times2\times2$ filters of the last backbone
downsampling layer are tiled to each extra-block shape and rescaled to He-initialization
magnitude ($\mathrm{std}=\sqrt{2/\mathrm{fan\_in}}$), preserving filter directions while keeping
activations within fp16 range; biases are tiled likewise and the extra-block LayerNorms start at
$(1,0)$.}
\label{fig:wilson_arch}
\end{figure}

\begin{table}[t]
\centering
\caption{\textbf{WILSON vision encoder (\texttt{ConvNeXtV3Extended}) --- parameter and compute
breakdown by component.} Output shapes, cumulative stride and MACs are given for the
$2048\times2048$ reference input (an $8\times8$ grid of 256-px tiles), the default resolution
for composite-based vision--language training. Each ``Block'' is a standard ConvNeXt residual
block (Fig.~\ref{fig:wilson_arch}b). Parameter counts were verified by instantiating the
model. The extended head holds 34.5\% of the parameters but
accounts for only 0.6\% of the compute, because it operates on
$\leq\!32\times32$ maps; conversely Stage~3 alone accounts for
72.7\% of the MACs.
$^{\dagger}$Blocks~2--3 also instantiate a $1\times1$ convolutional fallback
(2,099,200 and
6,294,528 params, included in the totals)
that is used only when the block's input is already $1\times1$; it is dormant for all inputs
$\geq 256$~px, leaving 125,327,488 active parameters
($\approx$125.3M).
$^{\ddagger}$Multiply--accumulates of convolution and linear layers only (LayerNorm, GELU,
layer scale and pooling omitted); 1~MAC $=$ 2~FLOPs. At $512\times512$ the total is
80.7~GMACs and at $256\times256$ (backbone only)
20.1~GMACs.}
\footnotesize
\setlength{\tabcolsep}{3pt}
\begin{tabular}{@{}llccrrrr@{}}
\toprule
Component & Operation & In$\to$Out ch. & Stride & Output ($2048^2$) & Params & \% & GMACs$^{\ddagger}$ \\
\midrule
\multicolumn{8}{l}{\textit{Backbone --- DINOv3 ConvNeXt-Base (87,566,464 params, 65.5\%; 1,283 GMACs, 99.4\%)}} \\
Stem & Conv $4\!\times\!4$ s4 + LN & $3\to128$ & 4 & $128\!\times\!512\!\times\!512$ & 6,528 & 0.0 & 1.6 \\
Stage 1 & 3$\times$ Block & $128$ & 4 & $128\!\times\!512\!\times\!512$ & 415,488 & 0.3 & 108 \\
Downsample 1 & LN + Conv $2\!\times\!2$ s2 & $128\to256$ & 8 & $256\!\times\!256\!\times\!256$ & 131,584 & 0.1 & 8.6 \\
Stage 2 & 3$\times$ Block & $256$ & 8 & $256\!\times\!256\!\times\!256$ & 1,617,408 & 1.2 & 106 \\
Downsample 2 & LN + Conv $2\!\times\!2$ s2 & $256\to512$ & 16 & $512\!\times\!128\!\times\!128$ & 525,312 & 0.4 & 8.6 \\
Stage 3 & 27$\times$ Block & $512$ & 16 & $512\!\times\!128\!\times\!128$ & 57,424,896 & 42.9 & 939 \\
Downsample 3 & LN + Conv $2\!\times\!2$ s2 & $512\to1024$ & 32 & $1024\!\times\!64\!\times\!64$ & 2,099,200 & 1.6 & 8.6 \\
Stage 4 & 3$\times$ Block & $1024$ & 32 & $1024\!\times\!64\!\times\!64$ & 25,344,000 & 19.0 & 104 \\
Final norm & LayerNorm & $1024$ & 32 & $1024\!\times\!64\!\times\!64$ & 2,048 & 0.0 & -- \\
\midrule
\multicolumn{8}{l}{\textit{Extended head --- three stride-2 blocks (46,154,752 params, 34.5\%; 8.1 GMACs, 0.6\%)}} \\
Extra block 1 & Conv $2\!\times\!2$ s2 + GELU (no LN) & $1024\to1024$ & 64 & $1024\!\times\!32\!\times\!32$ & 4,195,328 & 3.1 & 4.3 \\
Extra block 2 & LN + Conv $2\!\times\!2$ s2 + GELU\,$^{\dagger}$ & $1024\to2048$ & 128 & $2048\!\times\!16\!\times\!16$ & 10,491,904 & 7.8 & 2.1 \\
Extra block 3 & LN + Conv $2\!\times\!2$ s2 + GELU\,$^{\dagger}$ & $2048\to3072$ & 256 & $3072\!\times\!8\!\times\!8$ & 31,467,520 & 23.5 & 1.6 \\
Global avg. pool & AdaptiveAvgPool2d(1), Flatten & $3072$ & 256 & $3072$ & 0 & 0.0 & -- \\
\midrule
\textbf{Total} & & & 256 & \textbf{3072-d embedding} & \textbf{133,721,216} & 100 & \textbf{1,291} \\
\bottomrule
\end{tabular}
\label{tab:wilson_arch}
\end{table}

\section{End-to-end fine-tuning on MayoTNBC}
\label{sec:supp_finetune}

\subsection*{Data and splits}

MayoTNBC comprises 508 triple-negative breast carcinoma WSIs from 508
patients, one slide per patient, stored in Philips TIFF format
(Table~\ref{tab:wsi_datasets}). Each slide carries two independent four-class
labels, histologic subtype and stromal tumor-infiltrating lymphocyte (sTIL)
grade, and each label was fine-tuned as a separate run; slides with a missing
value for the label in question were dropped from that run, so the two tasks do
not use identical slide sets. Every WSI was rendered into 40 composites, giving
a training split of 407 WSIs (16,280 composites) and a validation split of 101
WSIs (4,040 composites), an approximately 80/20 partition. The two splits were
fixed in advance as separate manifest files and verified to share no composite
directory, so they are disjoint at the slide level and, given the one-to-one
slide-to-patient mapping, at the patient level. Because the pretraining corpus
was scanned on Leica GT450 and Pramana SpectralHT systems, the Philips TIFF
slides in this cohort were acquired on different hardware, so the experiment
measures adaptation under scanner domain shift rather than in-domain adaptation
alone.

The two tasks differ substantially in class balance. Histologic subtype is
severely imbalanced: the majority class accounts for 71.3\% of training and
65.3\% of validation composites, against 5.4\% and 5.9\% for the smallest
class, a ratio of roughly 13:1, although the imbalance itself is consistent
between splits. sTIL grade is close to uniform, with every class occupying
between 20\% and 31\% of each split, but its train and validation distributions
are less well matched, the validation split carrying about 8.5 percentage
points more of the lowest grade and correspondingly less of the next. No class
weighting, resampling or stratification was applied in either run; the loss is
unweighted cross-entropy throughout. For histologic subtype this means accuracy
is inflated relative to macro-averaged F1 by the dominant class, and the
macro-averaged F1 column of Table~\ref{tab:finetune} is the more informative
measure of whether fine-tuning improved performance on the minority subtypes.

\subsection*{Optimization}

A linear classification head was attached to the encoder's 3072-dimensional
global-average-pooled output and the network was optimized with a cross-entropy
loss under a staged schedule: the backbone was held frozen for the first two
epochs so that the randomly initialized head could settle, then unfrozen for
full end-to-end fine-tuning over the remaining three, for five epochs in total.
The head and backbone used separate learning rates of $10^{-3}$ and $10^{-6}$
respectively, with weight decay $10^{-4}$, a per-GPU batch size of 4 and 8
gradient-accumulation steps across a single node of 8 H200 GPUs, giving an
effective batch of 256, under mixed-precision training. No checkpoint selection was performed: the weights from the final epoch were used for all evaluations, so no information from the validation split influenced the reported model. The validation split therefore functions as a held-out test set for the fine-tuning experiment.

\subsection*{Row-shuffle augmentation}

Training composites were augmented by online row-strip shuffling. For each
training sample, two composites belonging to the same WSI were drawn (or one
composite reused twice when only one was available), each sliced into its eight
horizontal $256\times2048$ row-strips, and eight of the resulting sixteen strips
sampled without replacement and stacked vertically into a new synthetic
$2048\times2048$ composite. Only row order and row provenance are altered, so
each row retains its intact multi-magnification column structure (one
low-magnification context tile followed by its spatially linked higher-power
views, \S\ref{sec:phase3}). Each WSI contributed 40 augmented items per epoch,
matching the number of composites rendered per WSI so that epoch length is
unchanged relative to training on the unaugmented composites directly.
Validation and all retrieval evaluation used the original, unaugmented
composites.

\subsection*{Evaluation settings}

The unit of evaluation is the individual composite rather than the slide: every
composite of a training slide enters the retrieval gallery and every composite
of a validation slide is issued as a separate query. Base Zero-shot and
FT-Zero-shot both use top-1 cosine-similarity nearest-neighbor retrieval between
L2-normalized embeddings, with the query taking the label of its single nearest
gallery neighbor, and differ only in whether the encoder has been fine-tuned;
FT-Zero-shot discards the classification head entirely, so the comparison
isolates the effect of fine-tuning on the representation. FT-classification
instead evaluates the trained linear head as an ordinary classifier. Because the
head is a learned decision boundary and nearest-neighbor retrieval is
memoryless, the two are not expected to agree numerically, and the intermediate
setting is the one that speaks to representation quality.

This protocol is deliberately simpler than the zero-shot retrieval protocol used
elsewhere in this work (\S\ref{sec:supp_caption_metrics}
, \S\ref{subse:ZSRE}), which
uses five nearest neighbors under majority voting, excludes all slides from the
query's own patient, and reports the mean and standard deviation across the ten
composites generated per slide. Here a single fixed train gallery and validation
query set are used with one neighbor, and all composites are pooled into one
evaluation, so Table~\ref{tab:finetune} reports a single value per cell without
composite-to-composite variance.

\begin{table}[h]
  \centering
  \caption{WILSON + CoCa vision-language training hyperparameters.}
  \label{tab:wilson-coca-hparams}
  \small
  \begin{tabular}{lc}
    \toprule
    Hyperparameter & Value \\
    \midrule
    Input resolution & $2048 \times 2048$ px \\
    Vision spatial tokens & $8\times8=64$, 3072-dim \\
    Vision-encoder trainable params & 133.72M \\
    Attentional-pooler queries / heads & 256 / 8 \\
    Joint embedding dim & 768 \\
    Decoder width & 768 \\
    Decoder loss weight & 1.0 \\
    Initial logit scale / bias & 10.0 / $-10.0$ \\
    Negative bank steps & 8 \\
    LR (vision / text / proj.+pooler) & $10^{-5}$ / $10^{-5}$ / $10^{-3}$ \\
    Weight decay & $10^{-4}$ \\
    Warmup & 10\% of schedule \\
    Min.\ LR fraction & 1\% of peak \\
    Gradient clipping & 1.0 \\
    Per-GPU batch size & 4 \\
    Gradient accumulation & 8 steps \\
    Effective batch size & 1{,}024 pairs \\
    Epochs & 20 \\
    Precision & bfloat16 \\
    Hardware & 4 nodes $\times$ 8 GPUs (32 total) \\
    Validation fraction (patients) & 2\% \\
    Row-shuffle repeats/epoch & 30 \\
    Max caption length & 76 tokens \\
    \bottomrule
  \end{tabular}
\end{table}

\clearpage

\section{Quantitative evaluation of report generation}
\label{sec:supp_caption_metrics}

\subsection*{Evaluation protocol}

Report generation was evaluated by comparing each model's generated caption
with a reference caption derived from the slide's clinical pathology report
(Fig.~\ref{fig:wilsonCaptioning}a). For every held-out WSI, one
multi-magnification composite was encoded by the model's vision tower and
decoded autoregressively into free text; for WILSON this is the CoCa multimodal
decoder (\S\ref{sec:vlm_coca}), and for PRISM and PRISM2 the released
report-generation heads, used off the shelf with their published checkpoints.
Decoding was deterministic for all three models (no sampling). WILSON-CoCa
decodes greedily to a maximum of 64 tokens; PRISM decodes from its BioGPT
decoder with beam search (5 beams, maximum 100 tokens); and PRISM2 decodes from
its Phi-3 decoder with beam search (5 beams, maximum 200 new tokens) in
response to the prompt ``Write a report describing this whole slide image.''.
Each model's published defaults were retained rather than harmonized, so the
generated captions differ in typical length; because BLEU-4 applies a brevity
penalty and both ROUGE variants are length-sensitive, absolute values partly
reflect these differing generation budgets. Evaluation covered three cohorts: a
held-out subset of the Mayo Clinic corpus (1{,}000 image--caption pairs sampled
with a fixed random seed shared across models, spanning multiple organs and
diagnoses), HistAI (48 pairs; Table~\ref{tab:histAI_wsi_list}) and TCGA
(50 pairs; Table~\ref{tab:tcga_wsi_list}). No evaluation slide or patient
overlapped the training corpus.

All metrics were computed against a single reference caption per slide. Mayo
Clinic slides carry three paraphrased caption variants per report
(\S\ref{sec:report_derived_supervision}); the first variant was used, so that every
cohort is scored under the same single-reference protocol as HistAI and TCGA,
for which only one reference caption per slide is available. Metrics that
natively support a reference set (BLEU-4, CIDEr, ROUGE) were therefore run in
single-reference mode throughout, which lowers their absolute values --- CIDEr
especially, since it was designed for consensus against many references --- but
keeps the three cohorts and the three models directly comparable. All metrics
are per-pair (caption, reference) scores averaged over pairs within a cohort;
error bars in Fig.~\ref{fig:caption_metrics_supp} denote one standard deviation
across pairs.

\subsection*{Metric families}

The eight metrics differ in what they treat as a match, and consequently in
their usable dynamic range on long, freely worded diagnostic text. They fall
into three families.

\noindent\textbf{Surface $n$-gram overlap.}

\noindent\textbf{BLEU-4} takes the geometric mean of 1- to 4-gram precisions against the
reference and multiplies it by a brevity penalty for captions shorter than the
reference. It rewards only exact, contiguous token sequences and gives no
credit for synonymy, paraphrase or reordering. Designed for machine
translation, it approaches zero on long single-reference captions unless the
wording closely mirrors the reference, so absolute BLEU-4 values here index
lexical mimicry rather than diagnostic correctness. \textbf{ROUGE-1} is the
$F$-measure of unigram precision and recall after stemming; it credits shared
vocabulary irrespective of order, making it the most forgiving of the overlap
metrics but blind to grammar and phrase structure. \textbf{ROUGE-L} scores the
longest common subsequence of the two captions --- tokens shared in the same
relative order but not necessarily adjacent --- again as an $F$-measure after
stemming, and therefore sits between ROUGE-1's bag-of-words leniency and
BLEU-4's strict adjacency.

\noindent\textbf{Weighted $n$-gram overlap.} 

\noindent\textbf{CIDEr} represents both captions as TF--IDF-weighted 1- to 4-gram
vectors, so that corpus-frequent $n$-grams are down-weighted and distinctive,
content-bearing $n$-grams dominate, and averages the cosine similarity across
$n$-gram orders. CIDEr was designed as a corpus-level consensus measure against
many references per image; the per-pair, single-reference computation used here
is a substantially weaker signal and sits near zero even for clinically
accurate captions.

\noindent\textbf{Embedding-based similarity.}

\noindent\textbf{BERTScore} embeds every token of both captions with a pretrained
contextual language model (RoBERTa-large) and greedily matches each token to
its most cosine-similar counterpart in the other caption: precision averages
the best-match similarity over generated tokens, recall over reference tokens,
and F1 is their harmonic mean. Because matching occurs in embedding space,
reworded clinical phrasing and synonyms that BLEU, ROUGE and CIDEr treat as
total mismatches are credited, which is why BERTScore values occupy a
compressed high range ($\approx$0.83--0.89 here) even when two captions share
few literal tokens; differences of 0.01--0.02 in this range are therefore not
negligible. Within BERTScore, recall is the component most sensitive to whether
the generated caption covers the reference's diagnostic content, and precision
the component most sensitive to unsupported additions. \textbf{CLIPScore}
projects the slide's image embedding --- PRISM's tile-pooled slide
representation, or WILSON's own vision-tower encoding of the composite --- and
the generated caption's text embedding into that model's shared contrastive
space and takes their cosine similarity. Unlike every other metric, it never
inspects the reference: it measures whether the caption is consistent with what
the model itself encodes from the slide, making it an image-grounding rather
than a report-similarity measure. Because each model's contrastive space has
its own learned scale and temperature, a CLIPScore is interpretable only for
ranking or comparing WSIs \emph{within} one model; the raw cross-model
difference is not evidence that one model is better grounded than another.
CLIPScore is not reported for PRISM2, which exposes no general-purpose text
encoder capable of embedding an arbitrary string independently of an image.

\subsection*{Aggregate score}

Because the eight metrics have incompatible scales and effective ranges, the
summary column of Table~\ref{tab:caption_metrics} (``Avg'') reports, for each
model and cohort, the mean of its metric values after min--max normalization
within each metric and cohort, rescaled to 0--100. Normalization is performed
across the models compared in that cohort, so the aggregate is a relative
ranking within a cohort and carries no absolute meaning and no cross-cohort
comparability. CLIPScore is excluded from the aggregate, and from bolding,
because it is not comparable across models and is undefined for PRISM2.

\subsection*{Interpretation caveats}

\begin{itemize}
  \item \textbf{Floor effects are expected, not diagnostic of failure.}
        BLEU-4 and CIDEr were designed for short, multi-reference captioning
        benchmarks; under the single-reference protocol used here all three
        models sit near zero on long diagnostic prose, and the ordering between
        them is more informative than the magnitudes.
  \item \textbf{Ceiling compression in BERTScore.} All three models exceed 0.83
        on every cohort because clinical text shares substantial contextual
        structure regardless of diagnostic agreement; BERTScore separates
        models weakly and should be read alongside the overlap metrics rather
        than in place of them.
  \item \textbf{No metric here credits paraphrase directly.} BLEU-4, CIDEr and
        both ROUGE variants match surface tokens only, so a clinically
        equivalent caption worded differently from the reference is penalized;
        BERTScore is the only reference-based metric in this set that tolerates
        rewording, which is why it should not be read in isolation from the
        overlap metrics or from the qualitative comparisons.
  \item \textbf{CLIPScore is not a cross-model comparison.} See above.
  \item \textbf{Reference provenance differs between cohorts.} Mayo Clinic
        references were summarized from pathology reports with Gemini~2.5 Pro,
        TCGA references from the machine-readable reports of Kefeli et
        al.~\cite{kefeli2024tcga} with DeepSeek V4 Pro, and HistAI
        references are the diagnostic conclusion field supplied with the
        dataset's own metadata, used verbatim without LLM summarization.
        Because the overlap metrics reward literal token agreement,
        differences in reference style and length affect absolute values
        independently of caption quality, so metric values are comparable
        across models within a cohort but not across cohorts.
  \item \textbf{Domain adaptation is not controlled.} WILSON-CoCa's decoder and
        text tower were trained on Mayo Clinic captions generated by
        Gemini~2.5 Pro, whereas PRISM and PRISM2 were evaluated as released.
        The Mayo Clinic column is therefore the only one in which the reference
        style matches the style WILSON-CoCa learned to produce, and reflects
        both architecture and adaptation to the target reporting domain; the
        HistAI and TCGA columns, where no model was adapted to the reference
        style, are the more architecture-like comparisons.
  \item \textbf{Lexical similarity is not diagnostic correctness.} A caption
        can score well by reproducing reporting boilerplate while naming the
        wrong entity, and can score poorly while being clinically correct, as
        the qualitative comparisons in Fig.~\ref{fig:caption_comparison_supp}
        illustrate.
\end{itemize}
\end{appendices}


\end{document}